\documentclass{aa}  

\usepackage{graphicx}
\usepackage{txfonts}

\usepackage{graphicx}	
\usepackage{amsmath}	
\usepackage{amssymb}	
\usepackage[dvipsnames]{xcolor} 
\usepackage{nicefrac,xfrac}
\usepackage{verbatim}
\usepackage{hyperref}
\usepackage{booktabs, threeparttable}
\usepackage{gensymb}

\hypersetup{
    colorlinks=true,
    citecolor=blue,
    linkcolor=blue,
    filecolor=magenta,      
    urlcolor=cyan,
    pdftitle={Overleaf Example},
    pdfpagemode=FullScreen,
    }

\newcommand{\kms}{km~s$^{-1}$}
\newcommand{\Msun}{M$_\odot$}
\newcommand{\Reff}{$R_{\rm e}$}
\newcommand{\deVau}{de Vaucouleurs}
\newcommand{\sersic}{S\'{e}rsic}
\newcommand{\logM}{$\log M_{200}$}
\newcommand{\ML}{$M_\star/L_{\rm B}$}

\begin{document} 

\defcitealias{Taylor+17}{T17}
\defcitealias{Hughes+21}{H21}

   \title{Shapes of dark matter haloes with discrete globular cluster dynamics:  The example of NGC~5128 (Centaurus~A)}

   \author{T. Ver\v{s}i\v{c}
          \inst{1,2}\fnmsep\thanks{Corresponding author}
          \and
          M. Rejkuba\inst{2}
          \and
          M. Arnaboldi\inst{2}
          \and
          O. Gerhard\inst{3}
          \and
          C. Pulsoni\inst{3}
          \and
          L. M. Valenzuela\inst{4}
          \and
          J. Hartke\inst{5,6}
          \and
          L. L. Watkins\inst{7}
          \and
          G. van de Ven\inst{1}
          \and
          S. Thater\inst{1}
          }

   \institute{Department of Astrophysics, University of Vienna, T\"urkenschanzstra{\ss}e 17, 1180 Vienna, Austria\\
        \email{tadeja.versic@univie.ac.at}
         \and
        European Southern Observatory, Karl-Schwarzschild-Str. 2, 85748 Garching, Germany
         \and
         Max-Planck-Institut für extraterrestrische Physik, Giessenbachstraße, 85748 Garching, Germany
         \and
         Universitäts-Sternwarte, Fakultät für Physik, Ludwig-Maximilians-Universität München, Scheinerstr. 1, 81679 München, Germany
         \and
         Finnish Centre for Astronomy with ESO (FINCA), University of Turku, FI-20014 Turku, Finland
         \and
         Tuorla Observatory, Department of Physics and Astronomy, FI-20014 University of Turku, Finland
         \and
        AURA for the European Space Agency, ESA Office, Space Telescope Science Institute, 3700 San Martin Drive, Baltimore MD 21218, USA}

   \date{Received Month XX, 2023; accepted Month XX, 202X}
   \titlerunning{Dark matter halo shape of NGC~5128}
   \authorrunning{Ver\v{s}i\v{c} et al.}
 
  \abstract
  {Within the $\Lambda$CDM cosmology, dark matter haloes are expected to deviate from spherical symmetry. Different shapes of galactic haloes reflect their hosts' environment, mass assembly history and the nature of dark matter. Constraining the halo shapes at large galactocentric distances is challenging due to the low density of luminous tracers. The well-studied massive early-type galaxy NGC~5128, also known as Centaurus A (CenA), has a large number of radial velocity measurements for globular clusters (GCs) and planetary nebulae (PNe) extending over a vast area of its extended low surface brightness stellar halo. } 
   {In this work we aim to determine the deviation from spherical symmetry of the dark matter halo of CenA at 5~\Reff\ using its GCs as kinematic tracers of the gravitational potential.}
   {We investigated the largest photometric catalogue of GC candidates to accurately characterise the spatial distribution of the relaxed population of GCs. 
   To investigate the presence of non-relaxed structures in the kinematic catalogue of GCs we used the relaxed point-symmetric velocity field as determined by the host's PNe population. 
   We used anisotropic Jeans modelling under axisymmetric assumptions together with the Gaussian likelihood and GCs as discrete tracers.
   The gravitational potential is generated by flattened stellar and dark matter distributions.
   We leveraged different orbital properties of the blue and red GCs, such as rotation and velocity anisotropy, to model both populations separately.
   By minimising $\chi^2$ we iteratively find the best fit parameters.}
   {We find that discrete kinematics of the GCs are consistent with being drawn from an underlying relaxed velocity field determined from PNe. 
   The best-fit parameters of the gravitational potential recovered from the blue and red GCs separately agree well and we use them to compute the final results: $M_{200} = 1.86^{1.61}_{-0.69}\times 10^{12}$~\Msun, \ML $= 2.98^{+0.96}_{-0.78}$ and the flattening $q_{\rm DM} = 1.45^{+0.78}_{-0.53}$.
   Both GC populations show mild rotation, with red having a slightly stronger rotational signature and radially biased orbits, and blue GCs preferring negative velocity anisotropy.}
   {An oblate or a spherical dark matter halo of NGC~5128 is strongly disfavoured by our modelling.}

   \keywords{galaxies: elliptical and lenticular, cD -- galaxies: individual: NGC~5128 -- 
                galaxies: kinematics and dynamics -- 
   galaxies: halos 
               }

   \maketitle
%

\section{Introduction}\label{ch:Intro}

Galaxies are gravitationally-bound objects made of luminous and non-luminous matter and dynamical modelling enables us to map their spatial distributions \citep{BinneyAndTremaine08}. 
Understanding the intrinsic shape of galaxies is essential to prevent biased conclusions when interpreting results derived from such modelling.
The shape is quantified by the axis ratios of a 3D ellipsoid: $p = b/a$ and $q = c/a$, where $a>b>c$ are semi-major, intermediate and minor axes; $a>b=c$ is considered prolate, $a=b>c$ is considered oblate and, otherwise, the shape is triaxial.
Observations of the luminous components of early-type galaxies (ETGs) have revealed that stars show oblate, prolate, or even triaxial distributions \citep{Weijmans+14_ATLAS3D_intrinsic_shapes, Foster+16_SLUGGS_shapes, Foster+17_SAMI_intrinsic_shapes_of_stellar_ETGs, Pulsoni+18, Ene+18_Kinematics_MASSIVE_survey_shapes}. 
These studies used photometry and kinematics to constrain the shapes and found a link between the amount of rotational support and the degree of triaxiality, where more rotation favouring oblate distributions, but not excluding triaxial shapes.
Further, more oblate-axisymmetric distribution of stars can be embedded within a triaxial dark matter (DM) halo \citep[see for example the results from the TNG simulations by ][]{Pulsoni+20_ETGs_sims_photo_kin}. 
The Milky Way and its satellites are the only galaxies where we have detailed 6D phase space information for a sufficiently-large sample of tracers with which we can reconstruct the shape of the total mass distribution, luminous and dark matter \citep[e.g.][]{Sesar+11, Debattista+13, Posti+Helmi19, Iorio+Belokurov19, Naidu+21, Han+22}. 
Beyond the Local Group, we rely on projected quantities only. 
Because of the random orientation of galaxies, we observe these intrinsic distributions projected along a line-of-sight (LOS), which makes the recovery of intrinsic shape very challenging, especially for ETGs with non-regular rotation \citep{Krajnovic+05_regular_data_inclination_not_constrained, vdBosch_vdVen09_triaxiality}.

While great advances have been made in recent years on the intrinsic shapes of the central stellar components of ETGs with dynamical modelling \citep{Jin+20_Manga_shapes_ETGs, Santucci+22_Sami_galaxies_mass_distribution, Thater+23_Atlas3D_shapes}, much less is known about the haloes, where non-luminous matter dominates.
%
%
%
In this domain, cosmological simulations have been the source of insight into the intrinsic shapes of the dark matter haloes \citep[e.g.\ ][]{Cole+Lacey96, Allgood+06, Prada+19, Chua2019_halo_shapes}.
Simulations based on $\Lambda$ Cold Dark Matter ($\Lambda$CDM) cosmology, which agrees best with the observed properties of galaxies, have been used most extensively.
N-body simulations in the $\Lambda$CDM framework have shown strong radial dependence on the halo shape, with a preference for prolate haloes in the inner regions $r<0.1r_{\rm vir}$, that become increasingly triaxial in the outer regions \cite[e.g.][]{Bilin+05_shapes_orientation_of_DM_haloes, Hayashi+07_DM_shapes}.
The inclusion of baryonic physics acts to make the haloes rounder at all radii and the exact prescription of the baryonic feedback impacts the triaxiality of galaxies \citep[e.g.][]{Kazantzidis+04_gas_cooling_on_DM_shapes, Bryan+13_Shapes_of_DM_haloes_effect_of_baryons,Tenneti+15_baryon_fb_on_shapes, Chua2019_halo_shapes}.
Studies of N-body dissipationless simulations found a strong dependence of the halo shape with the different cosmological parameters and the dark matter fraction \citep{Allgood+06}. 
For example, haloes in warm DM cosmology (WDM) are found to be rounder than cold DM (CDM) haloes \cite[e.g.][]{Bullock02_WDM_CDM_shapes_proceedings, Mayer+2002_halo_shapes_WDM_FDM_CDM}.
Directly observed measurements of the dark matter halo shape can thus provide additional constraints for cosmological simulations.

The inner regions of luminous galaxies are typically dominated by baryons, while in their outer regions, dark matter becomes the dominant contributor to the mass distribution \cite[e.g.][]{Deason+12_DM_fraction_5Re, Cappellari+13_dynamics_mass_DM_fraction}.
In the case of ETGs, the DM fraction at 5~\Reff\ has been found to be 30-80\% \cite[and references therein]{Gerhard13_DM_shapes_masses_proceeding}. 
\citet{Harris+20_DM_fraction_5Re} found a median DM fraction of 80-90\% for galaxies with $M_\star >10^{10}$~\Msun.
For the recovery of galaxy DM distribution and shapes, we need luminous tracers in the regions where DM dominates.
Specific features such as stellar streams and polar rings have been used to put observational constraints on the 3D distribution of the dark matter \cite[e.g.][]{Iodice+03_polar_ring_galaxies_shape_constraints, Varghese+11_Streams_DM_morphology, Khoperskov+14, Pearson+15_tidal_stream_morphology_constraints_on_DM_shape, Leung+19}.
Weak lensing provides constraints on the projected flattening of DM haloes on a statistical basis \citep[e.g.][]{Robison+23_Shapes_of_DM_weak_lensing}.

Recovery of the mass distribution and de-projection of galaxy orientation of gas-poor ETGs can best be done through dynamical modelling of the luminous tracers that are distributed throughout the galaxy \cite[e.g.][]{Cappellari+08}.
In the inner few \Reff, stars can be used as kinematic tracers but, further out, other halo tracers are needed such as globular clusters (GCs) and planetary nebulae (PNe).
Globular clusters are old dense stellar systems populating the haloes of every large galaxy \cite[and references therein]{Brodie_Strader06_GC_review}. 
There is empirical evidence that the total mass in GCs is a better tracer of the total halo mass of its host galaxy rather than its stellar mass \citep{Blakeslee97,Spitler+Forbes09,Georgiev+10,Hudson+14,Harris+17}. 
This makes GCs suitable tracers of the overall mass of the galaxies. 
Thanks to their compact sizes and relatively high luminosities, the photometry of individual GCs can be observed in galaxies up to a few 100~Mpc away from us with the \textit{Hubble Space Telescope (HST)} \citep[e.g.][]{Jordan+09_ACSVCS_GCs, Jordan+15_ACSFCS_GCs, Harris23_GCS_BrightestGalaxies} and in nearby groups and clusters from the ground-based surveys \cite[e.g.][]{Taylor+16_SCABS_dataset}.
Spectroscopic observations needed for radial velocities (RVs) are however limited to a few nearby groups and clusters \citep{Brodie+14_SLUGGS_survey_Description, Avinash+22_Fornax_kin}.

NGC~5128 is the closest ETG to us classified as giant Elliptical or S0 \citep{Harris_G10_GiantBeneath} and has clear evidence of a recent accretion history or possibly a major merger \citep[][and references therein]{Wang+20_CenAmergerhistory}.
It has a stellar mass of $1.6 \times 10^{11}$\Msun\ \citep{Hui95_PNe_kin_sys_vel_mass_shape} and is estimated to host up to 2000 GCs \citep[e.g][]{Harris+06_HST_obs_of_CenA_GCs}.
It hosts a radio source Centaurus A and, for simplicity, we will refer to the galaxy as CenA.
Its proximity has enabled detailed studies of the resolved red giant branch (RGB) stars out to more than 25~\Reff\ (140~kpc) \citep{Rejkuba+22_RGBs}, providing constraints on the density distribution of the stellar halo.
Spectroscopic observations of a large population of PNe \citep{Peng+04_PNe_dyna_outer_halo, Walsh+15} and GCs \citep{Peng+04_GCS_CenA_kin_formation_metal, Woodley+10_Kin_and_mass, Hughes+23_Kinematics_dyn_estimate} over a vast halo area provide a necessary sample of kinematic tracers for dynamical modelling that we are exploiting in this work.
Several literature studies took advantage of these kinematic samples to measure the spherical mass distribution of this galaxy \citep{Peng+04_PNe_dyna_outer_halo, Woodley+07_PNe_GC_mass_estimates, Woodley+10_Kin_and_mass, Hughes+21_dyn_modeling, Dumont+23_CJAM_modelling_Spherical_mass}.
Observational evidence of the PNe kinematics \citep{Hui95_PNe_kin_sys_vel_mass_shape, Peng+04_PNe_dyna_outer_halo, Pulsoni+18} however supports strong triaxiality of the CenA stellar component and the gravitational potential.
In this work, we aim to test the validity of the assumption of spherical symmetry of the CenA DM halo with axisymetric dynamical modelling.

We utilise the large photometric and kinematics datasets of GCs recently compiled to determine the shape of DM halo of NGC~5128. 
Throughout the paper, we adopt the following properties: distance  D=3.8~Mpc \citep{Harris_G+10_distance_to_CenA}\footnote{$1$~arcmin corresponds to $1.1$~kpc at the distance of CenA.}, stellar effective radius of \Reff$ = 5$~arcmin and position angle (PA) of $35\degree$ \citep{Dufour79+stellar_profile} and systemic velocity $v_{\rm sys} = 541$~\kms\ \citep{Hui95_PNe_kin_sys_vel_mass_shape}.
In Sect.~\ref{ch:data} we introduce the photometric and kinematic datasets used in this work. 
In Sect.~\ref{ch:tracer_density_GCs} we present a careful analysis of both datasets.
The aim of the analysis is to determine the population of GCs that is consistent with being dynamically relaxed in the gravitational potential of CenA.
We describe the dynamical modelling tool used in this work in Sect.~\ref{ch:Modeling} and the statistical approach to find the best-fit parameters in Sect.~\ref{ch:statistical_method}.
We present the results of the modelling in Sect.~\ref{ch:Results} and discuss the findings in Sect.~\ref{ch:Discussion}.


\section{Data}\label{ch:data}

\subsection{Photometric dataset}\label{ch:DGCs_Literature_Photo_Catalogues}

In this work, we use the photometric catalogue from \citet{Taylor+17} (hereafter \citetalias{Taylor+17}) with multi-band \emph{u'g'r'i'z'} photometry
based on the observational programme aiming to characterise the baryon structures around CenA with the \emph{Dark Energy Camera} \citep[DECam,][]{Taylor+16_SCABS_dataset}.
\citetalias{Taylor+17} used 500,000 detected sources within $\sim$140~kpc of CenA and covering $\sim$21~deg$^2$.
To remove fore- and background contamination, the authors 
compared the distribution of their candidates with the 
kinematically-confirmed GCs from \cite{Woodley+10_Ages_metallicities} in the $(u'-r')_0$ - $(r'-z')_0$ plane.
The GCs are well separated from Milky Way stars at the red end of the colour-colour plane, however, the confusion increases for blue GCs (see their Fig.2).
The final dataset of $\sim 3200$ candidate GCs is therefore still affected by contamination, which we discuss in more detail in Sect~\ref{ch:tracer_density_GCs}.

We excluded GCs that did not have photometry in all of the photometric bands and selected those within $5<R<130$~arcmin, to remove the impact of the inner polar dust ring and observational footprint.
Additionally, we corrected the online catalogue for extinction.
We use the online tool IRSA\footnote{\url{https://irsa.ipac.caltech.edu/applications/DUST/}} to obtain the reddening correction $E(B-V)_{\rm SFD}$\footnote{SFD stands for the dust corrections from \citet{Schlegel+98_SFD}.} at the location of each GC.
We combine these with $A_b/E(B-V)_{\rm SFD}$ values from \cite[see Table 6]{Schlafly&Finkbeiner11_reddening} for $R_V=3.1$ in relevant SDSS bandpasses ($b$) to compute the colour excess $A_b$. 
Throughout the paper we also correct all the photometry for a distance modulus value $\mu = 27.88$ \citep{Harris_G+10_distance_to_CenA} to bring it to the absolute plane.
The reddening and distance corrected magnitude is then $b_0 = b 
- \mu
- A_b^{\rm SFD}$.

\subsection{Kinematic catalogue}\label{ch:Data_kinematics}

In this work, we use a heterogeneous compilation of literature kinematic measurements of GCs as presented in 
\citet[hereafter \citetalias{Hughes+21}]{Hughes+21}
(see their Sect.~3 and Table~4).
Together with the positions and velocities the authors also report their $g$ and $r$-band photometry.
Out of the whole sample of 557 GCs only 482 have both photometric and kinematic information and we use this subset of GCs in the following analysis and dynamical modelling.
This is the largest sample of kinematic measurements for CenA GCs in the literature and the authors carefully matched the different datasets.
The 482 GCs span a magnitude range $-11<r_0<-6$~mag and have a median radial extent of 8~arcmin (9~kpc, 1.6~\Reff) with 95\% of the GCs within 25~arcmin (27~kpc, 5~\Reff).
The median velocity uncertainty is 33~\kms.

\begin{figure}
    \centering
    \includegraphics[width=\columnwidth]{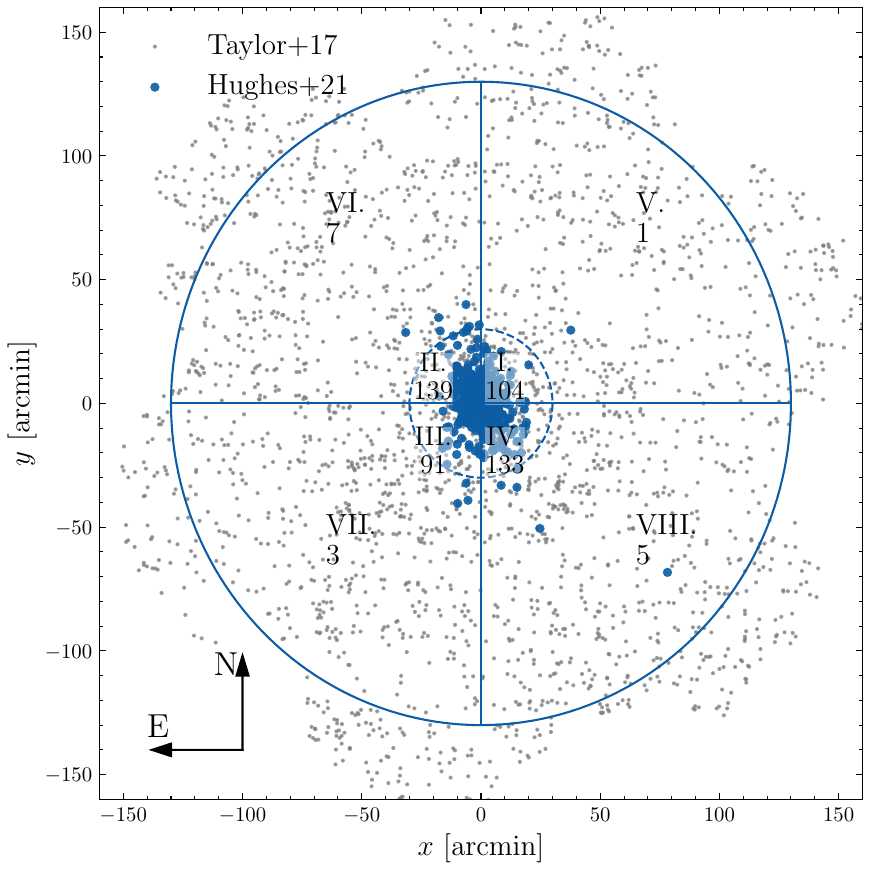}
    \caption{The spatial distribution of the photometric GC candidates from \citetalias{Taylor+17} in small grey points and of the kinematic dataset from \citetalias{Hughes+21} shown with larger blue dots.
    The dashed and solid blue circles mark the radius of 30~arcmin and 130~arcmin respectively. 
    The blue lines delimit the quadrants used in Sect.~\ref{ch:photmetric_completness} to determine the magnitude-selected subsample of GC candidates from \citetalias{Taylor+17} that are consistent with a relaxed axisymmetric distribution.
    The Roman numerals indicate the number of the respective quadrant and the Arabic numbers below the number of GCs with kinematic measurements in the corresponding quadrant.
}
    \label{fig:Data}
\end{figure}

The spatial distributions of both the kinematic and photometric catalogues used in this work are shown in Fig.~\ref{fig:Data}.
The photometric dataset extends much further than the kinematic.
The photometric dataset at $R>130$~arcmin clearly shows the observational footprint.
The kinematic sample is concentrated in the inner 30~arcmin ($\sim$6~\Reff), with a few GCs located in the outer halo.
It also shows an elongation in the NE - SW direction, with a larger number of GCs in the II. and IV. quadrants compared with I. and III. in the inner halo.
The same trend is also seen in the outer halo, but the number of GCs there is significantly lower.

\section{Selection of the relaxed tracer component}\label{ch:tracer_density_GCs}

The dynamical models used in this work (see Sect.~\ref{ch:JAM_formalism}) assume that the kinematic tracers are relaxed and fully phase-mixed in the gravitational potential. 
Additionally, they require an accurate characterization of their spatial distribution, the tracer density profile.

The presence of shells, streams and a young population of halo stars all support a recent merger scenario of CenA \citep[e.g.][]{Malin+83_CenA_Shells, Rejkuba+05_HST_observations_of_halo_stars, Rejkuba+11_ages_of_halo_stars_CenA, PISCeS_Crnojevic16}.
Recently \cite{Wang+20_CenAmergerhistory} used an N-body simulation that can reproduce several observed features of the galaxy.
The major merger scenario dates the merger back to $\sim$6~Gyrs, with the final fusion of the progenitors $\sim$2~Gyrs ago. 
\cite{Woodley+10_Ages_metallicities} found that CenA's globular cluster system (GCS) is dominated by old GCs ($\tau>8$Gyrs) and has evidence of intermediate age ($5<\tau<8$Gyrs) and young ($\tau<5$Gyrs) populations.
The metallicities and ages of the GCs and halo stars suggest that few minor mergers took place after the bulk of the galaxy was assembled $\sim$10~Gyrs ago \citep[e.g.][]{Beasley+08_met_distribution_of_GCs_in_CenA, Woodley+10_Ages_metallicities}.
Such timescales suggest that a population of GCs may not be completely phase-mixed in the CenA potential and necessitates a careful photometric and kinematic analysis to determine the relaxed population of GCs for the modelling. 

In this section, we present our analysis of the photometric GC candidates from \citetalias{Taylor+17} to identify the relaxed population.
Furthermore, we analysed the kinematic GC catalogue from \citetalias{Hughes+21} and used the smooth component of the PN velocity field as determined by \cite{Pulsoni+18} to investigate the possible presence of prominent non-relaxed features.
Lastly, we present how we build the tracer density profile used for dynamical modelling. 

\subsection{Analysis of the photometric dataset}\label{ch:photmetric_completness}

To build up the tracer density profile of the relaxed GC population we followed the two-step approach presented in \citet{Versic+23_SLUGGS}. 
In the first step, we select a subsample of GCs that is complete up to a limiting magnitude and radius and in the second step we fit an analytic function to the complete subsample.
A hallmark of a clean and dynamically-relaxed sample of GCs is that the magnitude distribution should closely follow a lognormal globular cluster luminosity function (GCLF) up to the completeness limit of the survey and the spatial distribution should be close to point symmetric.

\subsubsection*{Comparison between observed and theoretical GCLF}

Deep observations of GCs in the nearby Universe have revealed that the magnitude distribution of GCs around galaxies follows a nearly universal shape \citep[e.g.][]{Jordan+06_GCLF, Rejkuba12_GCLF, Harris+14_GCLF}.
The Gaussian shape of the GCLF is characterised by a peak turnover magnitude (TOM$_r$) and a spread ($\sigma_r$).
The values of these parameters mildly depend on the mass of the host galaxy \citep{Villegas+10_ACSFCS_ACSVCS_GCLF} and, for CenA, we adopt literature values ${\rm TOM}_r =$-7.36 and  $\sigma_r =$1.2.
In the first step of the analysis, we leverage the universality of the GCLF to further remove contaminants from the \citetalias{Taylor+17} catalogue of GC candidates.

\begin{figure}
    \centering
    \includegraphics[width=\columnwidth/6*5]{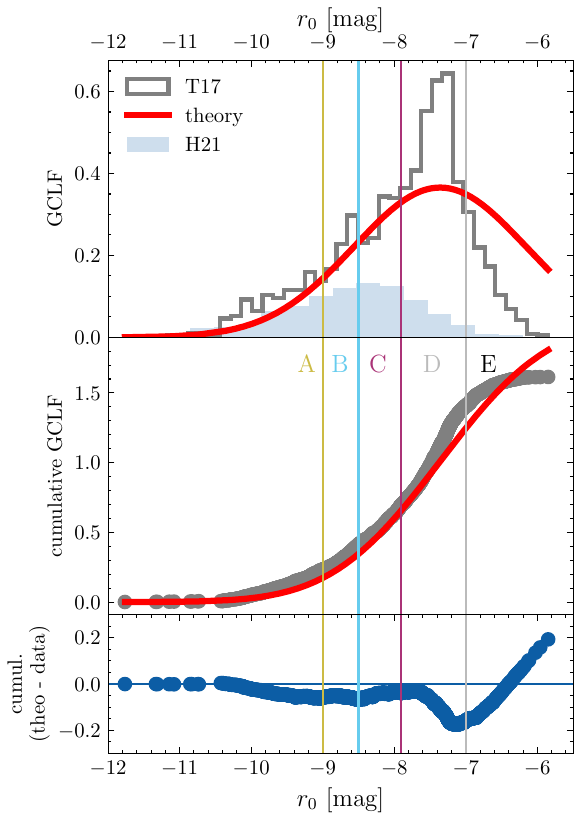}
    \caption{Comparison between the observed GCLF from the \citetalias{Taylor+17} candidate GCs in grey and theoretical GCLF in red.
    The top panel shows the GCLF and the middle panel the cumulative GCLF, which allows for more precise determination of the magnitude subsamples.
    Both theoretical GCLFs are scaled, to match the total observed number of GCs in regions A-C.
    The bottom panel is based on the cumulative GCLF from the middle panel and it shows the difference between the theoretical and observed distributions.
    The coloured vertical lines delineate different magnitude subsamples labelled with A-E in corresponding colours.
    The light blue histogram on the top panel shows the magnitude distribution of the kinematic \citetalias{Hughes+21} dataset and is arbitrarily scaled for clarity.}
    \label{fig:GCLF}
\end{figure}

To remove the effect of the central polar dust ring and the observational footprint of DECAM (as seen in Fig.~\ref{fig:Data}), we additionally limited the sample to $4{\rm arcmin} \leq R \leq 130 $~arcmin (4.4~kpc $\leq R \leq$ 143~kpc).
By comparing the theoretical GCLF and the observed magnitude distribution, we identified 5 different photometric subsamples, which we labelled A-E. 
The observed GCLF corrected for distance and extinction is compared with the theoretical GCLF.
Fig.~\ref{fig:GCLF} shows the two GCLFs together with the delineation of the different magnitude subsamples.
The limiting magnitudes that correspond to the different subsamples are shown in \autoref{tab:Mag_subsamples}.

\begin{table}
    \caption{Overview of the different magnitude ranges defined in this work. 
    The subsample name is shown in Col.~1. The $r_0$ band magnitude range of the corresponding subsample is in Col.~2 with the number of candidate GCs from the \citetalias{Taylor+17} dataset in Col.~3 and theoretical prediction based on the work by  \citetalias{Hughes+21} in Col.~4.
    The ratio between the number of theoretically predicted GCs and candidate GCs ($N_{\rm GC}^{\rm theo}/N_{\rm GC}$) is shown in Col.~5.
    The last column shows the fraction of red GCs based on the separating magnitude of $(u-z) = 2.6$.}
    \label{tab:Mag_subsamples}
\begin{threeparttable}
\small
\begin{tabular}{lccccc}
\hline \hline
 & Mag cut & $N_{\rm GC}$ & $N_{\rm GC}^{\rm theo}$ & $N_{\rm GC}^{\rm theo}/N_{\rm GC}$ & $f_r^\dagger$ \\
\hline
all & -- & 2341 & $1450\pm160$ & $0.62\pm0.07$ & 0.48 \\
A & $r_0 <-9$ &  338 & $124\pm13$ & $0.37\pm0.04$ & 0.36 \\
B & $-9\leq r_0 <-8.5$ & 259 & $123\pm13$ & $0.48\pm0.05$ & 0.41 \\
C$^\star$  & $-8.5\leq r_0 <-7.9$ & 412 & $225\pm24$ & $0.55\pm0.06$ & 0.53 \\
D & $-7.9\leq r_0 <-7$ & 1029 & $422\pm46$ & $0.41\pm0.05$ & 0.56 \\
E$^\ddagger$ & $r_0\geq -7$& 303 & $554\pm61$ & $1.83\pm0.2$ & 0.35 \\
\hline
\end{tabular}
    \begin{tablenotes}[para, flushleft]\footnotesize
\item[$\dagger$] Fraction of GCs with $(u-z)_0 \geq 2.6$. \\
\item[$\star$] The most relaxed sample of GCs. \\
\item[$\ddagger$] This subsample is strongly affected by the magnitude completeness limit of the survey.
\end{tablenotes}
\end{threeparttable}
\end{table}
Column~2 shows the number of candidate GCs in the corresponding magnitude range and in Col.~3 we computed the estimated number of GCs based on the lognormal GCLF and a total of  $N_{\rm CG} = 1450\pm160$ as estimated by \citetalias{Hughes+21}.

Based on the strong deviations from the theoretical GCLF, we refer to all sources with $r_0 > -7.9$ (D, E) as \textbf{faint} and $r_0 \leq -7.9$ (A-C) as \textbf{bright} in the following discussion.
The brightest GC bin A is expected to have the lowest contamination fraction, with the contamination expected to increase in fainter magnitude bins; a trend that is observed in subsamples D and E.
Subsample D shows the largest excess of observed GCs wrt the theoretical GCLF, and the completeness limit of the survey is evident in subsample E, with the number of GCs dropping rapidly towards the fainter magnitudes.
The faint sources are progressively affected by the survey's incompleteness and contamination, therefore we excluded them from the subsequent analysis.

On the bright end, we observe less strong deviations between the observed and theoretical GCLF.
The brightest magnitude bin A has potential contributions from ultra-compact dwarfs (UCDs) \citep{Dumont+22_UCDs_CenA}, that have distinct origins from GCs \citep{Hacegan+05_ACSVCS_GCs_UCDs, Mieske+08_UCDs_GC_difference, Voggel+20_UCDs_GCs_differences_CenA}, therefore they could bias the GC tracer density profile. 
There is an unexpected overdensity of sources in subsample B that deviates from the theoretical GCLF.
Identifying whether this is a real feature or the result of the GC candidate selection process is beyond the scope of this work.

Among all of the samples, subsample C shows the best agreement between the theoretical and observed GC numbers.
The values in Cols.~4 and 5 of \autoref{tab:Mag_subsamples} should be taken as guidelines and not as exact contamination fractions, since different studies estimate the total number of GCs in CenA from 1000 to 2000 GCs \citep{Harris+06_HST_obs_of_CenA_GCs, Harris_G+10_distance_to_CenA}, which would alter the exact values of the fraction, although the general trends would remain.

\subsubsection*{Point symmetry of magnitude selected subsamples}  

To further investigate whether the bright GC subsamples are consistent with being dynamically relaxed, we investigate the azimuthal variation of GCs in 8 quadrants as shown in Fig.~\ref{fig:Data}.
Guided by the findings of \cite{Rejkuba+22_RGBs}, who reported a transition between the inner and outer halo of RGB stars at $R\sim 30$~arcmin, we separately inspected the number of sources in 4  quadrants of the inner ($5<R\leq 30$~arcmin) and outer ($30<R \leq 130$~arcmin) regions.
These regions are shown in Fig.~\ref{fig:Data} as solid and dashed lines for the outer and inner halo regions, respectively.

\begin{figure}
    \centering
    \includegraphics[width=\columnwidth/5*4]{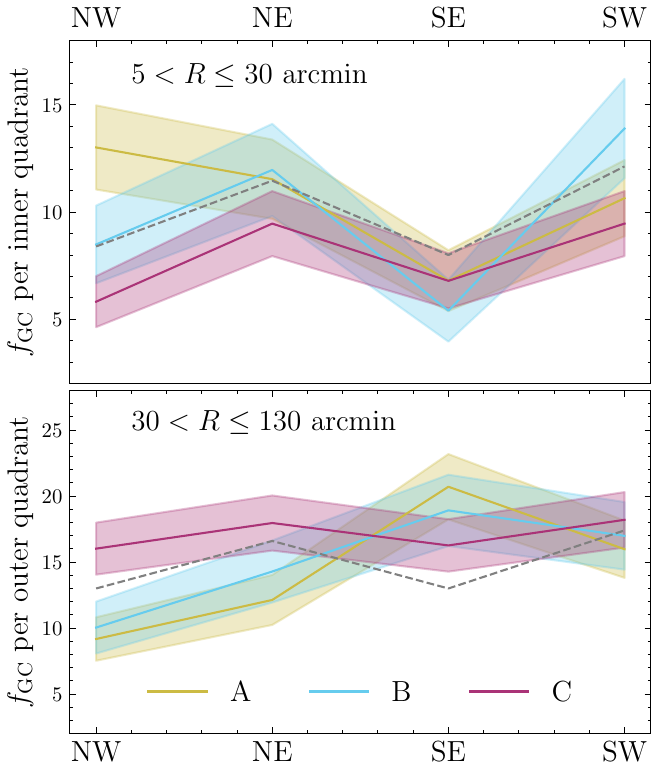}
    \caption{The fraction of candidate GCs in each of the 8 angular bins as shown in Fig.~\ref{fig:Data}.
    We focus on the subsamples A, B and C in yellow, light blue and purple, respectively as defined in Fig.~\ref{fig:GCLF} and \autoref{tab:Mag_subsamples}. 
    The top panel shows the azimuthal variation of the fraction of GCs in the inner halo (5<R<30~arcmin) and the bottom panel the outer halo (30<R<130~arcmin).
    The shaded regions show the associated 1$\sigma$ uncertainties.
    The stellar position angle is along NE - SW.
    To guide the eye, the grey dashed lines show an azimuthal variation for a mock ellipsoidal distribution of points aligned along the CenA stellar PA.
    }
    \label{fig:quadrant_analysis}
\end{figure}

The azimuthal variation of the fraction of GCs for the subsamples A, B and C in each of the 8 quadrants is shown in Fig.~\ref{fig:quadrant_analysis}.
A flattened point-symmetric distribution of tracers will show an increased fraction along the major axis of the flattened system.
For comparison, we also sampled positions of mock GCs from an elliptical distribution aligned with the stellar PA. 
The azimuthal variation of such a point-symmetric distribution is shown with a grey dashed line.
The coloured lines show the fraction of GCs compared with the subsample as $f_{\rm GC} = N_{i, \rm quad.} / N_{i}~[\%]$, where $i$ stands for A, B and C, coloured yellow, light blue and purple, respectively.
The shaded regions highlight the corresponding 1$\sigma$ uncertainty region.
The orientation of the stellar major axis is along the NE - SW direction and the magnitude-selected population C agrees best with the same alignment both in the inner and outer halo regions.
This subsample exhibits a higher than 1$\sigma$ difference in the fraction of the sources along the major and minor axis.

Subsample A shows deviations from point symmetry both in the inner and outer halo indicating a strong presence of non-relaxed features.
Subsample B does show signs of point symmetry in the inner regions, however in the outer regions the behaviour mimics that of subsample A.
There has been recent observational evidence supporting the presence of a younger and brighter population of GCs that might not be fully relaxed, which might explain the trends we observe \citep{Voggel+20_UCDs_GCs_differences_CenA, Dumont+22_UCDs_CenA}.
However, as is evident from \autoref{tab:Mag_subsamples}, there might still be substantial contamination and it is beyond the scope of this work to identify whether the source of the deviation from point symmetry in subsamples A and B is intrinsic or observational.
Therefore, we adopt subsample C as the most relaxed sample of candidate GCs from \citetalias{Taylor+17} and use it to determine the tracer density profile.

\subsection{PN vs GC velocity field}\label{ch:Kin_substructure}

Radial velocity measurements in the \citetalias{Hughes+21} dataset offer the possibility to investigate whether this sample of kinematic tracers is phase-mixed in the galactic potential.
Non-relaxed structures, originating in a recent interaction, appear as overdensities in phase space or perturbations of a point-symmetric velocity field \citep[e.g.][]{Coccato+13_Kin_signatures_of_accretion_events}. 
In this section, we investigated whether the kinematic sample of GCs is consistent with a smooth, point-symmetric velocity field or whether there are substructures that sample a broader range of magnitudes.
We investigate both the whole kinematic sample as well as the magnitude selected subsample C.
The magnitude distribution of the kinematic sample is shown in Fig.~\ref{fig:GCLF} as a light blue histogram on the top panel.
Compared with the photometric dataset it samples a narrower range in magnitude, due to a brighter limiting magnitude of the spectroscopic observations.

To investigate the presence of such features, we take advantage of the results from \citet{Pulsoni+18}. 
The authors used the PN catalogue from \citet{Walsh+15} as tracers of the stellar halo kinematics to determine the minimum relaxed velocity and velocity dispersion fields.
Non-relaxed structures are higher-level perturbations of the relaxed field and the authors found evidence for deviation from a point-symmetric velocity field at all radii.
In this work, we consider the point-symmetric model fitted to the PNe as a good proxy for the mean velocity field of the relaxed population and thus use it for comparison with the GCs. 

We carry out this analysis by using the entire sample of GCs with velocity measurements as well as separating them by colour.
A more detailed discussion on the colour distribution follows in the next section and here we adopt a colour cut of  $(g-r)_0$ = 0.74 to separate the blue and red GCs. 
We used this combination of photometric bands because it was available for the whole kinematic sample and a two-component Gaussian mixture model to determine the colour cut.

To compare the GC velocities with the PN velocity field, we used the radial variations of the harmonic expansion coefficients to generate the moments of the smooth PN LOS velocity field at the position of each GC.
The coefficients were computed in fixed annular bins and we used a linear interpolator to compute the values of these parameters for arbitrary radii within the extent of the literature measurements.
Using higher-order interpolators resulted in unrealistic radial variations of the coefficients.

\begin{figure}
    \includegraphics[width=\columnwidth]{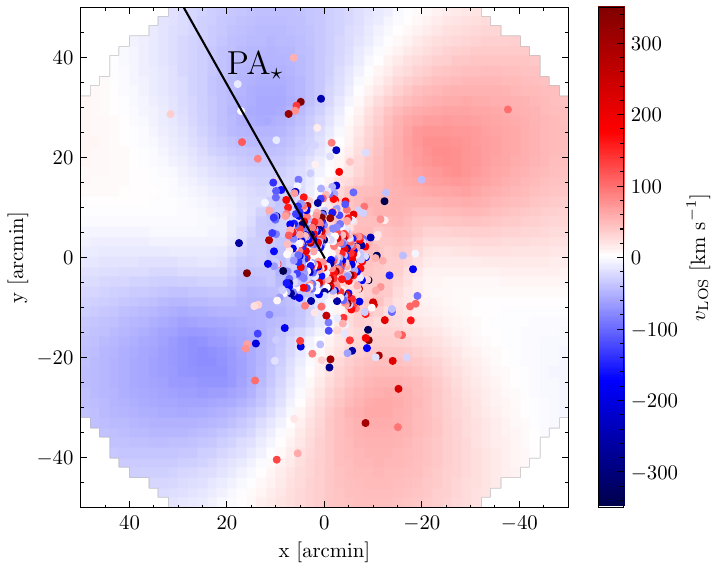}
      \caption{The mean velocity field of the PNe in the background obtained with harmonic expansion \citep{Pulsoni+18} in comparison with the individual kinematic measurements of GCs.
      The photometric position angle of the stellar component is shown with a black solid line \citep{Dufour79+stellar_profile}. 
              }
         \label{fig:Velocity_field_GCs}
   \end{figure}

The point-symmetric PN velocity field from \citet{Pulsoni+18} is plotted in Fig.~\ref{fig:Velocity_field_GCs} in the background and the individual GC velocity measurements used in this work are shown with coloured points.
The stellar photometric PA and the kinematic PA show clear signs of misalignment which is a signature of triaxiality.
However the amplitude of rotation $\sim 50$~\kms\ is much smaller than the velocity dispersion $\sim150$~\kms, which supports the axisymmetry assumption in the dynamical modelling approach we use in this work (Sect.~\ref{ch:JAM_formalism}). 

For each GC, we use the interpolated moments of the velocity field to compute the local LOS velocity field (LOSVD).
Then, we determine the probability of observing a GC with a velocity as extreme as measured assuming a local Gaussian LOSVD.
The latter assumption is reasonable for discrete halo tracers and has been used in the literature \citep[e.g.][]{Zhu+16}.
By investigating the spatial distribution of outliers greater than 1, 2 and 3$\sigma$ from the PN velocity field we find no discernible overdensity or trend with magnitude or colour out to $R\sim 40$~kpc.
We find that the individual GCs are consistent with being randomly drawn from a local Gaussian LOSVD, where the mean and $\sigma$ are informed by the PN velocity field and convolved with the measurement error.
The fractions of the GCs with velocities more extreme than 1, 2 and 3$\sigma$ from the local LOSVD are given in \autoref{tab:outliers_vel_field}.
We repeated the analysis for the whole kinematic sample and again for the kinematic sample in the magnitude range of subsample C.
In both cases, we redid the analysis for the whole subsample as well as the colour-selected subsample.
Given that the local Gaussian LOSVD is informed by the point-symmetric velocity and dispersion field our investigation justifies dynamical modelling of the whole kinematics sample of GC based on the assumption of dynamical equilibrium.

\begin{table}
\caption{Percentage and absolute number of GCs outside of the n$\times\sigma$ range of the PN velocity field. 1$\sigma\ [32\%]$  2$\sigma\ [5\%]$  3$\sigma\ [0.2\%]$}\label{tab:outliers_vel_field}
\footnotesize
\begin{tabular}{c|ccc|ccc|c}
\hline\hline
        sub sample & \multicolumn{3}{c}{Per. of n$\times\sigma$ outliers} & \multicolumn{3}{c}{No. of n$\times\sigma$ outliers} &  $N_{\rm GC, tot}$ \\
         & 1$\sigma$ & 2$\sigma$ & 3$\sigma$ & 1$\sigma$ & 2$\sigma$ & 3$\sigma$ & \\
        \hline
    All & 30.3 & 4.1 & 0.2 & 146 & 20 & 1 & 482 \\
    All blue & 30.4 & 3.5 & 0.4 & 78 & 9 & 1 & 257 \\
    All red & 30.2 & 4.9 & 0.0 & 68 & 11 & 0 & 225 \\
    \hline
    C & 34.7 & 6.6 & 0.8 & 42 & 8 & 1 & 121 \\
    C blue & 32.3 & 3.1 & 1.5 & 21 & 2 & 1 & 65 \\
    C red & 37.5 & 10.7 & 0.0 & 21 & 6 & 0 & 56 \\
    \hline
\end{tabular}
\end{table}

These results expand the findings of \cite{Peng+04_GCS_CenA_kin_formation_metal} to $R\sim40$~kpc.
These authors determined a significant correlation between the GC and PN velocity field, finding that the red GCs correlate more strongly with the velocity field of the PN.
They also observed on the mild rotational signature both around the major and minor axis for the GCs.
A rotation signature was also identified in \cite{Hughes+23_Kinematics_dyn_estimate}, where the velocity field was investigated with a constant kinematic position angle and a constant rotation.
The authors similarly find a stronger rotation signature for red GCs than blue GCs further motivating the modelling of the blue and red populations separately.

%
\subsection{Summary of our input observables and the use of GCS subsample C for the dynamical modelling}
As illustrated in the previous sections we use photometric subsample C to build the tracer density profile to use in dynamical modelling throughout this work.
Among the bright GCs only the subsample C shows a point-symmetric spatial distribution which is expected for a relaxed system.
While all bright GCs are expected to be photometrically complete and clean, subsamples A and B nonetheless show deviation from point symmetry. 
This could be due to the observational biases and presence of non-GC contaminants, such as Milky Way stars, or other observational-related effects, which we cannot ascertain here.
%
Concerning the kinematics, such magnitude selection is not necessary. 
In \autoref{tab:outliers_vel_field} we demonstrated that the entire kinematic dataset of GCs is consistent with the relaxed point symmetric velocity field from PNe, as velocity measurements are independent of photometric biases.
This holds both for the entire sample as well as for the analogously selected kinematic subsample C. 
%
Therefore, in the next steps, we use the spatial distributions of \citetalias{Taylor+17} subsample C, to constrain the spatial distribution of our tracers, and the kinematics available for the \citetalias{Hughes+21} GCs sample.

\subsection{Tracer density profile}\label{ch:tracer_density_profile}

In the dynamical modelling used in this work (see Sect.~\ref{ch:JAM_formalism}), GCs are used as massless tracers of the gravitational potential. 
An essential input for the modelling is an accurate description of their spatial distribution. 
In this section, we will build the tracer density profile of the relaxed GCs subsample (the subsample C) in the form of multi-Gaussian expansion \citep[MGE, ][]{Emsellem94_MGE}.
We also considered separately GCs according to their colour as independent tracer populations as described in the following section.

\subsubsection{Color distribution}\label{ch:color_bimodality}

%
%

GCSs of all large galaxies show bi- or multi-modal colour distributions \citep{Brodie_Strader06_GC_review}.
The different peaks in the colour distribution correspond to different populations of GCs with distinct origins, hence different spatial distribution and kinematics.
The latter has been used in the literature in the dynamical modelling of massive ETGs \citep[e.g.][]{Romanowsky+09_NGC1407_modelling, Zhu+16}
and can introduce biases in the modelling results if not accounted for \citep{Versic+23_SLUGGS}.
Here we determine the colour separation for the magnitude-selected population C.

To separate the GCs into red and blue populations we use Gaussian mixture model algorithm implemented in Python package \textit{sklearn}. 
As already pointed out in \citetalias{Taylor+17} the largest separation is in  $(u-z)_0$ colour, owing to the large wavelength baseline, therefore, we use this combination of magnitudes.
We determine the separation of the 2 components as the colour where the probability of belonging to either component is equal.
We use bootstrap with replacements to provide uncertainties and determine the separation to be  $(u-z)_0 =  2.69_{-0.03}^{+0.04}$.
Reassuringly, a similar separating colour was found by \cite{Hughes+21}.

\subsubsection{Position angle and flattening }\label{ch:MTD_GCs_PA_q}

In the next sections, we investigate the position angle and projected flattening ($q' = 1-e = b/a $)\footnote{Deprojected, intrinsic flattening, is denoted with $q$ throughout this paper. } of the red and blue GCs.
To measure the PA and flattening, we used the algorithm from \emph{mgefit}\footnote{A Python implementation of the MGE algorithm by \citet{Cappellari02_mgefit}} as summarised in \citet{Versic+23_SLUGGS}.
    
We binned the GCs in radial annuli and computed the PA and $q$ in each.
We use the different bins to compute the mean parameters and their uncertainties.
We explored different binning strategies and found that they have a negligible effect on the final result.
For different magnitude regions, we compared the recovered flattening in the 4 radial regions informed by the overdensity in GC candidates observed in \citetalias{Taylor+17} ($R/R_e\leq9$ and $9<R/R_e\leq17$), as well as the radial regions noted in the RGBs by \cite{Rejkuba+22_RGBs}   ($R\leq30$~kpc and $R>30$~kpc).

For subpopulation C, we observe that the flattening and PA are consistent within all the radial regions.
A comparison of the red and blue GCs shows that the red GCs prefer a more flattened distribution with $q'_{\rm GCS,\ red} = 0.72_{-0.08}^{+0.08}$ and blue a more spherical distribution with $q'_{\rm GCS,\ blue} = 0.82_{-0.02}^{+0.01}$.
Within the uncertainties, the PAs of the two components agree with the photometric position angle of the main body stellar component of the galaxy at 35$\degree$: ${\rm PA}_{\rm GCS,\ red} = 46.0_{-10.5}^{+10.5}$ and ${\rm PA}_{\rm GCS,\ blue} = 30.2_{-8.0}^{+8.0}$.

To correct for the PA and rotate the sample to a common PA, we use the literature results from the main stellar component of the galaxy ${\rm PA} = 35^\circ$.
This ensures the potential generated by the stars and the kinematics of the GCs are aligned with the same PA as required by the axisymmetric dynamical models used in this work.

\subsubsection{Fitting an analytic tracer density profile}

Based on the colour, and as discussed in Sect.~\ref{ch:color_bimodality}, we split the GCs in subsample C into blue and red.
To build the tracer density profile that will be an input to the dynamical modelling, we first fit the observed number density profile of the GCs with an analytic function and then decompose it in a series of Gaussians utilising \emph{mgefit}.
Due to the low number of sources in red and blue populations, we find that the most robust way to determine the MGE components is to construct the binned profile, fit an analytic function to that binned profile, and then decompose the fitted function into a series of Gaussians.

Assuming the PA of 35$^\circ$, we align the major axis of the GCS with that of the main body of the galaxy.
To account for the flattened spatial distribution of GCs, we assume the density is constant on confocal ellipses $R_{\rm ell}^2 = x^2 + (y/q)^2$ and use this to make a binned profile with 16 logarithmically-spaced bins.
Bins with fewer than 3 GCs are removed because of low statistical significance.
We explored different binning strategies to ensure the best-fit profile is minimally affected by it.

We found that a sum of a power law $\Sigma_P$ and \sersic\  $\Sigma_S$ profile \citep{Sersic+68_book_reference} can best represent the data:
\begin{equation}
    \Sigma_P (R_{\rm ell}) = A_i\ \Bigg[\bigg( \frac{R_{\rm ell}}{1 {\rm arcsec}}\bigg)^{-\alpha_i}\Bigg] \label{eq:PowerLaw_profile}
\end{equation}
\begin{equation}
    \Sigma_S (R_{\rm ell}) = \Sigma_{e,i} \exp\Bigg[ - b_n \bigg[\bigg( \frac{R_{\rm ell}}{R_{e,i}}\bigg)^{\nicefrac{1}{n_i}} - 1\bigg] \Bigg] \label{eq:Sersic_profile}
\end{equation}
where $R_{e,i}$ is the effective radius of the $i$th subpopulation, $\Sigma_{e,i}$ is the corresponding number density, $n_i$ the \sersic\ index and we use approximation of $b_n = 2n - 1/3 + 4/405 \times 1/n + 46/25515 \times 1/n^2$ \citep{Ciotti_Bertin99_Sersic_bn_expansion}.
The power law was computed in units of arcsec, $A_i$ is its scale and $\alpha_i$ is the slope.
$\Sigma_S$ and $\Sigma_P$ were computed in units of $[N_{\rm GC}~{\rm arcsec}^{-2}]$.

To find the best-fit parameters we use a Gaussian likelihood and uninformative (flat) priors on the modelled parameters ($\Sigma_{e,i},\ R_{e,i},\ n_i,\ A_i$ and $\alpha_i$), the priors are given in \autoref{ch:App_TD_MGEs}.
The slope and \sersic\ index were sampled in linear space and $\Sigma_{e,i},\ R_{e,i} $ and $ A_i$ in logarithmic space as they can span multiple orders of magnitude and such transformation results in more efficient sampling of the parameter space.

We sample the posterior, find the maximum likelihood region and explore the parameter space with \textsc{emcee}, an affine-invariant Markov Chain Monte Carlo (MCMC) ensemble sampler \citep{emcee}.
The posteriors on each of the 5 free parameters are unimodal with the weakest constraints on the \sersic\ index.
The chains were allowed to evolve for 12000 steps to ensure convergence of the best-fit parameters, the first 1000 were discarded as burn-in.

\begin{table}
    \centering
    \caption{Best fit results of the analytic tracer density profile for the blue and red GCs of the magnitude selected subsample C, corresponding to Fig.~\ref{fig:TD_GC_fit}. The results show 16th and 84th percentile uncertainties.}
    \label{tab:Best_fit_TD_parameters}
    \footnotesize
    \begin{tabular}{cccccc}
    \hline \hline
        $i$ & $\Sigma_{e,i}$ & $R_{e,i}$ & $n_i$ & $A_i$ & $\alpha_i$ \\
         & $[N_{\rm GC}~{\rm arcsec}^{-2}]$ & [arcmin] &  & $[N_{\rm GC}~{\rm arcsec}^{-2}]$ &  \\
        \hline
        blue & $10.01^{+6.61}_{-7.06}$ & $4.44^{+2.27}_{-2.65}$ & $0.67^{+0.28}_{-0.29}$ & $0.004^{+0.008}_{-0.003}$ & $1.08^{+0.12}_{-0.13}$ \\
        red & $6.95^{+4.95}_{-5.1}$ & $3.65^{+4.17}_{-2.37}$ & $0.7^{+0.28}_{-0.3}$ & $0.005^{+0.012}_{-0.004}$ & $1.1^{+0.14}_{-0.15}$ \\
        \hline
    \end{tabular}
\end{table}

The best-fit parameters for different subpopulations are shown in \autoref{tab:Best_fit_TD_parameters}.
The \sersic\ profile can better capture the behaviour of the density profile in the inner regions and the single power law in the outer regions.
Because of the choice of the fitting function, the interpretation of the individual best-fit parameters is not straightforward, especially in reference to literature studies.
We note that the blue population shows a slightly shallower \sersic\ and power law slope and a larger $R_e$ compared to the red population, which is consistent with the literature.

\begin{figure}[h]
    \centering
    \includegraphics[width=0.9\columnwidth]{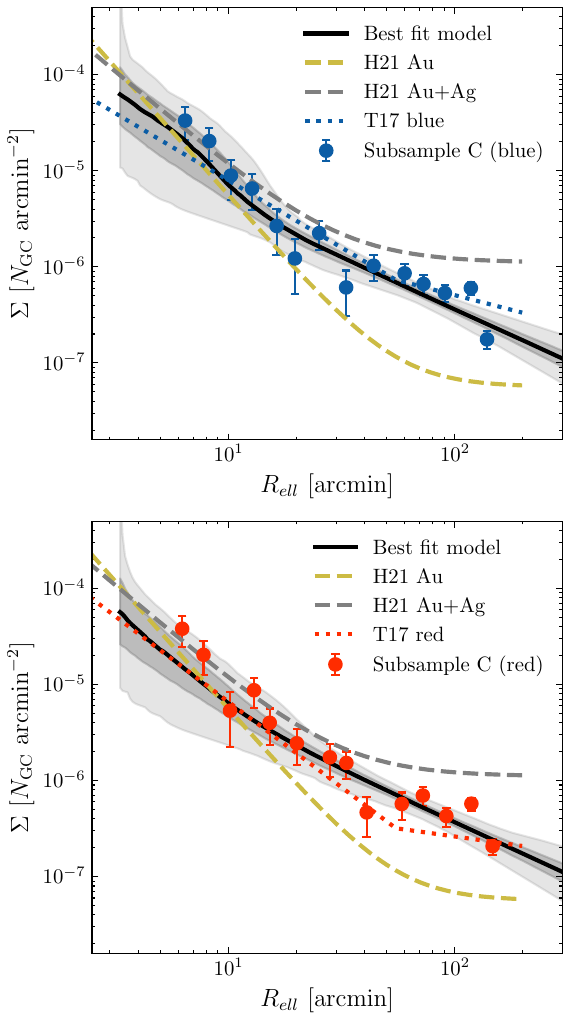}
    \caption{
    Tracer density distribution for population C of GCs. The top figure shows the blue and the bottom the red population.
    The blue and red points with associated uncertainties indicate the binned profile computed for the respective sub-population GCs. 
    A solid black line shows the best-fit model to the data, and dark and light grey shadings show 1 and 3$\sigma$ uncertainties computed as the 16th and 99.7th percentiles from the posterior. 
    The dotted blue and red lines show the best-fit double power law from \citetalias{Taylor+17} for each population.
    \citetalias{Hughes+21} used different datasets to determine three different stages of the clean photometric datasets.
    They refer to the best sample as golden (Au) and the next best sample as the silver (Ag) sample. 
    The dashed yellow line shows the best-fit model of the golden sample from \citetalias{Hughes+21} and the dashed grey line the silver and golden combined. 
    The two literature results are arbitrarily scaled for comparison.
    }
    \label{fig:TD_GC_fit}
\end{figure}

The resulting profile for the blue and red C subpopulations are shown as a solid black line in Fig.~\ref{fig:TD_GC_fit}, with dark and light shaded regions corresponding to 1 and 3 $\sigma$ uncertainties.
The top panel shows the blue GC data as blue points and the associated best-fit model.
The bottom shows the same for the red population.
Arbitrarily-scaled dashed and dotted profiles show the literature results.

The dashed \citetalias{Hughes+21} results were computed without the colour selection.
Our profile in the inner regions ($R\leq 30$~arcmin) is steeper than their silver sample (Au+Ag) for both the blue and red populations.
The blue population profile shows a slope more consistent with their higher purity (gold) sample, which is shown as a dashed orange line.
In the outer regions ($R > 30$~arcmin) our best-fit profile drops off faster than the literature results.
The latter might not be surprising as they impose a constant background.

The dotted lines show the broken double power-law fits presented in \citetalias{Taylor+17}.
In the inner region, we find a consistent slope for the red GCs and a steeper slope for the blue GCs.
While in the outer region, our best-fit profile shows a steeper slope compared with \citetalias{Taylor+17} profiles.

In general, our profiles lie within the literature results and show the main characteristics that have been observed for the GCS of CenA.
The differences stem from the different methods of selecting a high-purity and high-completeness sample of GCs and different fitting functions.

\subsubsection{MGE decomposition}

In the last step, we constructed an MGE representation of the number density profiles separately for the blue and red GCs of subsample C (as defined in Fig.~\ref{fig:GCLF}).
Such a representation is necessary for the dynamical modelling used in this work.
We drew samples from the posterior of the parameters from the tracer density fit and used them to evaluate the profile at each radius.
We used the \emph{mgefit} implementation of the MGE algorithm \citep{Cappellari02_mgefit} on such profiles that mimic observational noise following the procedure in \citet{Versic+23_SLUGGS}.
Lastly, we also accounted for the flattening of both the blue and red GC systems.

\section{Dynamical modelling with GCs}\label{ch:Modeling}

To model the GCs, we use the axisymmetric Jeans Anisotropic MGE (JAM) formalism introduced by \citet{Cappellari+08}. 
In this section, we introduce the models and then describe how we use them to reproduce the GC kinematics and spatial distribution, and, thus, derive the mass distribution of CenA.
The JAM formalism has been extensively used to model ETGs with stars, gas, PNe and GCs as tracers \citep[e.g.][]{Cappellari+13_dynamics_mass_DM_fraction, Poci+17_JAM_modelling, Bellstedt+18}.

\subsection{CJAM}\label{ch:JAM_formalism}

The JAM models are available assuming both spherical symmetry and cylindrical ($R,z,\phi$) axisymmetry. 
As we need to reproduce the flattening of the tracer density distribution and the stellar and dark matter distributions, we use the axisymmetric models here. 
We use the C implementation of the JAM models \citep[CJAM][]{Watkins+13}.

In principle, these models have two free angular parameters that describe the orientation of the galaxy: the PA and the inclination angle $i$. 
The PA defines the location of the photometric major axis. 
In Sect.~\ref{ch:MTD_GCs_PA_q}, we discussed the PA, and we henceforth keep this fixed. 
The main rotation of the galaxy is along the photometric major axis, which supports our choice of symmetry axis coinciding with the minor axis of the stellar and GC distribution.

The inclination angle is used to connect the intrinsic 3D quantities from the model with the observables (density profiles and velocity components) projected on the sky, where $i = 0\degree$ for a face-on system and $i = 90\degree$ for an edge-on system.
We will keep this parameter free in some of our models.


Both the tracer density and mass density distributions are incorporated into the model through MGE components.
By scaling the stellar surface brightness with mass-to-light ratio \ML\footnote{In this work, we used the stellar surface brightness in the photometric B band. The mass-to-light ratio is in solar units [\Msun L$_\odot^{-1}$] and for clarity of the text we omit it.} we derive the stellar mass density and add it together with the assumed dark matter density to define the gravitational potential:
\begin{equation}
    \rho_{\rm tot} = (M_\star /L_{\rm B}) \times \nu_{\rm B} + \rho_{\rm DM} 
\end{equation}
where $\nu_{\rm B}$ is the MGE representation of the deprojected luminosity density and $\rho_{\rm DM}$ MGE representation of the DM component.

The orbital properties of the kinematic tracers are incorporated through the anisotropy parameter $\beta_z$ and rotation parameter $\kappa$.
\textsc{CJAM} assumes a cylindrically-aligned velocity ellipsoid and that the anisotropy is defined in the meridional plane as:
\begin{equation}
    \beta_z = 1-\frac{\overline{v_z^2}}{\overline{v_R^2}}
\end{equation}
where $\overline{v_z^2}$ is the second velocity moment along the $z$-direction and $\overline{v_R^2}$ along the cylindrical radial $R$ direction.
The rotation parameter modulates the ratio of ordered to disordered motion:
\begin{equation}
    \kappa = \frac{\overline{v_\phi}}{\sqrt{\overline{v_\phi^2} - \overline{v_R^2}}}
\end{equation}
where $\overline{v_\phi}$, $\overline{v_\phi^2}$ are first and second order moments in the tangential coordinate $\phi$.

\subsection{Stellar surface density}\label{ch:stellar_surface_density}

To construct the stellar surface density profile, we combine the stellar surface brightness in the inner 1~\Reff\ from \cite{Dufour79+stellar_profile} with the appropriately scaled halo RGB number counts from \cite{Rejkuba+22_RGBs}.

Accounting for the inner dust disk the stellar surface brightness in the range 70~arcsec$\leq R\leq 255$~arcsec (0.2 - 0.8~\Reff) is best represented by the \deVau\, profile in the B-band magnitude \citep[eq. 12]{Dufour79+stellar_profile} :
\begin{equation}
    I_{\rm B} = 8.32 \Bigg[ \bigg( \frac{R}{R_{e, {\rm B}}} \bigg)^{1/4} - 1 \Bigg] +22.94 \quad [{\rm mag\ arcsec}^{-2}] \label{eq:DeVaucouleurs}
\end{equation}
where $I_{\rm B}$ is in mag arcsec$^{-2}$  and ${R_{e, {\rm B}}} = 305$~arcsec is the effective radius in the B band.
We adopt a position angle of $35\degree$ East of North and a flattening of q$_{\star, \ {\rm main}}$ = 0.93 from \citet{Dufour79+stellar_profile}.
 
Assuming the distance modulus of 
$\mu = $
$27.88$~mag  and absolute Solar magnitude 
$M_{\odot, {\rm B}} = 5.48$ \citep{BinneyAndTremaine08}, 
The B-band magnitude profile is converted to surface brightness profile $[{\rm L}_\odot ~{\rm arcsec}^{-2}]$:
\begin{equation}
    L_{\rm B} = 10^  {-0.4 \big( I_{\rm B} -{\rm (m-M)} - M_{\odot, {\rm B}} \big)}\quad [{\rm L}_\odot\  {\rm arcsec}^{-2}] \label{eq:DeVaucouleurs_luminosity}
\end{equation}
\cite{Rejkuba+22_RGBs} used \textit{HST} images to study the properties of resolved stellar halo of Cen A out to 30\Reff.
Constructing deep colour-magnitude diagrams (CMDs) that contained individual bright RGB stars enabled the authors to remove fore- and background contamination and construct the stellar number density profile from 28 individual \textit{HST} pointings across the galaxy.
Using the RGB number counts in the range $0.23 < R/R_{\rm e} < 30$ they find the tightest fit to the number density profile of RGB stars by fitting both a single power-law and for \deVau\, profile 
with an increased flattening in the outer stellar halo corresponding to $q'_{\star, \ {\rm halo}} = 0.54\pm 0.02$.
We adopt the best fit \deVau\, profile from \citet{Rejkuba+22_RGBs}: 
\begin{equation}
    \log \phi_{\rm RGB} = 7.685 - 3.008 (a/R_{\rm e})^{1/4} \label{eq:DeVaucouleurs_RGB_num_counts}
\end{equation}
where $a$ is the semi-major axis position. 
The authors convert the number counts to luminosity and provide a shift to match the inner stellar surface brightness profile from \cite{Dufour79+stellar_profile}, thereby extending the inner stellar surface brightness out to 30\Reff.
By integrating the combined functions, we recover the total luminosity of  $L_{\rm B} \sim 0.3 \times 10^{11}\rm L_\odot$ corresponding to $M_{\rm B} = -20.7$~mag, consistent with $B_{\rm T} = 7.84$~mag ($M_{\rm B} = -20.6$~mag), the measurements in the Third Reference Catalogue of Bright Galaxies \citep{RC3}. 

\begin{figure}
    \includegraphics[width=\columnwidth]{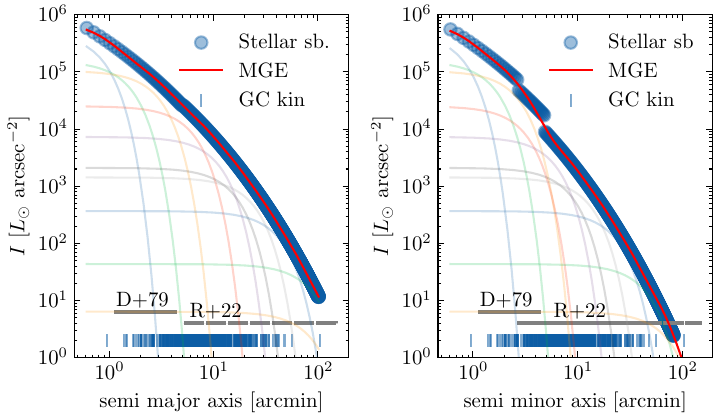}
      \caption{Combined stellar surface brightness profile of CenA from \cite{Dufour79+stellar_profile} and \cite{Rejkuba+22_RGBs} and its MGE decomposition.
      The left panel shows the profile along the semi-major axis and the right along the semi-minor axis.
      The blue points show the observed 2D profile we generated with Eq.~\ref{eq:DeVaucouleurs} in the region highlighted by the grey solid line (D+79 fitting region) and Eq.~\ref{eq:DeVaucouleurs_RGB_num_counts} in the outer regions highlighted with the grey dashed line (R+22 fitting region).
      The red line shows the resulting MGE fit to the profile and the faint coloured lines are the individual Gaussian components.
      The blue vertical bars at the bottom of the figure show the positions of the kinematic tracers.
              }
         \label{fig:stellar_MGE}
   \end{figure}

To obtain the stellar surface brightness we combine the profile from \cite{Dufour79+stellar_profile} and \cite{Rejkuba+22_RGBs}.
To account for the difference in the flattening when making the MGE of the galaxy's baryonic component, we generate a 2D image using Eq.~\ref{eq:DeVaucouleurs_luminosity} and Eq.~\ref{eq:DeVaucouleurs_RGB_num_counts} in their respective radial regions.
We do not have information on the radial variation of the flattening, so we kept $q'$ fixed to 0.93 within 0.8~\Reff\ (the inner region) and 0.54 in the outer region beyond 1.5~\Reff.
In the regions defined by both profiles, we assign the mean luminosity between both models.
We decompose the image with the \textit{mgefit} and the resulting components are shown in \autoref{tab:stellar_MGE}.
The profiles along the major and minor axes together with the MGE components are shown in Fig.~\ref{fig:stellar_MGE}.
The blue points show the observed stellar surface density along the major and minor axis in the left and right panels, respectively.
The individual Gaussian components are shown with faint coloured lines and the summed profile is shown with a solid red line.
The discontinuities along the minor axis reflect the transition region where both profiles are evaluated as a result of the different flattenings of the main stellar body and the halo RGBs.

\begin{table}

\caption{The moments of the MGE components for the combined stellar density in the form required by the modelling tool. The sorted number of a component is in Col.~1, the central surface brightness is in Col.~2, the major axis dispersion is in Col.~3 and the projected flattening is in the last column.}\label{tab:stellar_MGE}
\centering
\begin{tabular}{cccc}

\hline\hline
n & ${\rm L_{\odot}\,{\rm pc}^{-2}}$ & arcsec & ${\rm q'}$ \\
\hline

1 & 1432.364 & 34.54 & 0.93 \\
2 & 453.59 & 64.24 & 1.0 \\
3 & 307.05 & 119.78 & 0.89 \\
4 & 73.88 & 253.33 & 0.49 \\
5 & 21.45 & 444.91 & 0.54 \\
6 & 6.18 & 637.61 & 0.54 \\
7 & 4.19 & 962.74 & 0.54 \\
8 & 1.08 & 1710.16 & 0.54 \\
9 & 0.13 & 3515.16 & 0.51 \\
10 & 0.02 & 3515.16 & 0.83 \\
\hline
\end{tabular}
\end{table}

\subsection{Dark matter}\label{ch:NFW_formula}

For the dark matter potential, we assume a Navarro, Frenk and White profile \citep[NFW, ][]{NFW+97} in elliptical coordinates $m$: $m^2 = x^2 + y^2 + z^2/q_{\rm DM}^2$, where $q_{\rm DM}$ is the intrinsic flattening of the dark matter profile. 
It is important to emphasize that we fix the axis of symmetry to the stellar component.
Therefore, the oblate and prolate shapes of dark matter haloes have the same symmetry axis in our modelling approach. 
Then the DM density is
\begin{equation}
    \rho_{\rm DM} (m) = \frac{\rho_s}{\frac{m}{r_s}\ \Big( 1+\frac{m}{r_s} \Big)^2} ,  \label{eq:NFW}
\end{equation}
where $\rho_s$ and $r_s$ are the scale density and radius. 
We parameterise the profile with the total mass $M_{200} = (4\pi/3)~ 200 \rho_{\rm crit} r_{200}^3$, defined as the mass within a sphere of radius $r_{200}$ and an average density 200 times the critical density of the Universe ($\rho_{\rm crit}$), and concentration $c_{200} = r_{200}/r_s$.
We use the latest results from \cite{Planck_results18} for $\rho_{\rm crit} = (3H_o^2)/(8\pi G)$, where $H_o = 67.4~{\rm km~s}^{-1}~{\rm Mpc}^{-1}$ is the Hubble constant and $G$ is the gravitational constant.

Similar to the stellar density profile, the dark matter profile is parametrised as an MGE to generate the input for \textsc{CJAM}.
We decompose the 1D radial profile and account for the intrinsic flattening of the dark matter.
Finally, $q_{\rm DM}$ is projected based on the inclination $q_{\rm DM}' = \sqrt{q_{\rm DM }^2 \sin ^2i + \cos^2i}$ as well as the central surface density of the Gaussian components.

\section{Discrete likelihood analysis}\label{ch:statistical_method}

To compare our models with the data, we follow the discrete likelihood analysis approach described in \cite{Watkins+13}, which has been applied to different studies of discrete halo tracers of ETGs and simulated galaxies \citep[e.g.][]{Hughes+21_dyn_modeling, Versic+23_SLUGGS}.
\textsc{CJAM} provides projected velocity components of the tracer population given the gravitational potential and orbital parameters ($\beta_z, \kappa$).
Through the assumption of a local Gaussian LOSVD we compare the velocity of each of the tracers with model predictions.
We use this assumption to construct a likelihood function and constrain best-fit parameters.
Given the size of the dataset and its precision, a Gaussian LOSVD is a reasonable assumption.
For a tracer $k$ with LOS velocity $v_k$, the likelihood is given by
   \begin{equation}
      \ln\mathcal{L}_k = - \frac{1}{2}\ln \Big(2\pi \sigma_{{\rm LOS},k}^2\Big) - \frac{(\varv_k - \overline{\varv_k})^2}{2\sigma_{{\rm LOS},k}^2},\label{eq:likelihood}
   \end{equation}
where $\overline{v_k}$ and $\sigma_{{\rm LOS},k}$ are the mean LOS velocity and LOS velocity dispersion predicted by the model at the position of the $k$th tracer. 
The total likelihood of the whole sample is then given by
\begin{equation}
    \ln \mathcal{L} = \sum_k \ln \mathcal{L}_k .
    \label{eqn:total_likelihood}
\end{equation}
Since the measurement uncertainties are Gaussian there is a simple correlation between the likelihood  and $\chi^2$,
\begin{equation} 
\chi^2 = -2\ln \mathcal{L}. 
\end{equation} 

In our modelling, we use 7 free parameters: 3 to describe the dark matter ($M_{200}$, $c_{200}$ and $q_{\rm DM}$), one to scale the stellar surface density \ML, one to characterise the inclination (inc), and two to describe the orbital properties of the tracer population ($\beta_z$ and $\kappa$).

This 7-dimensional parameter space is explored by evaluating CJAM models on a fixed grid.
Each grid point represents a combination of model parameters.
Due to the computational time intensity of the CJAM models, we computed the likelihood and $\chi^2$ statistics on this grid instead of full posterior sampling.
We carried out the analysis in two steps.
First, to find the rough location of the global minimum, we varied only one parameter, keeping the others fixed.
Second, to find the best-fit parameters as well as to investigate their covariances and quantify uncertainties, we varied 2 parameters at a time keeping the others fixed.
%
We carried out the analysis separately for the blue and red GCs to get independent constraints on each of the parameters.


We found the best-fit solution through a series of iterations.
Not all 7 parameters were explored in each of the iterations.
For each parameter (or parameter pair) that is explored, we evaluate the likelihood on a grid; while one of the parameters (or two) was varied the rest were kept fixed to the initial guess at the corresponding iteration.
The initial guess of each successive iteration is determined from the output of the previous iteration based on the likelihood that was also used to determine the convergence of the best-fit solution.
As the last step of the analysis, we derive both the best-fit parameters and the uncertainties by combining the constraints from the individual 2D grids.

\subsection{Finding an approximate location of the global minimum}
\label{ch:1D_grid_setup}

At each iteration step $n$, we compute the $\chi^2$ surface for each of the parameters $\theta_i$ that is varied while keeping the other parameters $\theta_j^{( n, \rm fix)}$ (for $j\neq i$) fixed.
We adopted initial estimates for the two components of the gravitational potential from previous investigations: $\log M_{200}$ =12\footnote{In the following text $logM_{200}$ is in units [\Msun] and for clarity of the text we omit the unit.}, $c_{200}$=10 \citep{Peng+04_PNe_dyna_outer_halo}, and the dynamical $(M/L)_{\rm B}$ = 3.9 \citep{Hui95_PNe_kin_sys_vel_mass_shape}.
Our initial estimate for the inclination was inc$ = 75\degree$, and for the orbital parameters we used $\beta_z = 0$ and $\kappa = 0$.

\begin{figure}[]
    \centering
    \includegraphics[width=\columnwidth*3/4]{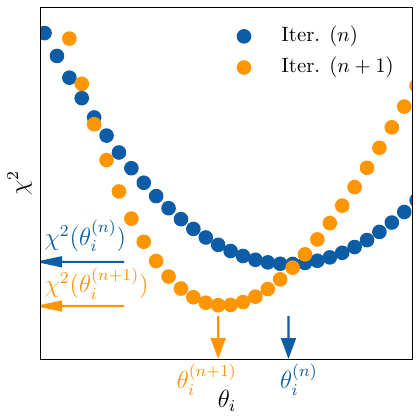}
    \caption{A sketch of a variation of $\chi^2$ statistics as a function of parameter $\theta_i$ for an iteration step $(n)$ in blue and $(n+1)$ in orange.
    The points demonstrate the grid points in $\theta_i$, with a grid step size of $\Delta \theta_i$, where the $\chi^2$ is computed.
    The coloured arrows highlight the value of the parameter at the minimum of the $\chi^2$ statistics on the $x$ axis and the absolute value of the statistics on the $y$ axis.}
    \label{fig:1D_Chi2Surface}
\end{figure}

We used 20 grid points for each parameter as a compromise between the computational time and the coarseness of the grid.
For each free parameter $\theta_i$, we evaluated the CJAM model, calculated the $\chi^2(\theta_i)$, and determined the minimum $(\chi^2)_i^{(n)}$ that corresponds to the  solution $\theta_i^{(n)}$.
We used two criteria to inform the location of the global minimum as shown in Fig.~\ref{fig:1D_Chi2Surface}.
One is related to the difference of the minimum $\chi^2$ of the two subsequent iterations. 
The second is related to the difference in the value of the solution $\theta_i^{(n)}$ between two subsequent iterations ($\Delta \theta_i =\theta_i^{(n+1)} -  \theta_i^{(n)}$).
\autoref{fig:1D_Chi2Surface} shows a sketch of $\chi^2$ for two iteration steps and demonstrates these two criteria.
Between the steps $(n)$ and $(n+1)$ we modified parameters $\theta_j^{n+1,\ \rm fix};\ (j\neq i)$ and the resulting minimum of $\chi^2$ statistic is lower than the minimum at the previous step: $\chi^2 (\theta_i^{(n+1)}) < \chi^2 (\theta_i^{(n)})$.  
This negative difference in the minimal $\chi^2$ value indicates that parameters $\theta_j^{n+1,\ \rm fix}$ are closer to the best-fit parameters than $\theta_j^{n,\ \rm fix}$.
Additionally, the value of the parameter at the minimum is different: $|\theta_i^{(n+1)} - \theta_i^{(n)}| >> \Delta \theta_i$, meaning that the $\theta_i^{(n+1)}$ is closer to the best-fit parameter than  $\theta_i^{(n)}$.
Therefore we considered the solution has converged once $\Delta \theta_i$ was a few grid step sizes and $(\chi^2)_i^{(n+1)}\approx(\chi^2)_i^{(n)}$.
Because of the correlations in 7D parameter space, $\theta_i^{(n)}$ will depend on the values of all other parameters $\theta_j^{(n)}$, j$\neq$i. 

As the result of the above iterative procedure, we identify the following best-fit parameters that are close to the global minimum and use them for the initial estimates for the next step of the analysis.
These were: $\log M_{200}=12$, $c_{200}=10$, $q_{\rm DM}=1.2$, $inc=85\degree$, $(M_\star/L_{\rm B}) = 4$, $\beta_z^{\rm GC,\ blue} = -0.2$, $\beta_z^{\rm GC,\ red} = 0.1$, $\kappa_z^{\rm GC,\ blue} = 0$, $\kappa_z^{\rm GC,\ red} =0$.

\subsection{Two parameter grid model}
\label{ch:2parameter_model}

\begin{table*}[t]

    \caption{Overview of the parameters used in different iteration steps for the two-parameter optimisation. 
    Column (1) contains the number of iterations (m).
    For each of the parameters used in this work, columns (2) - (10) contain the information on its best approximation and the range explored at the corresponding iteration step.
    For these columns, each row corresponding to a given iteration is split in two. 
    The top contains the values of the corresponding parameter ($\theta_k^{(\rm m, fix)}$) that was kept constant when exploring the $\chi^2$ surface in parameters $(i,j)$.
    The bottom contains the range of values we explored for a given parameter and is marked with (--) if that parameter is fixed at a given iteration step.
    When the parameters were changed between subsequent iteration steps we highlight them in \textbf{bold} and remark with $\dagger$ the change in range.
    The last column shows the number of grid points linearly distributed for each of the parameters in the corresponding range.}
    \label{tab:2D_grid_parameters}
    \centering
    \begin{threeparttable}
    \begin{tabular}{ p{0.02\linewidth} | p{0.07\linewidth} | p{0.075\linewidth} | p{0.07\linewidth} | p{0.07\linewidth} | p{0.07\linewidth} | p{0.07\linewidth} | p{0.07\linewidth} | p{0.07\linewidth} | p{0.07\linewidth} | p{0.06\linewidth} }

    \hline \hline
    it. & \logM & $c_{200}$ & $q_{\rm DM}$ & inc & \ML & $\beta_{\rm z, blue}$ & $\beta_{\rm z, red}$ & $\kappa_{\rm blue}$ & $\kappa_{\rm red}$ & no. grid \\
    &  $[{\rm M}_\odot]$ & & & [deg] & $[{\rm M}_\odot / {\rm L}_\odot]$ & & & & & points\\
    \hline
    1 & 12 & 10 & 1.2 & 85 & 4.5 & -0.12 &  0.1 & 0 & 0 &  \\
     & 10 - 13 & 2 - 20 & 0.6 - 2 & 60 - 90 & 1 - 8 & -0.3 - 0.3 &  -0.3 - 0.3 & -- & -- &  30 \\
     \hline
    2 & 12 & 10 & 1.2 & 85 & 4.5 & \textbf{-0.2} &  0.1 & 0 & 0 &  \\
     & 10 - 13 & 2 - 20 & 0.6 - 2 & 60 - 90 & 1 - 8 & -0.3 - 0.3 &  -0.3 - 0.3 & -- & -- & 30 \\
     \hline
    3 & \textbf{12.3} & \textbf{8.5} & 1.2 & 85 & 4.5 & -0.2 &  0.1 & 0 & 0 &  \\
     & 10 - 13 & 2 - 20 & 0.6 - 2 & 60 - 90 & 1 - 8 & -0.3 - 0.3 &  -0.3 - 0.3 & -- & -- & 30 \\
     \hline     
    4 & 12.3 & 8.5 & 1.2 & 85 &\textbf{ 3.5} & -0.2 &  0.1 & 0 & 0 &  \\
     & 10 - 13 & 2 - 20 & 0.6 - 2 & 60 - 90 & 1 - 8 & -0.3 - 0.3 &  -0.3 - 0.3 & -- & -- & 40 \\
     \hline  
    5 & 12.3 & \textsc{commah}\tnote{ a)} & 1.2 & \textbf{90}\tnote{ b)} & 3.5 & -0.2 &  0.1 & 0 & 0 &  \\
     & 10 - 13 & \textsc{commah}  & 0.6 - 2 & -- & 1 - 8 & -0.5 - 0.5\tnote{$\dagger$} &  -0.5 - 0.5\tnote{$\dagger$} & -- & -- & 40 \\
     \hline  
    6 & 12.3 & \textsc{commah}  & 1.2 & 90 & \textbf{3.0} & -0.2 &  \textbf{0.05} & 0 & 0 &  \\
     & 10 - 13 & \textsc{commah}  & 0.6 - 3\tnote{$\dagger$} & -- & 1 - 8 & -0.5 - 0.5 &  -0.5 - 0.5 & -- & -- & 40 \\
     \hline  
    7 & 12.3 & \textsc{commah}  & 1.2 & 90 & 3.0 & -0.2 &  0.05 & \textbf{-0.1} & \textbf{-0.20} &  \\
     & -- & \textsc{commah}  & 0.6 - 3 & -- & -- & -0.5 - 0.5 &  -0.5 - 0.5 &  -0.5 - 0.5\tnote{$\dagger$} &  -0.5 - 0.5\tnote{$\dagger$} & 10 \\
     \hline 
    8 & 12.3 & \textsc{commah}  & 1.2 & 90 & 3.0 & \textbf{-0.15} &  \textbf{0.15} & \textbf{-0.15} & -0.20 &  \\
     & 10 - 13 & \textsc{commah}  & 0.4 - 3.5\tnote{$\dagger$} & -- & 1 - 8 & -0.8 - 0.4\tnote{$\dagger$} &  -0.8 - 0.4\tnote{$\dagger$} &  -0.7 - 0.4\tnote{$\dagger$} &  -0.7 - 0.4\tnote{$\dagger$} & 30 \\
     \hline  
    \end{tabular}
    \begin{tablenotes}[para, flushleft]\footnotesize
\item[$\dagger$] Parameter range modified from the previous step,\\
\item[${\rm a)}$] \cite{Correa+15_M200-c200_relation} with \cite{Planck+15_cosmological_parameters} cosmology\\
\item[${\rm b)}$] \cite{Hui95_PNe_kin_sys_vel_mass_shape}
\end{tablenotes}
\end{threeparttable}
\end{table*}

Next, we explored the covariances by co-varying 2 parameters ($\theta_i$, $\theta_j$) and keeping the others fixed ($\{\theta\}_k$, where $i\neq j \neq k$).
The summary of the $\{\theta\}_k^{\rm (m, fix)}$, the ranges of parameters $(i,j)$ and number of grid points in each parameter is presented in \autoref{tab:2D_grid_parameters}.
The first column shows the iteration step - $m$, columns (2) - (10) contain the values of the $\theta_k^{m, \rm fix}$ in the top row and grid range for ($\theta_i$, $\theta_j$) in the bottom row of each iteration.
The last column shows the number of linearly-spaced grid points for each $\theta_i$ and $\theta_j$.
If the covariance is not explored for a given parameter at a given iteration step, then the range row is left empty (--).

In iterations 1-6, we set the rotation of both red and blue GCs to 0 ($\kappa = 0$) and we introduce rotation in iterations 7 and 8.
This is motivated by an increased computational time when $\kappa^{\rm GC}\neq0$ and a recent study that showed that $\beta_z$ and $\kappa_z$ have negligible impact on the enclosed mass measurements from GCS with similar methodology \citep{Versic+23_SLUGGS}.

In iteration step 5, we introduce the literature mass-concentration relation \citep{Correa+15_M200-c200_relation} using the Python package \textsc{commah} to assign $c_{200}$ given $M_{200}$ and the cosmological parameters from \citet{Planck+15_cosmological_parameters}.
This choice was driven by two observations of the 2D $\chi^2$ surfaces in the iterations 1-4.
First, the strong degeneracy of $c_{200}$ with both \logM\ and \ML\, and second, the minimum $\chi^2$ $c_{200}$ is consistent with the literature studies of mass-concentration relation as explained in \autoref{ch:App_m200_c200}.
In iteration step 5 we additionally introduce a constant inclination, $inc = 90\degree$ \citep{Hui95_PNe_kin_sys_vel_mass_shape}.
This choice was motivated because the inclination angle was not further constrained within the range 60$\degree$-90$\degree$ by our data in iterations 1-4 (see 2nd column of 1st-4th row
in Fig.~\ref{fig:2D_minChi2_sol_comparison_blue} an Fig.~\ref{fig:2D_minChi2_sol_comparison_red}). 
We explored the potential correlations between the inclination and other modelling parameters and no significant correlations were found with other parameters.

\begin{figure}[]
    \centering
    \includegraphics[width=\columnwidth]{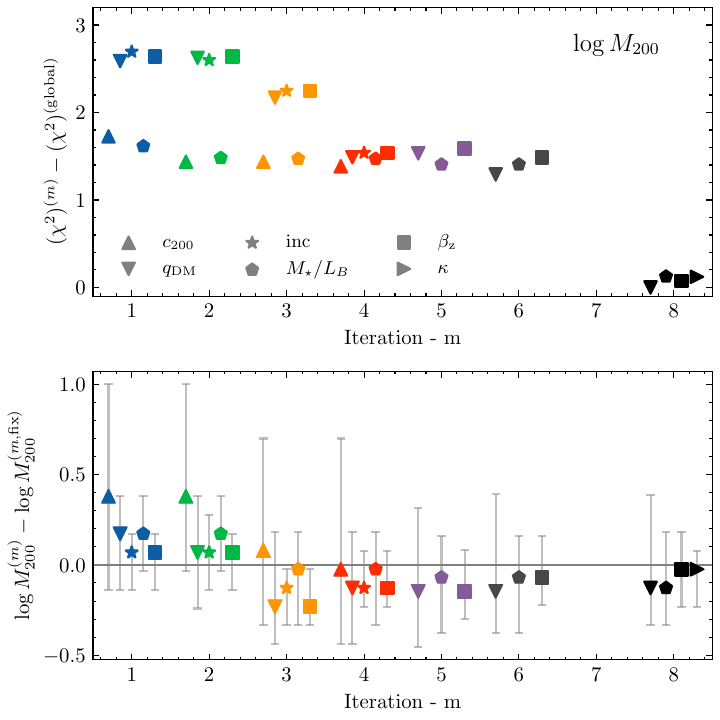}
    \caption{The evolution of the minimal $\chi^2$ and associated values of the parameter at the minimum, when different parameters were covarying for $\log M_{200}$.
    Each colour highlights the iteration step and the symbols correspond to the different parameters covarying in the 2D grid.
    In iteration 7 $\log M_{200}$ was kept constant as seen in \autoref{tab:2D_grid_parameters}.
    The top panel shows the evolution of the minimal $\chi^2$ value for the individual 2D covariation compared with the global minimum of $\chi^2$ for all grids and iterations.
    The bottom panel shows the difference between the value of the parameter at the minimum with 1$\sigma$ uncertainties and the value of the parameter shown in \autoref{tab:2D_grid_parameters} at a given iteration. 
    The grey horizontal line is shown as a guide.}
    \label{fig:2D_minChi2_sol_comparison}
\end{figure}

We set two criteria for the convergence of the solution. 
First, $\theta_i^{(m)} \approx \theta_i^{(m, \rm fix)}$ for a given parameter $i$ in all 2-dimensional slices of the $\chi^2$ surface.
Second, $\chi^2(\theta_i^{(m)}, \theta_j^{(m)}) \approx \chi^2(\theta_k^{(m)}, \theta_l^{(m)})$ for all ($i,j$), ($k,l$) parameter pairs as shown in Fig.~ \ref{fig:2D_minChi2_sol_comparison}.
Once both criteria were met, we are confident to have found the global minimum.
We used only the blue GCs to locate the minimum in this step.
The constraints from the red GCs were used only to inform the population-specific parameters: velocity anisotropy $\beta_{z}$ and rotation $\kappa$.
We chose this approach because, in this way, the convergence of the parameters from the red GCs provides an independent verification of our method and the location of the global minimum.

The iterative process for the blue GCs is demonstrated in Fig.~\ref{fig:2D_minChi2_sol_comparison} with \logM\ being a representative $\theta_i$ component (all of the parameters are shown in \autoref{ch:App_convergence_grid_set_up} for both blue and red GCs).
The top panel shows the difference between the minimum value of $\chi^2$ statistics at the iteration step $(m)$ and the global minimum value.
The bottom panel shows the difference in the best-fit value of $\log M_{200}$ at a given step and the fixed value $\log M_{200}^{(m, \rm fix)}$.
The $\theta_j$ components are shown with different symbols and each colour corresponds to a given iteration.
On the top panel the convergence trend is evident, with the minimum value of $\chi^2$ decreasing with each subsequent iteration.
Additionally, the spread of the minimum of the $\chi^2$ is decreasing, suggesting that all the parameter pairs are converging to the same global minimum.
A similar trend is seen in the bottom panel as the $\log M_{200}^{(m)}$ becomes more consistent among the different $(i,j)$ pairs and gets closer to the value assumed at a given iteration $\log M_{200}^{(m, \rm fix)}$.
A comparison of the initial run marked with blue symbols and the final set-up with black symbols shows improved consistency between parameter pairs.
Based on this, we conclude that we identified the global minimum in the 5D parameter space.

\subsection{Best fit parameters and uncertainties}
\label{ch:Method_marginalising}

For the quantification of the uncertainties, we start by taking the 2-D likelihood surface and, by marginalising over one of the dimensions, we come to the marginalised posterior, which is equivalent to the probability. 
For each of the $\theta_i$, we obtain 4 independent constraints on the parameter from the blue GCs and another 4 from red GCs. 
We use the joint posterior to quantify the best-fit solution and the 16th and 84th-percentile uncertainties.

\section{Results}\label{ch:Results}

In this section, we present the results from the final iteration -- iteration 8.
The convergence in both the minimum $\chi^2$ and the values of the parameter at the minimum support the conclusion that the iteration 8 has converged. 
This is, therefore, our final iteration.
We first focus on the $\Delta \chi^2$ surface followed by the combined constraints from independent modelling of the red and blue GCs and finally, we present the results of the modelling.

\subsection{$\Delta \chi^2$ surface}

\begin{figure*}[]
    \centering
    \includegraphics[width=\columnwidth*2]{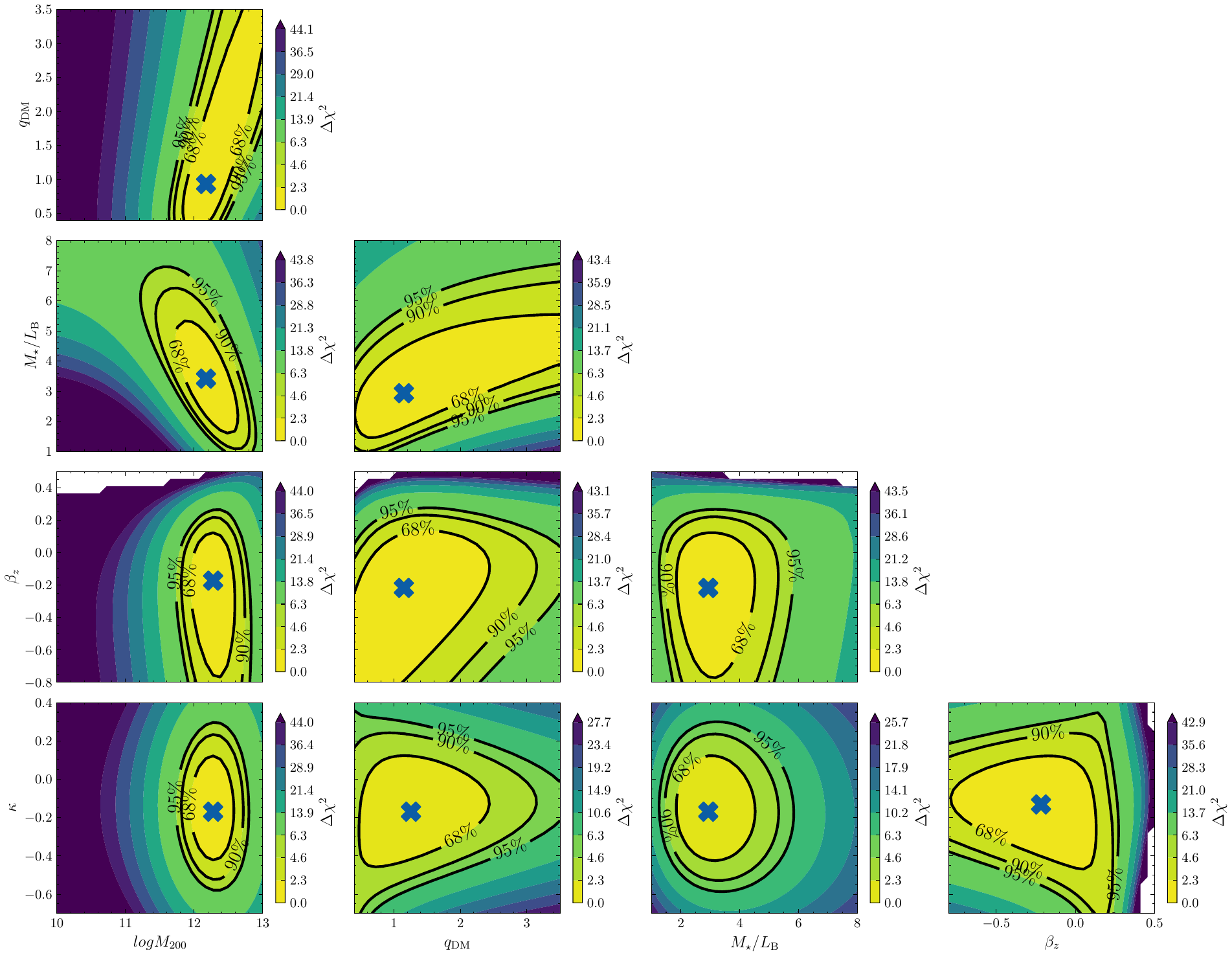}
    \caption{The smoothed $\Delta \chi^2$ surfaces for the different combinations of parameters from iteration 8 from the blue GCs. The black lines highlight the 68th, 90th and 95th percentile corresponding to $\Delta \chi^2$ levels of 2.3, 4.6 and 6.3 respectively.
    The minimal $\chi^2$ solution is highlighted with the blue cross.}
    \label{fig:Final_corner_blue}
\end{figure*}

The $\Delta \chi^2$ surface for the 5 parameters explored in iteration 8 (see \autoref{tab:2D_grid_parameters}) is shown in Fig.~\ref{fig:Final_corner_blue} for the constraints from the blue GCs and constraints from red GCs show similar trends.
Each panel shows the smoothed $\Delta \chi^2$ surface on a grid of parameter pairs.
The black contours highlight the 68.3th, 90th and 95th percentiles, where the 68.3th percentile approximately corresponds to the 1$\sigma$ region.
The blue cross highlights the location of the minimum $\chi^2$.

This figure provides useful insight into the correlations between the different parameters. 
However, combining the constraints on each of the parameters coming from the $\Delta \chi^2$ surface is not trivial and instead, we take advantage of the evaluated likelihoods.
We show this figure for demonstrative purposes and the black contours help to guide the eye.

There are several degeneracies between the parameters.
The parameter best constrained by the data is \logM, which has tight constraints and shows minimal degeneracy with the parameters of the velocity distribution of the GCs - $\beta_z$ and $\kappa$.
The total dark matter mass is strongly anticorrelated with the stellar \ML, in such a way that models with lower \logM\  prefer higher \ML.
The flattening of the dark matter halo is positively correlated with \logM, with more massive haloes preferring stronger elongation of the isodensity along the minor axis of the stellar and GC distribution of NGC~5128.

The flattening of the DM halo depends very weakly on both the virial dark matter mass and \ML.
The strongest constraints on this parameter come from the properties of the GC velocity distribution. 
As can be seen in the second column, $\kappa$ provides the strongest constraints on the flattening $q_{\rm DM}$, disfavouring an oblate or spherical halo.
The velocity anisotropy provides somewhat weaker constraints.
The parameters of the velocity distribution are not correlated.

\subsection{Combining constraints from red and blue GCs}

\begin{figure*}[]
    \centering
    \includegraphics[width=\columnwidth*2]{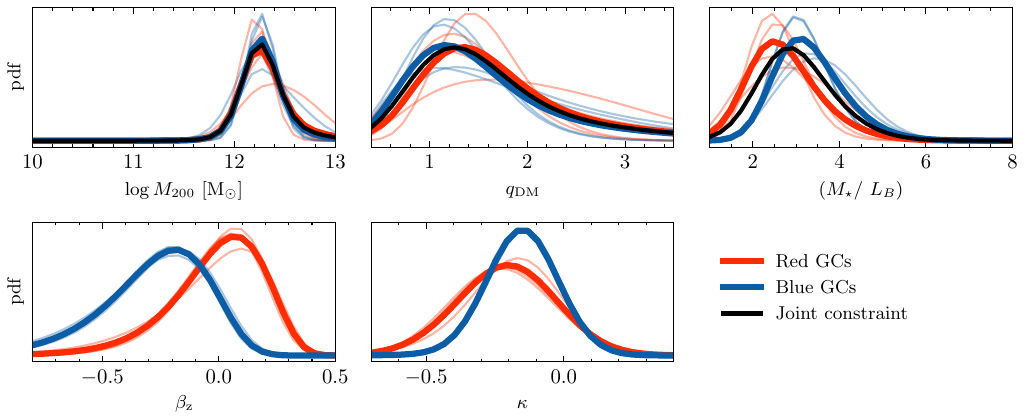}
    \caption{Posterior distributions of the individual parameters for the final iteration. 
    Red and blue coloured lines correspond to the constraints from the corresponding GC subpopulation and black shows the joint posterior. 
    Thin lines are the marginalised posteriors from the different parameter pairs shown in Fig.~\ref{fig:Final_corner_blue} and the thick lines are the joint probabilities from all of the parameter pairs.
    The parameters in the bottom row are intrinsic to the two GC populations and we do not combine them. 
    }
    \label{fig:final_results_marginalisation}
\end{figure*}

Following the methodology outlined in Sect.~\ref{ch:Method_marginalising}, we compute the marginalised probability for each of the parameters $\theta_i$. 
Combining the probabilities we arrive at the joint posterior distribution for all free parameters. 
The final results are shown in Fig.~\ref{fig:final_results_marginalisation}.
The red and blue lines correspond to the constraints from the corresponding population of GCs, and the black lines show the joint posterior.
The faint lines show the probability distributions coming from the individual parameters. 

The top row shows the parameters of the stellar and dark matter distributions, which should be consistent between both GC populations.
The bottom row shows population-specific parameters.
The median, 16th and 84th percentiles of the bold joint probabilities are presented in \autoref{tab:final_results_marginalisation}.

\begin{table}    
\caption{The final results from the marginalised probabilities seen in  Fig.~\ref{fig:final_results_marginalisation}. }\label{tab:final_results_marginalisation}
\centering
\begin{tabular}{lcc c}
\hline \hline
$\theta_i^{m}$ & Joint & red GCs & blue GCs \\
\hline
$\log M_{200}$ & $12.27_{-0.20}^{+0.21}$ & $12.27_{-0.20}^{+0.23}$ & $12.27_{-0.19}^{+0.20}$ \\
$q_{\rm DM}$ & $1.45_{-0.53}^{+0.78}$ & $1.52_{-0.53}^{+0.77}$ & $1.38_{-0.51}^{+0.79}$ \\
$(M/L)_{B}$ & $2.98_{-0.78}^{+0.96}$ & $2.66_{-0.72}^{+0.91}$ & $3.26_{-0.69}^{+0.90}$ \\
$\beta_{\rm z}$ & -- & $0.02_{-0.22}^{+0.16}$ & $-0.24_{-0.24}^{+0.19}$ \\
$\kappa$ & -- & $-0.21_{-0.18}^{+0.18}$ & $-0.15_{-0.13}^{+0.13}$ \\
\hline
\end{tabular}
\end{table}

The total dark matter mass is best constrained by the data, with very consistent posteriors coming from both GC populations.
The flattening of the dark matter halo in the top-middle panel shows a broader posterior, with the red population slightly favouring a more prolate halo.
The top-right panel shows that the stellar \ML\ results differ slightly in that the red population favours lower values than the blue population.
However, these differences are within the 1$\sigma$ uncertainties.

The bottom-left panel shows that the red population prefers an isotropic to mild radial velocity anisotropy, while the blue population strongly disfavours isotropy, showing a preference for negative cylindrical anisotropy.
The bottom-right panel shows strong consistency in the rotation parameters of both GC populations, with the red GCs preferring a slightly stronger rotation.

\subsection{Best fit model}

\begin{figure}[]
    \centering
    \includegraphics[width=\columnwidth]{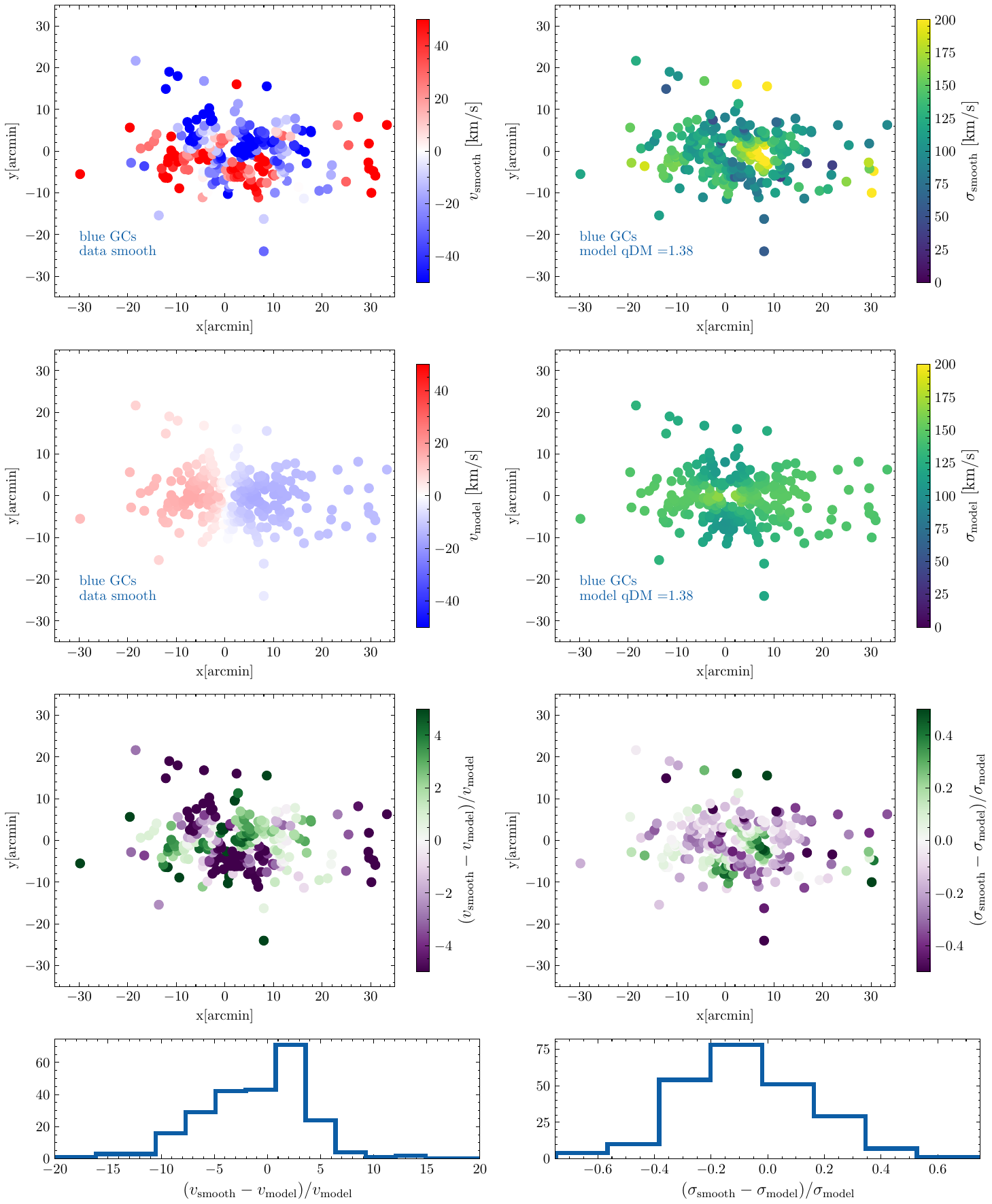}
    \caption{The comparison between the smooth velocity field of blue GCs and the model. The left column shows the mean velocity and the right column velocity dispersion.
    The first row shows the data and the second row the \textsc{CJAM} model evaluation for the best-fit values as given in \autoref{tab:final_results_marginalisation}. The third row shows the residuals between the smooth data and the model and the histogram of the residuals is shown in the bottom row.
    }
    \label{fig:vel_blue}
\end{figure}

Now that we have established the best-fit parameters, we can compare the resulting velocity fields of the globular clusters with that of the best-fitting model.
\autoref{fig:vel_blue} shows the velocity field of the blue GCs in the left column and the velocity dispersion field in the right column. 
The first row shows the smooth velocity and dispersion fields of the observed GCs, the second row shows the model predictions, the third shows the residuals, and the final row the histogram of the residuals.
The smoothing parameters are the same as discussed in Sect.~\ref{ch:Kin_substructure}.

In the top-left panel,  both major and minor axis rotation is strongly visible, indicative of a triaxial potential.
Axisymmetric modelling cannot reproduce this feature as can be seen in the panel below.
Instead, the model is sensitive to the average rotation between the major and minor axis rotation and compensates by lowering the rotational amplitude.
This can be seen as the negative residuals close to the minor axis, where the axisymmetric rotation has the opposite sign, and underestimation close to the major axis, where the residuals are positive.

The right column shows the velocity dispersion maps. 
The smoothed data shows signs of a drop in the velocity dispersion along the minor axis, which is even more prominently visible in the model panel below.
The velocity dispersion residuals, which are of the order of $\sim$25\%, do not show coherent patterns, unlike in the left panel, and their distribution is much lower compared with the velocity field residuals.

Both top panels show signs of clustering, which are partially driven by the smoothing procedure adopted in this work.
As the smooth velocity fields are only used for demonstration purposes and were not used in the model, a full investigation is beyond the scope of this work.

\section{Discussion}\label{ch:Discussion}

\subsection{Enclosed mass}\label{ch:enclosed_mass_discussion}

\begin{figure}[]
    \centering
    \includegraphics[width=\columnwidth]{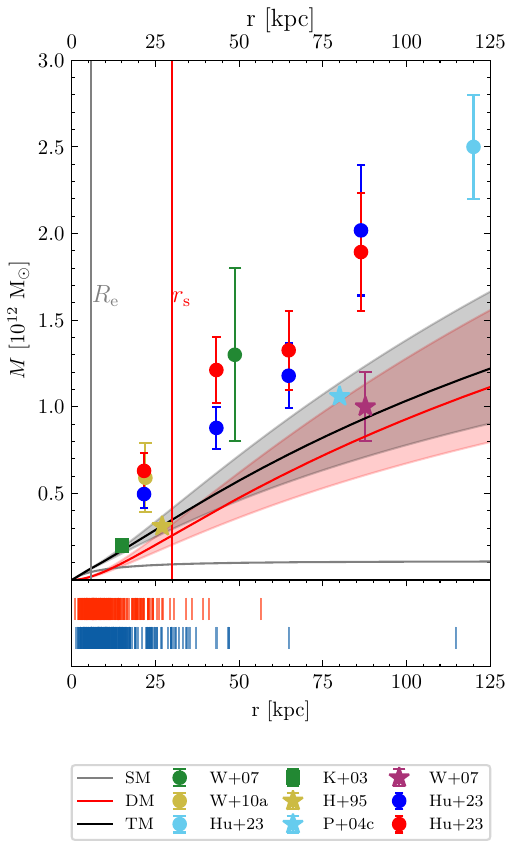}
    \caption{Enclosed mass measurements from this work are shown with lines and literature studies with symbols.
    The grey line shows the stellar enclosed mass, with \ML\ determined in this work. 
    The red line with the associated shaded 1$\sigma$ uncertainty region shows the enclosed dark matter mass integrated in spherical shells with $r_s$ the scale radius of an NFW halo.
    The black line shows the total mass.
    Circles show literature measurements of the total mass with GCs as dynamical tracers, stars show PN-based measurements, and the square shows measurements from X-ray gas from \cite{Kraft+03_X_ray_mass}.
    The green point shows the measurement with the whole sample of GCs from \cite{Woodley+07_PNe_GC_mass_estimates}, the yellow point shows the most robust measurement from \cite{Woodley+10_Kin_and_mass}, and the light blue point shows the measurement from \cite{Hughes+23_Kinematics_dyn_estimate}.
    Blue and red circles show the enclosed mass estimates from blue and red GCs respectively from \cite{Hughes+23_Kinematics_dyn_estimate}.
    The stars show constraints from PNe, with yellow showing estimate from \cite{Hui95_PNe_kin_sys_vel_mass_shape}, light blue from \cite{Peng+04_PNe_dyna_outer_halo}, with correction from \citet{Woodley+07_PNe_GC_mass_estimates}, and the measurement from the latter study is shown in purple.
    The location of the red and blue GCs used in this work is shown in the bottom panel.}
    \label{fig:Enclosed_mass_comparison}
\end{figure}

Based on the best-fit model, the parameters of the NFW distribution are: $M_{200} = 1.86^{+1.61}_{-0.69}\times 10^{12}$~\Msun, $ c_{200}= 8.58_{-0.41}^{+0.35}$, $r_s = 29.94_{-5.21}^{+6.95}~{\rm kpc}$, and 
$ r_{200}= 257.11_{-36.23}^{+44.10}~{\rm kpc}$.
\autoref{fig:Enclosed_mass_comparison} shows the comparison between the constraints for the enclosed mass from this work and literature studies.
The lines show results from this work with the stellar mass in green, dark matter mass in red and combined baryonic and non-baryonic in black. 
The shaded regions span the 16th to 84th percentiles.

Our enclosed mass measurement is on average lower than the literature estimates from other dynamical modelling of GCs.
The origin of the discrepancy could be the difference in the slope of the tracer density profile. 
\cite{Woodley+07_PNe_GC_mass_estimates} reanalysed the enclosed mass from \cite{Peng+04_PNe_dyna_outer_halo} and found a factor of 2 higher mass, when using the de-projected 3D tracer density profile instead of 2D, used in the original study.
The slope used in \cite{Woodley+07_PNe_GC_mass_estimates} is steeper than that used in our work.

\cite{Woodley+07_PNe_GC_mass_estimates} modelled the pressure- and rotationally-supported masses of PNe and GCs separately, making no separation based on GC colour.
To avoid biases in the inner regions obscured by the dust they excluded $R<5$arcmin.
They adopted a distance of 3.9~Mpc to CenA and we account for the different distances between our and the literature study. 
Later, \citet{Woodley+10_Kin_and_mass} applied a tracer mass estimator \citep[TME,][]{Evans+03_TME} to the whole sample of GCs to estimate the enclosed mass within 20~arcmin.
They found the total mass of $(5.9\pm2.0)\times 10^{11}$~\Msun, which agrees within 2$\sigma$ with our result at 20~arcmin.
They excluded eight outliers based on their projected velocity and radius ($v^2R$), which could inflate the mass estimate and their robust measurements exclude GCs within 1~\Reff\ ($R<5$~arcmin).
Relaxing the latter criteria results in increased mass estimate at all radii almost by a factor of 2, hinting at the sensitivity of this method to outliers.

\cite{Hughes+23_Kinematics_dyn_estimate} used the most spatially extended sample of GCs to provide enclosed mass estimates separately for the blue and red GCs. 
Our measurements agree within 2$\sigma$, but are systematically lower.
They also used a TME (corrected for rotational support) to estimate the total enclosed mass within a given radius assuming spherical symmetry in the total mass distribution.
In their modelling, GCs are spherically distributed with a single power-law profile with an isotropic distribution of orbits. 
Their measurements are in good agreement with mass estimates from GCs, but are higher than the estimates using PNe. 
They postulate that the simplified assumption of isotropy in their estimates could be one of the reasons for this, as also commented in \cite{Woodley+07_PNe_GC_mass_estimates}, but the mass-anisotropy degeneracy in spherical symmetry \citep{BinneyAndTremaine08} prevents them from making a conclusive determination. 

\citet{Dumont+23_CJAM_modelling_Spherical_mass} used literature datasets of PNe, GCs and satellite galaxies to trace the total mass distribution.
They employed a similar methodology, with separate luminous and dark matter components and used \textsc{cjam}. 
A spherical generalised NFW halo was used to characterise the dark matter halo and in their modelling parameterised with $\log (\rho_s)$ and $\log (\rho_s r_s^3)$. 
To characterise the spatial distribution of red and blue GCs they used single power power-law fits from \citet{Hughes+23_Kinematics_dyn_estimate}. 
In their fiducial model, they used Gaussian priors on the $c_{vir}$, concentration at the virial radius\footnote{The virial radius assumes the density contrast is greater than the value of the 200 we adopt in this work and $M_{200} \sim 0.8 M_{vir}$.}.
Their fiducial model results in a dark matter mass of $M_{200} = 3.6_{-1.0}^{+1.7} \times 10^{12}$~\Msun. 
This value agrees within uncertainty with our measurements, however, it is systematically higher. 
Some of the differences in our results could arise from the different tracer density profiles adopted, as well as the different data used.
The authors carried out robustness tests by constraining the parameters of the gravitational potential with individual tracer groups.
This analysis results in very different values of $M_{vir}$ (see their Table~3) when different tracers are used.
The constraints from GCs and satellites agree within the uncertainties, however, when modelling PNe kinematics only they recover a virial mass 6 times lower than their Fiducial model. 
The implication of this result on the Fiducial model is not discussed.
%

\cite{Pearson+22_StreamModdelingMass} used simulations to model stream formation due to a disruption of a dwarf galaxy. 
They sampled different masses of the host galaxy and found a lower limit on the mass of $M_{200} > 4.7 \times 10^{12}~M_\odot$.
They compared the visual appearance of the stream observed by \cite{PISCeS_Crnojevic16} and a single RV along the stream with their simulation results to identify the gravitational potential needed for the observed stream morphology and kinematics.
Because of covariances between the mass of the disrupted galaxy, the infall velocity and the line-of-sight position, more RV measurements along the stream are needed to provide stronger constraints on the mass. 

\cite{Muller+22_dynmaical_mass_with_satellites} used the satellite galaxies of the CenA to determine the dynamical mass within $R_{\rm max}\sim 800$~kpc (the maximal extent of their tracers).
Using the mean circular velocity of 27 dwarfs, they estimated a spherical enclosed mass and determined $M_{200} = (5.3\pm 3.5) \times 10^{12}$~\Msun.
This agrees with our measurements within the uncertainties. 
The dynamical time scales at such large distances, where the satellites are distributed, are longer than the Hubble time and non-virialised motion could inflate the enclosed mass estimate.

The best fit \ML\ from this work is lower than that of \cite{Hui95_PNe_kin_sys_vel_mass_shape} and \cite{Mathieu+96_dynamical_model_CenA_PNe}, which could be attributed to the different treatment of the DM component in the modelling. 
\cite{Cappellari+09_CenA_central_SINFONI_kinenmatics} modelled the central region of CenA to determine the mass of the central BH and found $M_\star/L_K = 0.65 \pm 0.15$.

\subsection{Limitations of the modelling assumptions}

\begin{figure}[]
    \centering
    \includegraphics[width=\columnwidth]{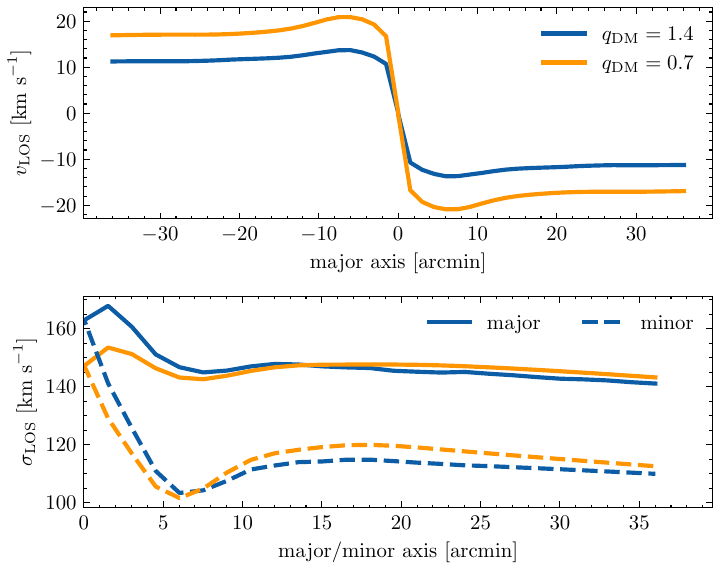}
    \caption{Comparing best-fit prolate (blue) and oblate (orange) CJAM model results.  
    The top panel shows the mean LOS velocity along the major axis. 
    The bottom panel shows the velocity dispersion along the major axis as solid lines and along the minor axis as dashed lines. 
    The LOS velocity is shown symmetrically along the major and minor axes, to highlight the asymmetric rotational signature. 
    However, the velocity dispersion on the bottom panel is shown for positive values of the axes due to the symmetry of $\sigma_{\rm LOS}$.
    }
    \label{fig:comparing_oblate_prolate_best_fit}
\end{figure}

In the context of CJAM, the flattening of $q_{\rm DM}>1$ that we recover corresponds to the elongation of the dark matter distribution along the minor axis of the stellar and GC distribution.
Literature investigations \citep{Hui95_PNe_kin_sys_vel_mass_shape, Mathieu+96_dynamical_model_CenA_PNe, Peng+04_PNe_dyna_outer_halo} have identified clear signs of a triaxial halo and, in this section, we compare the modelling results from a prolate and oblate halo to identify the features in the observing plane that favour $q_{\rm DM}>1$.

\autoref{fig:comparing_oblate_prolate_best_fit} compares the projected velocity moments from the minimum $\chi^2$ solution for our fiducial result $q_{\rm DM} = 1.45$ in blue and an oblate case with $q_{\rm DM} = 0.7$ in orange.
For the latter, we recomputed the other parameters from the 2D $\chi^2$ grids shown in Fig.~\ref{fig:Final_corner_blue}.
We chose a value of $q_{\rm DM} = 0.7$ as an intermediate flattening between the outer stellar halo flattening of 0.54 and the flattening of the GCS (0.72 for red and 0.82 for blue GCs).
Here we present simplified velocity moments along the major and minor axis only. 
A more detailed analysis of the 2D velocity moments is presented in \autoref{ap:2D_plots}.

The mean velocity along the major axis is higher in the oblate solution, however, the difference is smaller than the typical measurement uncertainty. 
Beyond 10~arcmin the velocity dispersion profiles of the oblate and prolate DM haloes are very similar for both major and minor axes, where the majority of the kinematic tracers lie.
To match the velocity dispersion at these radii, the resulting $M_{200}$ and \ML\ estimates for the oblate halo are lower compared with the prolate halo.
This anti-correlation between the flattening and mass can be seen in Fig.~\ref{fig:Final_corner_blue}. 
A lower total mass results in a central value of the velocity dispersion that is lower for the oblate solution ($\sim 150$~\kms), compared with the prolate solution ($\sim 165$~\kms).
The latter value is in better agreement with the observed central stellar root-mean-square velocity of $\sim 170$~\kms\ in the inner 0.2~arcsec of the CenA centre from \cite{Cappellari+09_CenA_central_SINFONI_kinenmatics}.

A drop in the velocity dispersion along the minor axis is also seen in the smooth velocity field in Fig.~\ref{fig:vel_blue}.
Such a drop is consistent with the projected dispersion map of a triaxial system with short and long-axis tube orbits as seen in \cite{vdBosch_vdVen09_triaxiality}.
The full details of how measured flattening constrained with axisymmetric modelling can be mapped to the intrinsic shape and orientation of the DM halo is a topic for a future study.

\subsection{Extremely misaligned dark matter haloes in cosmological simulations}\label{ch:meaning_of_q>1}

Within the axisymmetric modelling approach used in this work, the dark matter halo shape with $q_{\rm DM}>1$ corresponds to its elongation along the minor axis of the stellar distribution. 
This raises a question of how frequent are such misalignments between the minor axes of the stellar and DM profiles.
For example, in the Milky Way evidence disfavours a configuration in which the minor axis of the disc and DM halo are parallel \citep[e.g.][]{Law+09_Triaxial_tilted_DM_of_MW, Debattista+13, Han+23_tilted_MW_DM_halo}. 
In the Auriga analogues of the Milky Way, \citet[see their Fig.~6]{Prada+19} found that the minor axes of the DM haloes and angular momentum vector of the stellar discs are aligned for most of their 30 haloes, with 2 haloes showing misalignment $>70\degree$ at 1/16~$R_{200}$.

In this section, we conduct a qualitative investigation to examine the frequency of highly-misaligned distributions in two cosmological simulations. 
This analysis aims to enhance our understanding of the results by employing radial regimes analogous to the sensitivity observed in the CenA study.
For this purpose, we use IllustrisTNG \citep[][and references therein]{Nelson+19_TNGSims} and Magneticum pathfinder\footnote{\url{www.magneticum.org}} simulations to find analogues of CenA with a central oblate stellar component hosted in a prolate DM halo elongated along the stellar minor axis. 
We follow a procedure similar to \cite{Pulsoni+21} and Valenzuela et al.\ (in prep.) to measure the shapes of the simulated galaxies, corresponding to method S1 from \cite{Zemp+11_shapes_DM_halo_method}. 
This is not carried out through dynamical modelling, but instead by computing the eigenvalues of the inertial tensor calculated using the positions of the simulated particles and their masses within ellipsoidal shells.

In both simulations, we selected galaxies in the mass range of CenA as a first step: $M_{\rm vir} = 3\times 10^{12} -3\times 10^{13}~{\rm M}_\odot$ and $M_\star  = 1-5 \times 10^{11}~{\rm M}_\odot$.
Secondly, we selected those that have oblate stellar distributions embedded in prolate haloes. 
We then looked at the angle between the stellar minor axis and DM major axis and inspected the frequency of small misalignments between these two axes.
We use the definition for triaxiality: $ T \equiv (1-q^2)/(1-p^2) $, where $q = b/a$ and $p = c/a$ are the axis ratios between the major ($a$), intermediate ($b$) and minor ($c$) axis.
A completely prolate halo has $T = 1$ ($a>b = c$), and perfectly oblate halo has $T=0$ with ($a = b > c$).
We first selected galaxies with $T_{\star}< 1/3$ and that are hosted in a halo with $T_{\rm DM} > 2/3$.
To mimic the configuration of stellar and dark matter in line with CenA as presented in this work, we compute the shape of the stellar component at $3r_{\star, \rm half}$ (stellar half-mass radius) and the shape of the dark matter halo at $5r_{\star, \rm half}$.
For those, we compute the angle between the minor axis of the stellar distribution and the major axis of the dark matter.

\subsubsection*{Magneticum}

From Box4 (uhr) of the Magneticum Pathfinder simulation suite, which has side length of 68~Mpc, DM particle mass of $m_{\rm DM} = 5.1 \times 10^7$~\Msun\ and average stellar particle mass of $m_* = 1.8 \times 10^6$~\Msun\ \citep{Teklu+15, Valenzuela&Remus22}, we find 111 galaxies that match the mass criteria, of which 94 have a modest star formation rate of ${\rm SFR} < 3~{\rm M}_\odot \, {\rm yr}^{-1}$. 
Of these, 12 galaxies meet the shape criteria as laid out above, of which 10 have modest star formation rates.
Only one of those galaxies has a small angle of 17$\degree$ between the stellar minor axis and the DM major axis, whereas the rest all have such angles above 80$\degree$. 
That galaxy is at the lower end of the mass selection limits with $M_\star  = 1.0 \times 10^{11}~{\rm M}_\odot$ and $M_{\rm vir} = 4.6\times 10^{12}~{\rm M}_\odot$. 
It has no star formation and shows signs of recent interaction in the stellar halo. 
It should be noted, however, that the stellar component has a relatively spherical shape, such that the orientation of its axes are not very well defined.

\subsubsection*{IllustrisTNG}

We considered the TNG100 simulation run, with a simulation box of 110.7~Mpc on a side, DM particle mass of $7.5\times 10^6M_\odot$ and mean baryonic particle mass of $1.4\times10^6M_\odot$\ \citep{Nelson+19_TNGSims}. 
From the redshift $z=0$ snapshot, we identified 455 galaxies in the mass range around CenA.
Of these, 121 have a prolate DM halo ($\sim 27\%$), but only 17 have an oblate stellar component within $3r_{\star, \rm half}$. 
Most of the 121 prolate DM haloes have the major axis close to the plane defined by the intermediate and major axis of the stellar component within 20$\degree$. 
Out of the 17 galaxies selected, only in 1 case there is significant misalignment. 
This is a clear merger, as shown by the presence of substructures or strong asymmetries in the stellar distribution and velocities. 
This galaxy has a moderate SFR of $1.8 M_\odot/yr$.
The similarities between the simulated galaxy in the Magneticum simulation and CenA as well as the consistent merger imprints and low probability of such extreme misalignment found in both simulation suite suggest that it is likely a transient phenomenon.
This work opens up possibilities for further studies of how the dark matter halo shape is influenced by the merger history of its host and the environment.



\subsection{Connection with the CenA satellite distribution}

Recent studies of CenA satellites have found evidence of a corotating plane of satellites around CenA \citep{Mueller+18_Science}. 
In their latest study, \citet{Mueller+21_CenA_satellites_plane} showed that 21 out of 28 dwarf galaxies, which have both distance and radial velocity measurements that make them satellites of CenA, share a coherent motion within a flattened planar structure. 
This plane of satellites is almost perpendicular to the dust disk and lies very close to the line connecting CenA and its nearest massive neighbour M83.
\cite{Muller+19_Dwarfs_of_CenA_shape} used the accurate distances of CenA satellites using the tip of the RGB to determine the intrinsic 3D shape of the satellite system.
They find the ${\rm PA} = 60^{+21}_{-23}\degree$ and determine the axis ratios $b/a = 0.71$ and $c/a = 0.41$, which results in a triaxiality of $T = 0.6$, close to prolate shape.
They find that the long axis is aligned along the line-of-sight, in contrast with the stellar shape measurements traced by PNe within 25~kpc by \cite{Hui95_PNe_kin_sys_vel_mass_shape} and within 80~kpc by \cite{Peng+04_PNe_dyna_outer_halo}.
The former found $T = 0.4$ (triaxial, but more oblate) and determined the viewing angles, such that the intermediate axis lies along the line-of-sight.
Using these viewing angles, \cite{Peng+04_PNe_dyna_outer_halo} found that the kinematics of the PNe at larger distances favour a significantly more prolate shape of $T = 0.98$.

\citet{Mueller+21_CenA_satellites_plane} explored the frequency of similar flattened structures in TNG100 
\citep{Nelson+19_TNGSims}. 
Such flattened and coherently moving structures are found in only 0.3\% of Cen A analogues in dark-matter-only simulations and in 0.2\% cases in hydrodynamical simulations. 
More generally, the distribution of satellite galaxies in the $\Lambda$CDM is not expected to be completely random, i.e. isotropic \citep[e.g.][]{Zentner+05_Anisotropic_distribution_of_satellites}, due to the anisotropic accretion from the cosmic filaments. 
In this framework, the DM halo of CenA is expected to be elongated along the major axis of the PNe and the 
satellite system, which is in contrast with the results of our modelling under axisymmetric assumptions.
On the other hand, the triaxial shape of the satellite system with T=0.6 \citep{Muller+19_Dwarfs_of_CenA_shape} and of PNe with T=0.98 \citep{Peng+04_PNe_dyna_outer_halo} strongly disfavour spherical or oblate distributions, which is in line with our findings.

\section{Summary and conclusions}\label{ch:Conclussions}

We carried out axisymmetric dynamical modelling of the blue and red GCs in the CenA galaxy 
to measure the total mass and the intrinsic shape of its dark matter halo.
The axisymmetric JAM modelling used in this work requires a relaxed system.
Given that CenA shows signatures of past mergers, special care was taken to identify a sample of GCs that shows 
a point symmetry
indicative of a relaxed distribution 
when building the tracer density profile.
For the kinematic analysis, we use the entire available sample, which we verified to follow the relaxed point-symmetric velocity field from PNe.

The dataset that constrains the model is composed of the photometric dataset from \citepalias{Taylor+17} and the radial velocity sample from \cite{Hughes+21}, which extend out to 115~kpc and include 225 red and 257 blue GCs. 
We note that 90\% of the kinematic dataset is within 40~kpc.

We used axisymmetric JAM modelling together with a discrete Gaussian likelihood analysis to determine the best-fitting parameters of the gravitational potential and orbital properties of the GCs.
The gravitational potential is composed of both an axisymmetric stellar surface brightness out to 140~kpc (25~\Reff), a mass-to-light ratio \ML, and an axisymmetric dark matter distribution.
We assumed an NFW form for the dark matter profile, parametrised by $M_{200}$ and used the latest literature mass-concentration relation to determine the $c_{200}$. 
By varying $q_{\rm DM}$ we explored the deviation of DM halo from spherical symmetry.
To find the best-fit model, we maximise the Gaussian likelihood. 
The best-fit parameters were determined in an iterative approach, accounting for the covariances in the modelled parameters.
%
The only parameter that was fixed in the final iteration was inclination taken from \citet{Hui95_PNe_kin_sys_vel_mass_shape} and we verified that it does not correlate with other parameters.
%
We modelled the blue and red GCs separately and used them as two independent samples to verify our approach and determine the highest likelihood solution for the parameters of the gravitational potential.
The need to treat them separately was further motivated by the intrinsically different spatial distribution and orbital properties.

We derive a dark matter mass of $M_{200} = 1.86^{1.61}_{-0.69}\times 10^{12}$~\Msun\ and \ML $= 2.98^{+0.96}_{-0.78}$, resulting in a total stellar mass of $M_\star \approx 0.9\times 10^{11}$~\Msun.
The total halo mass is lower than literature investigations using GC kinematics, but in agreement with measurements from PNe as discussed in Sect.~\ref{ch:enclosed_mass_discussion}.
Our flattening of $q_{\rm DM} = 1.45^{+0.78}_{-0.53}$ disfavours an oblate or spherical DM haloes.
We explored the implications of $q_{\rm DM} > 1$ under the axisymmetry assumption of the JAM modelling by examining the frequency of prolate DM haloes that host an oblate stellar distribution in the Magneticum and TNG simulations.
We found that in both simulation suites, DM elongated along the minor axis of the stellar distribution is very rare and identified only in galaxies with strong signatures of active or recent mergers.
Comparing the smooth velocity field of the GCs, we identified a difference in the velocity dispersion along the major and minor axes that can also arise in intrinsically triaxial potentials (Sect.~\ref{ch:meaning_of_q>1}).
This suggests that GCs, and the modelling carried out in this work, are sensitive to deviations from spherical and cylindrical symmetry.

We find that the orbital properties of blue and red GCs are different.
While both populations show mild rotation, we find that the rotation parameter $\kappa$ is larger for red GCs than blue GCs.
The latter population also shows negative orbital anisotropy in cylindrical coordinates while the red population is consistent with being isotropic.
Both results highlight the need for incorporating rotation and anisotropy in future dynamical modelling.

Future work should include more complex modelling that can account for triaxial symmetry and non-relaxed structures such as streams.
To enable such modelling, RV measurements in the outer halo need to include the full extent of the galaxy potential. 
This will allow us to gain a better understanding of the global potential of the host and its assembly history. 

\begin{acknowledgements}
We thank the referee for the helpful comments and suggestions that have improved this manuscript. 
We thank Prashin Jethwa, Ryan Leaman and Laurane Fréour for their fruitful discussions and feedback on the manuscript.
This project has received funding from the European Research Council (ERC) under the European Union’s Horizon 2020 research and innovation programme under grant agreement No 724857 (Consolidator Grant ArcheoDyn). TV acknowledges the studentship support from the European Southern Observatory. 
LMV acknowledges support by the German Academic Scholarship Foundation (Studienstiftung des deutschen Volkes) and the Marianne-Plehn-Program of the Elite Network of Bavaria.
The computational results presented have been achieved (in part) using the Vienna Scientific Cluster (VSC).

This work made use of the following software:
\textsc{NumPy} \citep{Numpy2020}, \textsc{SciPy} \citep{Scipy}, \textsc{Matplotlib} \citep{Matplotlib}, \textsc{Astropy} \citep{astropy:2013, astropy:2018, astropy:2022}.
\end{acknowledgements}

%
\bibliographystyle{aa} 
\bibliography{biblio} 

\begin{thebibliography}{128}
\expandafter\ifx\csname natexlab\endcsname\relax\def\natexlab#1{#1}\fi

\bibitem[{{Allgood} {et~al.}(2006){Allgood}, {Flores}, {Primack}, {Kravtsov}, {Wechsler}, {Faltenbacher}, \& {Bullock}}]{Allgood+06}
{Allgood}, B., {Flores}, R.~A., {Primack}, J.~R., {et~al.} 2006, mnras, 367, 1781

\bibitem[{{Astropy Collaboration} {et~al.}(2022){Astropy Collaboration}, {Price-Whelan}, {Lim}, {Earl}, {Starkman}, {Bradley}, {Shupe}, {Patil}, {Corrales}, {Brasseur}, {N{"o}the}, {Donath}, {Tollerud}, {Morris}, {Ginsburg}, {Vaher}, {Weaver}, {Tocknell}, {Jamieson}, {van Kerkwijk}, {Robitaille}, {Merry}, {Bachetti}, {G{"u}nther}, {Aldcroft}, {Alvarado-Montes}, {Archibald}, {B{'o}di}, {Bapat}, {Barentsen}, {Baz{'a}n}, {Biswas}, {Boquien}, {Burke}, {Cara}, {Cara}, {Conroy}, {Conseil}, {Craig}, {Cross}, {Cruz}, {D'Eugenio}, {Dencheva}, {Devillepoix}, {Dietrich}, {Eigenbrot}, {Erben}, {Ferreira}, {Foreman-Mackey}, {Fox}, {Freij}, {Garg}, {Geda}, {Glattly}, {Gondhalekar}, {Gordon}, {Grant}, {Greenfield}, {Groener}, {Guest}, {Gurovich}, {Handberg}, {Hart}, {Hatfield-Dodds}, {Homeier}, {Hosseinzadeh}, {Jenness}, {Jones}, {Joseph}, {Kalmbach}, {Karamehmetoglu}, {Ka{l}uszy{'n}ski}, {Kelley}, {Kern}, {Kerzendorf}, {Koch}, {Kulumani}, {Lee}, {Ly}, {Ma}, {MacBride}, {Maljaars}, {Muna}, {Murphy}, {Norman}, {O'Steen},
  {Oman}, {Pacifici}, {Pascual}, {Pascual-Granado}, {Patil}, {Perren}, {Pickering}, {Rastogi}, {Roulston}, {Ryan}, {Rykoff}, {Sabater}, {Sakurikar}, {Salgado}, {Sanghi}, {Saunders}, {Savchenko}, {Schwardt}, {Seifert-Eckert}, {Shih}, {Jain}, {Shukla}, {Sick}, {Simpson}, {Singanamalla}, {Singer}, {Singhal}, {Sinha}, {Sip{H{o}}cz}, {Spitler}, {Stansby}, {Streicher}, {{{S}}umak}, {Swinbank}, {Taranu}, {Tewary}, {Tremblay}, {Val-Borro}, {Van Kooten}, {Vasovi{'c}}, {Verma}, {de Miranda Cardoso}, {Williams}, {Wilson}, {Winkel}, {Wood-Vasey}, {Xue}, {Yoachim}, {Zhang}, {Zonca}, \& {Astropy Project Contributors}}]{astropy:2022}
{Astropy Collaboration}, {Price-Whelan}, A.~M., {Lim}, P.~L., {et~al.} 2022, apj, 935, 167

\bibitem[{{Astropy Collaboration} {et~al.}(2018){Astropy Collaboration}, {Price-Whelan}, {Sip{\H{o}}cz}, {G{\"u}nther}, {Lim}, {Crawford}, {Conseil}, {Shupe}, {Craig}, {Dencheva}, {Ginsburg}, {Vand erPlas}, {Bradley}, {P{\'e}rez-Su{\'a}rez}, {de Val-Borro}, {Aldcroft}, {Cruz}, {Robitaille}, {Tollerud}, {Ardelean}, {Babej}, {Bach}, {Bachetti}, {Bakanov}, {Bamford}, {Barentsen}, {Barmby}, {Baumbach}, {Berry}, {Biscani}, {Boquien}, {Bostroem}, {Bouma}, {Brammer}, {Bray}, {Breytenbach}, {Buddelmeijer}, {Burke}, {Calderone}, {Cano Rodr{\'\i}guez}, {Cara}, {Cardoso}, {Cheedella}, {Copin}, {Corrales}, {Crichton}, {D'Avella}, {Deil}, {Depagne}, {Dietrich}, {Donath}, {Droettboom}, {Earl}, {Erben}, {Fabbro}, {Ferreira}, {Finethy}, {Fox}, {Garrison}, {Gibbons}, {Goldstein}, {Gommers}, {Greco}, {Greenfield}, {Groener}, {Grollier}, {Hagen}, {Hirst}, {Homeier}, {Horton}, {Hosseinzadeh}, {Hu}, {Hunkeler}, {Ivezi{\'c}}, {Jain}, {Jenness}, {Kanarek}, {Kendrew}, {Kern}, {Kerzendorf}, {Khvalko}, {King}, {Kirkby}, {Kulkarni},
  {Kumar}, {Lee}, {Lenz}, {Littlefair}, {Ma}, {Macleod}, {Mastropietro}, {McCully}, {Montagnac}, {Morris}, {Mueller}, {Mumford}, {Muna}, {Murphy}, {Nelson}, {Nguyen}, {Ninan}, {N{\"o}the}, {Ogaz}, {Oh}, {Parejko}, {Parley}, {Pascual}, {Patil}, {Patil}, {Plunkett}, {Prochaska}, {Rastogi}, {Reddy Janga}, {Sabater}, {Sakurikar}, {Seifert}, {Sherbert}, {Sherwood-Taylor}, {Shih}, {Sick}, {Silbiger}, {Singanamalla}, {Singer}, {Sladen}, {Sooley}, {Sornarajah}, {Streicher}, {Teuben}, {Thomas}, {Tremblay}, {Turner}, {Terr{\'o}n}, {van Kerkwijk}, {de la Vega}, {Watkins}, {Weaver}, {Whitmore}, {Woillez}, {Zabalza}, \& {Astropy Contributors}}]{astropy:2018}
{Astropy Collaboration}, {Price-Whelan}, A.~M., {Sip{\H{o}}cz}, B.~M., {et~al.} 2018, \aj, 156, 123

\bibitem[{{Astropy Collaboration} {et~al.}(2013){Astropy Collaboration}, {Robitaille}, {Tollerud}, {Greenfield}, {Droettboom}, {Bray}, {Aldcroft}, {Davis}, {Ginsburg}, {Price-Whelan}, {Kerzendorf}, {Conley}, {Crighton}, {Barbary}, {Muna}, {Ferguson}, {Grollier}, {Parikh}, {Nair}, {Unther}, {Deil}, {Woillez}, {Conseil}, {Kramer}, {Turner}, {Singer}, {Fox}, {Weaver}, {Zabalza}, {Edwards}, {Azalee Bostroem}, {Burke}, {Casey}, {Crawford}, {Dencheva}, {Ely}, {Jenness}, {Labrie}, {Lim}, {Pierfederici}, {Pontzen}, {Ptak}, {Refsdal}, {Servillat}, \& {Streicher}}]{astropy:2013}
{Astropy Collaboration}, {Robitaille}, T.~P., {Tollerud}, E.~J., {et~al.} 2013, \aap, 558, A33

\bibitem[{{Bailin} \& {Steinmetz}(2005)}]{Bilin+05_shapes_orientation_of_DM_haloes}
{Bailin}, J. \& {Steinmetz}, M. 2005, \apj, 627, 647

\bibitem[{{Beasley} {et~al.}(2008){Beasley}, {Bridges}, {Peng}, {Harris}, {Harris}, {Forbes}, \& {Mackie}}]{Beasley+08_met_distribution_of_GCs_in_CenA}
{Beasley}, M.~A., {Bridges}, T., {Peng}, E., {et~al.} 2008, \mnras, 386, 1443

\bibitem[{{Bellstedt} {et~al.}(2018){Bellstedt}, {Forbes}, {Romanowsky}, {Remus}, {Stevens}, {Brodie}, {Poci}, {McDermid}, {Alabi}, {Chevalier}, {Adams}, {Ferr{\'e}-Mateu}, {Wasserman}, \& {Pandya}}]{Bellstedt+18}
{Bellstedt}, S., {Forbes}, D.~A., {Romanowsky}, A.~J., {et~al.} 2018, \mnras, 476, 4543

\bibitem[{{Binney} \& {Tremaine}(2008)}]{BinneyAndTremaine08}
{Binney}, J. \& {Tremaine}, S. 2008, {Galactic Dynamics: Second Edition}

\bibitem[{{Blakeslee}(1997)}]{Blakeslee97}
{Blakeslee}, J.~P. 1997, \apjl, 481, L59

\bibitem[{{Brodie} {et~al.}(2014){Brodie}, {Romanowsky}, {Strader}, {Forbes}, {Foster}, {Jennings}, {Pastorello}, {Pota}, {Usher}, {Blom}, {Kader}, {Roediger}, {Spitler}, {Villaume}, {Arnold}, {Kartha}, \& {Woodley}}]{Brodie+14_SLUGGS_survey_Description}
{Brodie}, J.~P., {Romanowsky}, A.~J., {Strader}, J., {et~al.} 2014, \apj, 796, 52

\bibitem[{{Brodie} \& {Strader}(2006)}]{Brodie_Strader06_GC_review}
{Brodie}, J.~P. \& {Strader}, J. 2006, \araa, 44, 193

\bibitem[{{Bryan} {et~al.}(2013){Bryan}, {Kay}, {Duffy}, {Schaye}, {Dalla Vecchia}, \& {Booth}}]{Bryan+13_Shapes_of_DM_haloes_effect_of_baryons}
{Bryan}, S.~E., {Kay}, S.~T., {Duffy}, A.~R., {et~al.} 2013, \mnras, 429, 3316

\bibitem[{{Bullock}(2002)}]{Bullock02_WDM_CDM_shapes_proceedings}
{Bullock}, J.~S. 2002, in The Shapes of Galaxies and their Dark Halos, ed. P.~{Natarajan}, 109--113

\bibitem[{{Cappellari}(2002)}]{Cappellari02_mgefit}
{Cappellari}, M. 2002, \mnras, 333, 400

\bibitem[{{Cappellari}(2008)}]{Cappellari+08}
{Cappellari}, M. 2008, mnras, 390, 71

\bibitem[{{Cappellari} {et~al.}(2009){Cappellari}, {Neumayer}, {Reunanen}, {van der Werf}, {de Zeeuw}, \& {Rix}}]{Cappellari+09_CenA_central_SINFONI_kinenmatics}
{Cappellari}, M., {Neumayer}, N., {Reunanen}, J., {et~al.} 2009, \mnras, 394, 660

\bibitem[{{Cappellari} {et~al.}(2013){Cappellari}, {Scott}, {Alatalo}, {Blitz}, {Bois}, {Bournaud}, {Bureau}, {Crocker}, {Davies}, {Davis}, {de Zeeuw}, {Duc}, {Emsellem}, {Khochfar}, {Krajnovi{\'c}}, {Kuntschner}, {McDermid}, {Morganti}, {Naab}, {Oosterloo}, {Sarzi}, {Serra}, {Weijmans}, \& {Young}}]{Cappellari+13_dynamics_mass_DM_fraction}
{Cappellari}, M., {Scott}, N., {Alatalo}, K., {et~al.} 2013, \mnras, 432, 1709

\bibitem[{{Chaturvedi} {et~al.}(2022){Chaturvedi}, {Hilker}, {Cantiello}, {Napolitano}, {van de Ven}, {Spiniello}, {Fahrion}, {Paolillo}, {Gatto}, \& {Puzia}}]{Avinash+22_Fornax_kin}
{Chaturvedi}, A., {Hilker}, M., {Cantiello}, M., {et~al.} 2022, \aap, 657, A93

\bibitem[{{Chua} {et~al.}(2019){Chua}, {Pillepich}, {Vogelsberger}, \& {Hernquist}}]{Chua2019_halo_shapes}
{Chua}, K. T.~E., {Pillepich}, A., {Vogelsberger}, M., \& {Hernquist}, L. 2019, mnras, 484, 476

\bibitem[{{Ciotti} \& {Bertin}(1999)}]{Ciotti_Bertin99_Sersic_bn_expansion}
{Ciotti}, L. \& {Bertin}, G. 1999, \aap, 352, 447

\bibitem[{{Coccato} {et~al.}(2013){Coccato}, {Arnaboldi}, \& {Gerhard}}]{Coccato+13_Kin_signatures_of_accretion_events}
{Coccato}, L., {Arnaboldi}, M., \& {Gerhard}, O. 2013, \mnras, 436, 1322

\bibitem[{{Cole} \& {Lacey}(1996)}]{Cole+Lacey96}
{Cole}, S. \& {Lacey}, C. 1996, \mnras, 281, 716

\bibitem[{{Correa} {et~al.}(2015){Correa}, {Wyithe}, {Schaye}, \& {Duffy}}]{Correa+15_M200-c200_relation}
{Correa}, C.~A., {Wyithe}, J. S.~B., {Schaye}, J., \& {Duffy}, A.~R. 2015, \mnras, 452, 1217

\bibitem[{{Crnojevi{\'c}} {et~al.}(2016){Crnojevi{\'c}}, {Sand}, {Spekkens}, {Caldwell}, {Guhathakurta}, {McLeod}, {Seth}, {Simon}, {Strader}, \& {Toloba}}]{PISCeS_Crnojevic16}
{Crnojevi{\'c}}, D., {Sand}, D.~J., {Spekkens}, K., {et~al.} 2016, \apj, 823, 19

\bibitem[{{de Vaucouleurs} {et~al.}(1991){de Vaucouleurs}, {de Vaucouleurs}, {Corwin}, {Buta}, {Paturel}, \& {Fouque}}]{RC3}
{de Vaucouleurs}, G., {de Vaucouleurs}, A., {Corwin}, Herold~G., J., {et~al.} 1991, {Third Reference Catalogue of Bright Galaxies}

\bibitem[{{Deason} {et~al.}(2012){Deason}, {Belokurov}, {Evans}, \& {McCarthy}}]{Deason+12_DM_fraction_5Re}
{Deason}, A.~J., {Belokurov}, V., {Evans}, N.~W., \& {McCarthy}, I.~G. 2012, \apj, 748, 2

\bibitem[{{Debattista} {et~al.}(2013){Debattista}, {Ro{\v{s}}kar}, {Valluri}, {Quinn}, {Moore}, \& {Wadsley}}]{Debattista+13}
{Debattista}, V.~P., {Ro{\v{s}}kar}, R., {Valluri}, M., {et~al.} 2013, \mnras, 434, 2971

\bibitem[{{Dufour} {et~al.}(1979){Dufour}, {van den Bergh}, {Harvel}, {Martins}, {Schiffer}, {Talbot}, {Talent}, \& {Wells}}]{Dufour79+stellar_profile}
{Dufour}, R.~J., {van den Bergh}, S., {Harvel}, C.~A., {et~al.} 1979, \aj, 84, 284

\bibitem[{{Dumont} {et~al.}(2023){Dumont}, {Seth}, {Strader}, {Sand}, {Voggel}, {Hughes}, {Crnojevi{\'c}}, {Forbes}, {Mateo}, \& {Pearson}}]{Dumont+23_CJAM_modelling_Spherical_mass}
{Dumont}, A., {Seth}, A.~C., {Strader}, J., {et~al.} 2023, arXiv e-prints, arXiv:2306.11786

\bibitem[{{Dumont} {et~al.}(2022){Dumont}, {Seth}, {Strader}, {Voggel}, {Sand}, {Hughes}, {Caldwell}, {Crnojevi{\'c}}, {Mateo}, {Bailey}, \& {Forbes}}]{Dumont+22_UCDs_CenA}
{Dumont}, A., {Seth}, A.~C., {Strader}, J., {et~al.} 2022, \apj, 929, 147

\bibitem[{{Emsellem} {et~al.}(1994){Emsellem}, {Monnet}, \& {Bacon}}]{Emsellem94_MGE}
{Emsellem}, E., {Monnet}, G., \& {Bacon}, R. 1994, \aap, 285, 723

\bibitem[{{Ene} {et~al.}(2018){Ene}, {Ma}, {Veale}, {Greene}, {Thomas}, {Blakeslee}, {Foster}, {Walsh}, {Ito}, \& {Goulding}}]{Ene+18_Kinematics_MASSIVE_survey_shapes}
{Ene}, I., {Ma}, C.-P., {Veale}, M., {et~al.} 2018, \mnras, 479, 2810

\bibitem[{{Evans} {et~al.}(2003){Evans}, {Wilkinson}, {Perrett}, \& {Bridges}}]{Evans+03_TME}
{Evans}, N.~W., {Wilkinson}, M.~I., {Perrett}, K.~M., \& {Bridges}, T.~J. 2003, \apj, 583, 752

\bibitem[{{Foreman-Mackey} {et~al.}(2013){Foreman-Mackey}, {Conley}, {Meierjurgen Farr}, {Hogg}, {Lang}, {Marshall}, {Price-Whelan}, {Sanders}, \& {Zuntz}}]{emcee}
{Foreman-Mackey}, D., {Conley}, A., {Meierjurgen Farr}, W., {et~al.} 2013, {emcee: The MCMC Hammer}, Astrophysics Source Code Library, record ascl:1303.002

\bibitem[{{Foster} {et~al.}(2016){Foster}, {Pastorello}, {Roediger}, {Brodie}, {Forbes}, {Kartha}, {Pota}, {Romanowsky}, {Spitler}, {Strader}, {Usher}, \& {Arnold}}]{Foster+16_SLUGGS_shapes}
{Foster}, C., {Pastorello}, N., {Roediger}, J., {et~al.} 2016, \mnras, 457, 147

\bibitem[{{Foster} {et~al.}(2017){Foster}, {van de Sande}, {D'Eugenio}, {Cortese}, {McDermid}, {Bland-Hawthorn}, {Brough}, {Bryant}, {Croom}, {Goodwin}, {Konstantopoulos}, {Lawrence}, {L{\'o}pez-S{\'a}nchez}, {Medling}, {Owers}, {Richards}, {Scott}, {Taranu}, {Tonini}, \& {Zafar}}]{Foster+17_SAMI_intrinsic_shapes_of_stellar_ETGs}
{Foster}, C., {van de Sande}, J., {D'Eugenio}, F., {et~al.} 2017, \mnras, 472, 966

\bibitem[{{Georgiev} {et~al.}(2010){Georgiev}, {Puzia}, {Goudfrooij}, \& {Hilker}}]{Georgiev+10}
{Georgiev}, I.~Y., {Puzia}, T.~H., {Goudfrooij}, P., \& {Hilker}, M. 2010, \mnras, 406, 1967

\bibitem[{{Gerhard}(2013)}]{Gerhard13_DM_shapes_masses_proceeding}
{Gerhard}, O. 2013, in The Intriguing Life of Massive Galaxies, ed. D.~{Thomas}, A.~{Pasquali}, \& I.~{Ferreras}, Vol. 295, 211--220

\bibitem[{{Ha{\c{s}}egan} {et~al.}(2005){Ha{\c{s}}egan}, {Jord{\'a}n}, {C{\^o}t{\'e}}, {Djorgovski}, {McLaughlin}, {Blakeslee}, {Mei}, {West}, {Peng}, {Ferrarese}, {Milosavljevi{\'c}}, {Tonry}, \& {Merritt}}]{Hacegan+05_ACSVCS_GCs_UCDs}
{Ha{\c{s}}egan}, M., {Jord{\'a}n}, A., {C{\^o}t{\'e}}, P., {et~al.} 2005, \apj, 627, 203

\bibitem[{{Han} {et~al.}(2023){Han}, {Conroy}, \& {Hernquist}}]{Han+23_tilted_MW_DM_halo}
{Han}, J.~J., {Conroy}, C., \& {Hernquist}, L. 2023, Nature Astronomy [\eprint[arXiv]{2309.07209}]

\bibitem[{{Han} {et~al.}(2022){Han}, {Conroy}, {Johnson}, {Speagle}, {Bonaca}, {Chandra}, {Naidu}, {Ting}, {Woody}, \& {Zaritsky}}]{Han+22}
{Han}, J.~J., {Conroy}, C., {Johnson}, B.~D., {et~al.} 2022, \aj, 164, 249

\bibitem[{{Harris} {et~al.}(2020{\natexlab{a}}){Harris}, {Millman}, {van der Walt}, {Gommers}, {Virtanen}, {Cournapeau}, {Wieser}, {Taylor}, {Berg}, {Smith}, {Kern}, {Picus}, {Hoyer}, {van Kerkwijk}, {Brett}, {Haldane}, {del R{\'\i}o}, {Wiebe}, {Peterson}, {G{\'e}rard-Marchant}, {Sheppard}, {Reddy}, {Weckesser}, {Abbasi}, {Gohlke}, \& {Oliphant}}]{Numpy2020}
{Harris}, C.~R., {Millman}, K.~J., {van der Walt}, S.~J., {et~al.} 2020{\natexlab{a}}, \nat, 585, 357

\bibitem[{{Harris}(2010)}]{Harris_G10_GiantBeneath}
{Harris}, G. L.~H. 2010, \pasa, 27, 475

\bibitem[{{Harris} {et~al.}(2010){Harris}, {Rejkuba}, \& {Harris}}]{Harris_G+10_distance_to_CenA}
{Harris}, G. L.~H., {Rejkuba}, M., \& {Harris}, W.~E. 2010, pasa, 27, 457

\bibitem[{{Harris}(2023)}]{Harris23_GCS_BrightestGalaxies}
{Harris}, W.~E. 2023, \apjs, 265, 9

\bibitem[{{Harris} {et~al.}(2017){Harris}, {Blakeslee}, \& {Harris}}]{Harris+17}
{Harris}, W.~E., {Blakeslee}, J.~P., \& {Harris}, G. L.~H. 2017, \apj, 836, 67

\bibitem[{{Harris} {et~al.}(2006){Harris}, {Harris}, {Barmby}, {McLaughlin}, \& {Forbes}}]{Harris+06_HST_obs_of_CenA_GCs}
{Harris}, W.~E., {Harris}, G. L.~H., {Barmby}, P., {McLaughlin}, D.~E., \& {Forbes}, D.~A. 2006, \aj, 132, 2187

\bibitem[{{Harris} {et~al.}(2014){Harris}, {Morningstar}, {Gnedin}, {O'Halloran}, {Blakeslee}, {Whitmore}, {C{\^o}t{\'e}}, {Geisler}, {Peng}, {Bailin}, {Rothberg}, {Cockcroft}, \& {Barber DeGraaff}}]{Harris+14_GCLF}
{Harris}, W.~E., {Morningstar}, W., {Gnedin}, O.~Y., {et~al.} 2014, \apj, 797, 128

\bibitem[{{Harris} {et~al.}(2020{\natexlab{b}}){Harris}, {Remus}, {Harris}, \& {Babyk}}]{Harris+20_DM_fraction_5Re}
{Harris}, W.~E., {Remus}, R.-S., {Harris}, G. L.~H., \& {Babyk}, I.~V. 2020{\natexlab{b}}, \apj, 905, 28

\bibitem[{{Hayashi} {et~al.}(2007){Hayashi}, {Navarro}, \& {Springel}}]{Hayashi+07_DM_shapes}
{Hayashi}, E., {Navarro}, J.~F., \& {Springel}, V. 2007, \mnras, 377, 50

\bibitem[{{Hudson} {et~al.}(2014){Hudson}, {Harris}, \& {Harris}}]{Hudson+14}
{Hudson}, M.~J., {Harris}, G.~L., \& {Harris}, W.~E. 2014, \apjl, 787, L5

\bibitem[{{Hughes} {et~al.}(2023){Hughes}, {Sand}, {Seth}, {Strader}, {Lidman}, {Voggel}, {Dumont}, {Crnojevi{\'c}}, {Mateo}, {Caldwell}, {Forbes}, {Pearson}, {Guhathakurta}, \& {Toloba}}]{Hughes+23_Kinematics_dyn_estimate}
{Hughes}, A.~K., {Sand}, D.~J., {Seth}, A., {et~al.} 2023, \apj, 947, 34

\bibitem[{{Hughes} {et~al.}(2021{\natexlab{a}}){Hughes}, {Sand}, {Seth}, {Strader}, {Voggel}, {Dumont}, {Crnojevi{\'c}}, {Caldwell}, {Forbes}, {Simon}, {Guhathakurta}, \& {Toloba}}]{Hughes+21}
{Hughes}, A.~K., {Sand}, D.~J., {Seth}, A., {et~al.} 2021{\natexlab{a}}, \apj, 914, 16

\bibitem[{{Hughes} {et~al.}(2021{\natexlab{b}}){Hughes}, {Jethwa}, {Hilker}, {van de Ven}, {Martig}, {Pfeffer}, {Bastian}, {Kruijssen}, {Trujillo-Gomez}, {Reina-Campos}, \& {Crain}}]{Hughes+21_dyn_modeling}
{Hughes}, M.~E., {Jethwa}, P., {Hilker}, M., {et~al.} 2021{\natexlab{b}}, mnras, 502, 2828

\bibitem[{{Hui} {et~al.}(1995){Hui}, {Ford}, {Freeman}, \& {Dopita}}]{Hui95_PNe_kin_sys_vel_mass_shape}
{Hui}, X., {Ford}, H.~C., {Freeman}, K.~C., \& {Dopita}, M.~A. 1995, \apj, 449, 592

\bibitem[{Hunter(2007)}]{Matplotlib}
Hunter, J.~D. 2007, Computing in Science \& Engineering, 9, 90

\bibitem[{{Iodice} {et~al.}(2003){Iodice}, {Arnaboldi}, {Bournaud}, {Combes}, {Sparke}, {van Driel}, \& {Capaccioli}}]{Iodice+03_polar_ring_galaxies_shape_constraints}
{Iodice}, E., {Arnaboldi}, M., {Bournaud}, F., {et~al.} 2003, \apj, 585, 730

\bibitem[{{Iorio} \& {Belokurov}(2019)}]{Iorio+Belokurov19}
{Iorio}, G. \& {Belokurov}, V. 2019, \mnras, 482, 3868

\bibitem[{{Jin} {et~al.}(2020){Jin}, {Zhu}, {Long}, {Mao}, {Wang}, \& {van de Ven}}]{Jin+20_Manga_shapes_ETGs}
{Jin}, Y., {Zhu}, L., {Long}, R.~J., {et~al.} 2020, \mnras, 491, 1690

\bibitem[{{Jord{\'a}n} {et~al.}(2006){Jord{\'a}n}, {McLaughlin}, {C{\^o}t{\'e}}, {Ferrarese}, {Peng}, {Blakeslee}, {Mei}, {Villegas}, {Merritt}, {Tonry}, \& {West}}]{Jordan+06_GCLF}
{Jord{\'a}n}, A., {McLaughlin}, D.~E., {C{\^o}t{\'e}}, P., {et~al.} 2006, \apjl, 651, L25

\bibitem[{{Jord{\'a}n} {et~al.}(2015){Jord{\'a}n}, {Peng}, {Blakeslee}, {C{\^o}t{\'e}}, {Eyheramendy}, \& {Ferrarese}}]{Jordan+15_ACSFCS_GCs}
{Jord{\'a}n}, A., {Peng}, E.~W., {Blakeslee}, J.~P., {et~al.} 2015, \apjs, 221, 13

\bibitem[{{Jord{\'a}n} {et~al.}(2009){Jord{\'a}n}, {Peng}, {Blakeslee}, {C{\^o}t{\'e}}, {Eyheramendy}, {Ferrarese}, {Mei}, {Tonry}, \& {West}}]{Jordan+09_ACSVCS_GCs}
{Jord{\'a}n}, A., {Peng}, E.~W., {Blakeslee}, J.~P., {et~al.} 2009, \apjs, 180, 54

\bibitem[{{Kazantzidis} {et~al.}(2004){Kazantzidis}, {Kravtsov}, {Zentner}, {Allgood}, {Nagai}, \& {Moore}}]{Kazantzidis+04_gas_cooling_on_DM_shapes}
{Kazantzidis}, S., {Kravtsov}, A.~V., {Zentner}, A.~R., {et~al.} 2004, \apjl, 611, L73

\bibitem[{{Khoperskov} {et~al.}(2014){Khoperskov}, {Moiseev}, {Khoperskov}, \& {Saburova}}]{Khoperskov+14}
{Khoperskov}, S.~A., {Moiseev}, A.~V., {Khoperskov}, A.~V., \& {Saburova}, A.~S. 2014, mnras, 441, 2650

\bibitem[{{Komatsu} {et~al.}(2009){Komatsu}, {Dunkley}, {Nolta}, {Bennett}, {Gold}, {Hinshaw}, {Jarosik}, {Larson}, {Limon}, {Page}, {Spergel}, {Halpern}, {Hill}, {Kogut}, {Meyer}, {Tucker}, {Weiland}, {Wollack}, \& {Wright}}]{WMAP+09}
{Komatsu}, E., {Dunkley}, J., {Nolta}, M.~R., {et~al.} 2009, \apjs, 180, 330

\bibitem[{{Kraft} {et~al.}(2003){Kraft}, {V{\'a}zquez}, {Forman}, {Jones}, {Murray}, {Hardcastle}, {Worrall}, \& {Churazov}}]{Kraft+03_X_ray_mass}
{Kraft}, R.~P., {V{\'a}zquez}, S.~E., {Forman}, W.~R., {et~al.} 2003, \apj, 592, 129

\bibitem[{{Krajnovi{\'c}} {et~al.}(2005){Krajnovi{\'c}}, {Cappellari}, {Emsellem}, {McDermid}, \& {de Zeeuw}}]{Krajnovic+05_regular_data_inclination_not_constrained}
{Krajnovi{\'c}}, D., {Cappellari}, M., {Emsellem}, E., {McDermid}, R.~M., \& {de Zeeuw}, P.~T. 2005, \mnras, 357, 1113

\bibitem[{{Law} {et~al.}(2009){Law}, {Majewski}, \& {Johnston}}]{Law+09_Triaxial_tilted_DM_of_MW}
{Law}, D.~R., {Majewski}, S.~R., \& {Johnston}, K.~V. 2009, \apjl, 703, L67

\bibitem[{{Leung} {et~al.}(2021){Leung}, {Leaman}, {Battaglia}, {van de Ven}, {Brooks}, {Pe{\~n}arrubia}, \& {Venn}}]{Leung+19}
{Leung}, G. Y.~C., {Leaman}, R., {Battaglia}, G., {et~al.} 2021, mnras, 500, 410

\bibitem[{{Malin} {et~al.}(1983){Malin}, {Quinn}, \& {Graham}}]{Malin+83_CenA_Shells}
{Malin}, D.~F., {Quinn}, P.~J., \& {Graham}, J.~A. 1983, \apjl, 272, L5

\bibitem[{{Mathieu} {et~al.}(1996){Mathieu}, {Dejonghe}, \& {Hui}}]{Mathieu+96_dynamical_model_CenA_PNe}
{Mathieu}, A., {Dejonghe}, H., \& {Hui}, X. 1996, \aap, 309, 30

\bibitem[{{Mayer} {et~al.}(2002){Mayer}, {Moore}, {Quinn}, {Governato}, \& {Stadel}}]{Mayer+2002_halo_shapes_WDM_FDM_CDM}
{Mayer}, L., {Moore}, B., {Quinn}, T., {Governato}, F., \& {Stadel}, J. 2002, \mnras, 336, 119

\bibitem[{{Mieske} {et~al.}(2008){Mieske}, {Hilker}, {Jord{\'a}n}, {Infante}, {Kissler-Patig}, {Rejkuba}, {Richtler}, {C{\^o}t{\'e}}, {Baumgardt}, {West}, {Ferrarese}, \& {Peng}}]{Mieske+08_UCDs_GC_difference}
{Mieske}, S., {Hilker}, M., {Jord{\'a}n}, A., {et~al.} 2008, \aap, 487, 921

\bibitem[{{M{\"u}ller} {et~al.}(2022){M{\"u}ller}, {Lelli}, {Famaey}, {Pawlowski}, {Fahrion}, {Rejkuba}, {Hilker}, \& {Jerjen}}]{Muller+22_dynmaical_mass_with_satellites}
{M{\"u}ller}, O., {Lelli}, F., {Famaey}, B., {et~al.} 2022, \aap, 662, A57

\bibitem[{{M{\"u}ller} {et~al.}(2018){M{\"u}ller}, {Pawlowski}, {Jerjen}, \& {Lelli}}]{Mueller+18_Science}
{M{\"u}ller}, O., {Pawlowski}, M.~S., {Jerjen}, H., \& {Lelli}, F. 2018, Science, 359, 534

\bibitem[{{M{\"u}ller} {et~al.}(2021){M{\"u}ller}, {Pawlowski}, {Lelli}, {Fahrion}, {Rejkuba}, {Hilker}, {Kanehisa}, {Libeskind}, \& {Jerjen}}]{Mueller+21_CenA_satellites_plane}
{M{\"u}ller}, O., {Pawlowski}, M.~S., {Lelli}, F., {et~al.} 2021, \aap, 645, L5

\bibitem[{{M{\"u}ller} {et~al.}(2019){M{\"u}ller}, {Rejkuba}, {Pawlowski}, {Ibata}, {Lelli}, {Hilker}, \& {Jerjen}}]{Muller+19_Dwarfs_of_CenA_shape}
{M{\"u}ller}, O., {Rejkuba}, M., {Pawlowski}, M.~S., {et~al.} 2019, \aap, 629, A18

\bibitem[{{Naidu} {et~al.}(2021){Naidu}, {Conroy}, {Bonaca}, {Zaritsky}, {Weinberger}, {Ting}, {Caldwell}, {Tacchella}, {Han}, {Speagle}, \& {Cargile}}]{Naidu+21}
{Naidu}, R.~P., {Conroy}, C., {Bonaca}, A., {et~al.} 2021, \apj, 923, 92

\bibitem[{{Navarro} {et~al.}(1997){Navarro}, {Frenk}, \& {White}}]{NFW+97}
{Navarro}, J.~F., {Frenk}, C.~S., \& {White}, S. D.~M. 1997, \apj, 490, 493

\bibitem[{{Nelson} {et~al.}(2019){Nelson}, {Springel}, {Pillepich}, {Rodriguez-Gomez}, {Torrey}, {Genel}, {Vogelsberger}, {Pakmor}, {Marinacci}, {Weinberger}, {Kelley}, {Lovell}, {Diemer}, \& {Hernquist}}]{Nelson+19_TNGSims}
{Nelson}, D., {Springel}, V., {Pillepich}, A., {et~al.} 2019, Computational Astrophysics and Cosmology, 6, 2

\bibitem[{{Neto} {et~al.}(2007){Neto}, {Gao}, {Bett}, {Cole}, {Navarro}, {Frenk}, {White}, {Springel}, \& {Jenkins}}]{Neto+07_c200-M200_uncertainties}
{Neto}, A.~F., {Gao}, L., {Bett}, P., {et~al.} 2007, \mnras, 381, 1450

\bibitem[{{Pearson} {et~al.}(2015){Pearson}, {K{\"u}pper}, {Johnston}, \& {Price-Whelan}}]{Pearson+15_tidal_stream_morphology_constraints_on_DM_shape}
{Pearson}, S., {K{\"u}pper}, A. H.~W., {Johnston}, K.~V., \& {Price-Whelan}, A.~M. 2015, \apj, 799, 28

\bibitem[{{Pearson} {et~al.}(2022){Pearson}, {Price-Whelan}, {Hogg}, {Seth}, {Sand}, {Hunt}, \& {Crnojevi{\'c}}}]{Pearson+22_StreamModdelingMass}
{Pearson}, S., {Price-Whelan}, A.~M., {Hogg}, D.~W., {et~al.} 2022, \apj, 941, 19

\bibitem[{{Peng} {et~al.}(2004{\natexlab{a}}){Peng}, {Ford}, \& {Freeman}}]{Peng+04_PNe_dyna_outer_halo}
{Peng}, E.~W., {Ford}, H.~C., \& {Freeman}, K.~C. 2004{\natexlab{a}}, apj, 602, 685

\bibitem[{{Peng} {et~al.}(2004{\natexlab{b}}){Peng}, {Ford}, \& {Freeman}}]{Peng+04_GCS_CenA_kin_formation_metal}
{Peng}, E.~W., {Ford}, H.~C., \& {Freeman}, K.~C. 2004{\natexlab{b}}, \apj, 602, 705

\bibitem[{{Planck Collaboration} {et~al.}(2016){Planck Collaboration}, {Ade}, {Aghanim}, {Arnaud}, {Ashdown}, {Aumont}, {Baccigalupi}, {Banday}, {Barreiro}, {Bartlett}, {Bartolo}, {Battaner}, {Battye}, {Benabed}, {Beno{\^\i}t}, {Benoit-L{\'e}vy}, {Bernard}, {Bersanelli}, {Bielewicz}, {Bock}, {Bonaldi}, {Bonavera}, {Bond}, {Borrill}, {Bouchet}, {Boulanger}, {Bucher}, {Burigana}, {Butler}, {Calabrese}, {Cardoso}, {Catalano}, {Challinor}, {Chamballu}, {Chary}, {Chiang}, {Chluba}, {Christensen}, {Church}, {Clements}, {Colombi}, {Colombo}, {Combet}, {Coulais}, {Crill}, {Curto}, {Cuttaia}, {Danese}, {Davies}, {Davis}, {de Bernardis}, {de Rosa}, {de Zotti}, {Delabrouille}, {D{\'e}sert}, {Di Valentino}, {Dickinson}, {Diego}, {Dolag}, {Dole}, {Donzelli}, {Dor{\'e}}, {Douspis}, {Ducout}, {Dunkley}, {Dupac}, {Efstathiou}, {Elsner}, {En{\ss}lin}, {Eriksen}, {Farhang}, {Fergusson}, {Finelli}, {Forni}, {Frailis}, {Fraisse}, {Franceschi}, {Frejsel}, {Galeotta}, {Galli}, {Ganga}, {Gauthier}, {Gerbino}, {Ghosh}, {Giard},
  {Giraud-H{\'e}raud}, {Giusarma}, {Gjerl{\o}w}, {Gonz{\'a}lez-Nuevo}, {G{\'o}rski}, {Gratton}, {Gregorio}, {Gruppuso}, {Gudmundsson}, {Hamann}, {Hansen}, {Hanson}, {Harrison}, {Helou}, {Henrot-Versill{\'e}}, {Hern{\'a}ndez-Monteagudo}, {Herranz}, {Hildebrandt}, {Hivon}, {Hobson}, {Holmes}, {Hornstrup}, {Hovest}, {Huang}, {Huffenberger}, {Hurier}, {Jaffe}, {Jaffe}, {Jones}, {Juvela}, {Keih{\"a}nen}, {Keskitalo}, {Kisner}, {Kneissl}, {Knoche}, {Knox}, {Kunz}, {Kurki-Suonio}, {Lagache}, {L{\"a}hteenm{\"a}ki}, {Lamarre}, {Lasenby}, {Lattanzi}, {Lawrence}, {Leahy}, {Leonardi}, {Lesgourgues}, {Levrier}, {Lewis}, {Liguori}, {Lilje}, {Linden-V{\o}rnle}, {L{\'o}pez-Caniego}, {Lubin}, {Mac{\'\i}as-P{\'e}rez}, {Maggio}, {Maino}, {Mandolesi}, {Mangilli}, {Marchini}, {Maris}, {Martin}, {Martinelli}, {Mart{\'\i}nez-Gonz{\'a}lez}, {Masi}, {Matarrese}, {McGehee}, {Meinhold}, {Melchiorri}, {Melin}, {Mendes}, {Mennella}, {Migliaccio}, {Millea}, {Mitra}, {Miville-Desch{\^e}nes}, {Moneti}, {Montier}, {Morgante}, {Mortlock},
  {Moss}, {Munshi}, {Murphy}, {Naselsky}, {Nati}, {Natoli}, {Netterfield}, {N{\o}rgaard-Nielsen}, {Noviello}, {Novikov}, {Novikov}, {Oxborrow}, {Paci}, {Pagano}, {Pajot}, {Paladini}, {Paoletti}, {Partridge}, {Pasian}, {Patanchon}, {Pearson}, {Perdereau}, {Perotto}, {Perrotta}, {Pettorino}, {Piacentini}, {Piat}, {Pierpaoli}, {Pietrobon}, {Plaszczynski}, {Pointecouteau}, {Polenta}, {Popa}, {Pratt}, {Pr{\'e}zeau}, {Prunet}, {Puget}, {Rachen}, {Reach}, {Rebolo}, {Reinecke}, {Remazeilles}, {Renault}, {Renzi}, {Ristorcelli}, {Rocha}, {Rosset}, {Rossetti}, {Roudier}, {Rouill{\'e} d'Orfeuil}, {Rowan-Robinson}, {Rubi{\~n}o-Mart{\'\i}n}, {Rusholme}, {Said}, {Salvatelli}, {Salvati}, {Sandri}, {Santos}, {Savelainen}, {Savini}, {Scott}, {Seiffert}, {Serra}, {Shellard}, {Spencer}, {Spinelli}, {Stolyarov}, {Stompor}, {Sudiwala}, {Sunyaev}, {Sutton}, {Suur-Uski}, {Sygnet}, {Tauber}, {Terenzi}, {Toffolatti}, {Tomasi}, {Tristram}, {Trombetti}, {Tucci}, {Tuovinen}, {T{\"u}rler}, {Umana}, {Valenziano}, {Valiviita}, {Van Tent},
  {Vielva}, {Villa}, {Wade}, {Wandelt}, {Wehus}, {White}, {White}, {Wilkinson}, {Yvon}, {Zacchei}, \& {Zonca}}]{Planck+15_cosmological_parameters}
{Planck Collaboration}, {Ade}, P.~A.~R., {Aghanim}, N., {et~al.} 2016, \aap, 594, A13

\bibitem[{{Planck Collaboration} {et~al.}(2020){Planck Collaboration}, {Aghanim}, {Akrami}, {Ashdown}, {Aumont}, {Baccigalupi}, {Ballardini}, {Banday}, {Barreiro}, {Bartolo}, {Basak}, {Battye}, {Benabed}, {Bernard}, {Bersanelli}, {Bielewicz}, {Bock}, {Bond}, {Borrill}, {Bouchet}, {Boulanger}, {Bucher}, {Burigana}, {Butler}, {Calabrese}, {Cardoso}, {Carron}, {Challinor}, {Chiang}, {Chluba}, {Colombo}, {Combet}, {Contreras}, {Crill}, {Cuttaia}, {de Bernardis}, {de Zotti}, {Delabrouille}, {Delouis}, {Di Valentino}, {Diego}, {Dor{\'e}}, {Douspis}, {Ducout}, {Dupac}, {Dusini}, {Efstathiou}, {Elsner}, {En{\ss}lin}, {Eriksen}, {Fantaye}, {Farhang}, {Fergusson}, {Fernandez-Cobos}, {Finelli}, {Forastieri}, {Frailis}, {Fraisse}, {Franceschi}, {Frolov}, {Galeotta}, {Galli}, {Ganga}, {G{\'e}nova-Santos}, {Gerbino}, {Ghosh}, {Gonz{\'a}lez-Nuevo}, {G{\'o}rski}, {Gratton}, {Gruppuso}, {Gudmundsson}, {Hamann}, {Handley}, {Hansen}, {Herranz}, {Hildebrandt}, {Hivon}, {Huang}, {Jaffe}, {Jones}, {Karakci}, {Keih{\"a}nen},
  {Keskitalo}, {Kiiveri}, {Kim}, {Kisner}, {Knox}, {Krachmalnicoff}, {Kunz}, {Kurki-Suonio}, {Lagache}, {Lamarre}, {Lasenby}, {Lattanzi}, {Lawrence}, {Le Jeune}, {Lemos}, {Lesgourgues}, {Levrier}, {Lewis}, {Liguori}, {Lilje}, {Lilley}, {Lindholm}, {L{\'o}pez-Caniego}, {Lubin}, {Ma}, {Mac{\'\i}as-P{\'e}rez}, {Maggio}, {Maino}, {Mandolesi}, {Mangilli}, {Marcos-Caballero}, {Maris}, {Martin}, {Martinelli}, {Mart{\'\i}nez-Gonz{\'a}lez}, {Matarrese}, {Mauri}, {McEwen}, {Meinhold}, {Melchiorri}, {Mennella}, {Migliaccio}, {Millea}, {Mitra}, {Miville-Desch{\^e}nes}, {Molinari}, {Montier}, {Morgante}, {Moss}, {Natoli}, {N{\o}rgaard-Nielsen}, {Pagano}, {Paoletti}, {Partridge}, {Patanchon}, {Peiris}, {Perrotta}, {Pettorino}, {Piacentini}, {Polastri}, {Polenta}, {Puget}, {Rachen}, {Reinecke}, {Remazeilles}, {Renzi}, {Rocha}, {Rosset}, {Roudier}, {Rubi{\~n}o-Mart{\'\i}n}, {Ruiz-Granados}, {Salvati}, {Sandri}, {Savelainen}, {Scott}, {Shellard}, {Sirignano}, {Sirri}, {Spencer}, {Sunyaev}, {Suur-Uski}, {Tauber}, {Tavagnacco},
  {Tenti}, {Toffolatti}, {Tomasi}, {Trombetti}, {Valenziano}, {Valiviita}, {Van Tent}, {Vibert}, {Vielva}, {Villa}, {Vittorio}, {Wandelt}, {Wehus}, {White}, {White}, {Zacchei}, \& {Zonca}}]{Planck_results18}
{Planck Collaboration}, {Aghanim}, N., {Akrami}, Y., {et~al.} 2020, \aap, 641, A6

\bibitem[{{Poci} {et~al.}(2017){Poci}, {Cappellari}, \& {McDermid}}]{Poci+17_JAM_modelling}
{Poci}, A., {Cappellari}, M., \& {McDermid}, R.~M. 2017, \mnras, 467, 1397

\bibitem[{{Posti} \& {Helmi}(2019)}]{Posti+Helmi19}
{Posti}, L. \& {Helmi}, A. 2019, \aap, 621, A56

\bibitem[{{Prada} {et~al.}(2012){Prada}, {Klypin}, {Cuesta}, {Betancort-Rijo}, \& {Primack}}]{Prada+12_M200-c200}
{Prada}, F., {Klypin}, A.~A., {Cuesta}, A.~J., {Betancort-Rijo}, J.~E., \& {Primack}, J. 2012, \mnras, 423, 3018

\bibitem[{{Prada} {et~al.}(2019){Prada}, {Forero-Romero}, {Grand}, {Pakmor}, \& {Springel}}]{Prada+19}
{Prada}, J., {Forero-Romero}, J.~E., {Grand}, R. J.~J., {Pakmor}, R., \& {Springel}, V. 2019, \mnras, 490, 4877

\bibitem[{{Pulsoni} {et~al.}(2018){Pulsoni}, {Gerhard}, {Arnaboldi}, {Coccato}, {Longobardi}, {Napolitano}, {Moylan}, {Narayan}, {Gupta}, {Burkert}, {Capaccioli}, {Chies-Santos}, {Cortesi}, {Freeman}, {Kuijken}, {Merrifield}, {Romanowsky}, \& {Tortora}}]{Pulsoni+18}
{Pulsoni}, C., {Gerhard}, O., {Arnaboldi}, M., {et~al.} 2018, \aap, 618, A94

\bibitem[{{Pulsoni} {et~al.}(2020){Pulsoni}, {Gerhard}, {Arnaboldi}, {Pillepich}, {Nelson}, {Hernquist}, \& {Springel}}]{Pulsoni+20_ETGs_sims_photo_kin}
{Pulsoni}, C., {Gerhard}, O., {Arnaboldi}, M., {et~al.} 2020, \aap, 641, A60

\bibitem[{{Pulsoni} {et~al.}(2021){Pulsoni}, {Gerhard}, {Arnaboldi}, {Pillepich}, {Rodriguez-Gomez}, {Nelson}, {Hernquist}, \& {Springel}}]{Pulsoni+21}
{Pulsoni}, C., {Gerhard}, O., {Arnaboldi}, M., {et~al.} 2021, \aap, 647, A95

\bibitem[{{Rejkuba}(2012)}]{Rejkuba12_GCLF}
{Rejkuba}, M. 2012, \apss, 341, 195

\bibitem[{{Rejkuba} {et~al.}(2005){Rejkuba}, {Greggio}, {Harris}, {Harris}, \& {Peng}}]{Rejkuba+05_HST_observations_of_halo_stars}
{Rejkuba}, M., {Greggio}, L., {Harris}, W.~E., {Harris}, G. L.~H., \& {Peng}, E.~W. 2005, \apj, 631, 262

\bibitem[{{Rejkuba} {et~al.}(2022){Rejkuba}, {Harris}, {Greggio}, {Crnojevi{\'c}}, \& {Harris}}]{Rejkuba+22_RGBs}
{Rejkuba}, M., {Harris}, W.~E., {Greggio}, L., {Crnojevi{\'c}}, D., \& {Harris}, G.~L.~H. 2022, \aap, 657, A41

\bibitem[{{Rejkuba} {et~al.}(2011){Rejkuba}, {Harris}, {Greggio}, \& {Harris}}]{Rejkuba+11_ages_of_halo_stars_CenA}
{Rejkuba}, M., {Harris}, W.~E., {Greggio}, L., \& {Harris}, G.~L.~H. 2011, \aap, 526, A123

\bibitem[{{Robison} {et~al.}(2023){Robison}, {Hudson}, {Cuillandre}, {Erben}, {Fabbro}, {Gavazzi}, {Guinot}, {Gwyn}, {Hildebrandt}, {Kilbinger}, {McConnachie}, {Miller}, {Spitzer}, \& {van Waerbeke}}]{Robison+23_Shapes_of_DM_weak_lensing}
{Robison}, B., {Hudson}, M.~J., {Cuillandre}, J.-C., {et~al.} 2023, \mnras, 523, 1614

\bibitem[{{Romanowsky} {et~al.}(2009){Romanowsky}, {Strader}, {Spitler}, {Johnson}, {Brodie}, {Forbes}, \& {Ponman}}]{Romanowsky+09_NGC1407_modelling}
{Romanowsky}, A.~J., {Strader}, J., {Spitler}, L.~R., {et~al.} 2009, \aj, 137, 4956

\bibitem[{{Santucci} {et~al.}(2022){Santucci}, {Brough}, {van de Sande}, {McDermid}, {van de Ven}, {Zhu}, {D'Eugenio}, {Bland-Hawthorn}, {Barsanti}, {Bryant}, {Croom}, {Davies}, {Green}, {Lawrence}, {Lorente}, {Owers}, {Poci}, {Richards}, {Thater}, \& {Yi}}]{Santucci+22_Sami_galaxies_mass_distribution}
{Santucci}, G., {Brough}, S., {van de Sande}, J., {et~al.} 2022, \apj, 930, 153

\bibitem[{{Schlafly} \& {Finkbeiner}(2011)}]{Schlafly&Finkbeiner11_reddening}
{Schlafly}, E.~F. \& {Finkbeiner}, D.~P. 2011, \apj, 737, 103

\bibitem[{{Schlegel} {et~al.}(1998){Schlegel}, {Finkbeiner}, \& {Davis}}]{Schlegel+98_SFD}
{Schlegel}, D.~J., {Finkbeiner}, D.~P., \& {Davis}, M. 1998, \apj, 500, 525

\bibitem[{{Sersic}(1968)}]{Sersic+68_book_reference}
{Sersic}, J.~L. 1968, {Atlas de Galaxias Australes}

\bibitem[{{Sesar} {et~al.}(2011){Sesar}, {Juri{\'c}}, \& {Ivezi{\'c}}}]{Sesar+11}
{Sesar}, B., {Juri{\'c}}, M., \& {Ivezi{\'c}}, {\v{Z}}. 2011, \apj, 731, 4

\bibitem[{{Spitler} \& {Forbes}(2009)}]{Spitler+Forbes09}
{Spitler}, L.~R. \& {Forbes}, D.~A. 2009, \mnras, 392, L1

\bibitem[{{Taylor} {et~al.}(2016){Taylor}, {Mu{\~n}oz}, {Puzia}, {Mieske}, {Eigenthaler}, \& {Bovill}}]{Taylor+16_SCABS_dataset}
{Taylor}, M.~A., {Mu{\~n}oz}, R.~P., {Puzia}, T.~H., {et~al.} 2016, arXiv e-prints, arXiv:1608.07285

\bibitem[{{Taylor} {et~al.}(2017){Taylor}, {Puzia}, {Mu{\~n}oz}, {Mieske}, {Lan{\c{c}}on}, {Zhang}, {Eigenthaler}, \& {Bovill}}]{Taylor+17}
{Taylor}, M.~A., {Puzia}, T.~H., {Mu{\~n}oz}, R.~P., {et~al.} 2017, \mnras, 469, 3444

\bibitem[{{Teklu} {et~al.}(2015){Teklu}, {Remus}, {Dolag}, {Beck}, {Burkert}, {Schmidt}, {Schulze}, \& {Steinborn}}]{Teklu+15}
{Teklu}, A.~F., {Remus}, R.-S., {Dolag}, K., {et~al.} 2015, \apj, 812, 29

\bibitem[{{Tenneti} {et~al.}(2015){Tenneti}, {Mandelbaum}, {Di Matteo}, {Kiessling}, \& {Khandai}}]{Tenneti+15_baryon_fb_on_shapes}
{Tenneti}, A., {Mandelbaum}, R., {Di Matteo}, T., {Kiessling}, A., \& {Khandai}, N. 2015, \mnras, 453, 469

\bibitem[{{Thater} {et~al.}(2023){Thater}, {Jethwa}, {Lilley}, {Zocchi}, {Santucci}, \& {van de Ven}}]{Thater+23_Atlas3D_shapes}
{Thater}, S., {Jethwa}, P., {Lilley}, E.~J., {et~al.} 2023, arXiv e-prints, arXiv:2305.09344

\bibitem[{{Valenzuela} \& {Remus}(2022)}]{Valenzuela&Remus22}
{Valenzuela}, L.~M. \& {Remus}, R.-S. 2022, arXiv e-prints, arXiv:2208.08443

\bibitem[{{van den Bosch} \& {van de Ven}(2009)}]{vdBosch_vdVen09_triaxiality}
{van den Bosch}, R. C.~E. \& {van de Ven}, G. 2009, \mnras, 398, 1117

\bibitem[{{Varghese} {et~al.}(2011){Varghese}, {Ibata}, \& {Lewis}}]{Varghese+11_Streams_DM_morphology}
{Varghese}, A., {Ibata}, R., \& {Lewis}, G.~F. 2011, \mnras, 417, 198

\bibitem[{{Ver{\v{s}}i{\v{c}}} {et~al.}(2024){Ver{\v{s}}i{\v{c}}}, {Thater}, {van de Ven}, {Watkins}, {Jethwa}, {Leaman}, \& {Zocchi}}]{Versic+23_SLUGGS}
{Ver{\v{s}}i{\v{c}}}, T., {Thater}, S., {van de Ven}, G., {et~al.} 2024, \aap, 681, A46

\bibitem[{{Villegas} {et~al.}(2010){Villegas}, {Jord{\'a}n}, {Peng}, {Blakeslee}, {C{\^o}t{\'e}}, {Ferrarese}, {Kissler-Patig}, {Mei}, {Infante}, {Tonry}, \& {West}}]{Villegas+10_ACSFCS_ACSVCS_GCLF}
{Villegas}, D., {Jord{\'a}n}, A., {Peng}, E.~W., {et~al.} 2010, \apj, 717, 603

\bibitem[{{Virtanen} {et~al.}(2020){Virtanen}, {Gommers}, {Oliphant}, {Haberland}, {Reddy}, {Cournapeau}, {Burovski}, {Peterson}, {Weckesser}, {Bright}, {van der Walt}, {Brett}, {Wilson}, {Millman}, {Mayorov}, {Nelson}, {Jones}, {Kern}, {Larson}, {Carey}, {Polat}, {Feng}, {Moore}, {VanderPlas}, {Laxalde}, {Perktold}, {Cimrman}, {Henriksen}, {Quintero}, {Harris}, {Archibald}, {Ribeiro}, {Pedregosa}, {van Mulbregt}, \& {SciPy 1. 0 Contributors}}]{Scipy}
{Virtanen}, P., {Gommers}, R., {Oliphant}, T.~E., {et~al.} 2020, Nature Methods, 17, 261

\bibitem[{{Voggel} {et~al.}(2020){Voggel}, {Seth}, {Sand}, {Hughes}, {Strader}, {Crnojevic}, \& {Caldwell}}]{Voggel+20_UCDs_GCs_differences_CenA}
{Voggel}, K.~T., {Seth}, A.~C., {Sand}, D.~J., {et~al.} 2020, \apj, 899, 140

\bibitem[{{Walsh} {et~al.}(2015){Walsh}, {Rejkuba}, \& {Walton}}]{Walsh+15}
{Walsh}, J.~R., {Rejkuba}, M., \& {Walton}, N.~A. 2015, aap, 574, A109

\bibitem[{{Wang} {et~al.}(2020){Wang}, {Hammer}, {Rejkuba}, {Crnojevi{\'c}}, \& {Yang}}]{Wang+20_CenAmergerhistory}
{Wang}, J., {Hammer}, F., {Rejkuba}, M., {Crnojevi{\'c}}, D., \& {Yang}, Y. 2020, mnras, 498, 2766

\bibitem[{{Watkins} {et~al.}(2013){Watkins}, {van de Ven}, {den Brok}, \& {van den Bosch}}]{Watkins+13}
{Watkins}, L.~L., {van de Ven}, G., {den Brok}, M., \& {van den Bosch}, R. C.~E. 2013, mnras, 436, 2598

\bibitem[{{Weijmans} {et~al.}(2014){Weijmans}, {de Zeeuw}, {Emsellem}, {Krajnovi{\'c}}, {Lablanche}, {Alatalo}, {Blitz}, {Bois}, {Bournaud}, {Bureau}, {Cappellari}, {Crocker}, {Davies}, {Davis}, {Duc}, {Khochfar}, {Kuntschner}, {McDermid}, {Morganti}, {Naab}, {Oosterloo}, {Sarzi}, {Scott}, {Serra}, {Verdoes Kleijn}, \& {Young}}]{Weijmans+14_ATLAS3D_intrinsic_shapes}
{Weijmans}, A.-M., {de Zeeuw}, P.~T., {Emsellem}, E., {et~al.} 2014, \mnras, 444, 3340

\bibitem[{{Woodley} {et~al.}(2010{\natexlab{a}}){Woodley}, {G{\'o}mez}, {Harris}, {Geisler}, \& {Harris}}]{Woodley+10_Kin_and_mass}
{Woodley}, K.~A., {G{\'o}mez}, M., {Harris}, W.~E., {Geisler}, D., \& {Harris}, G. L.~H. 2010{\natexlab{a}}, \aj, 139, 1871

\bibitem[{{Woodley} {et~al.}(2007){Woodley}, {Harris}, {Beasley}, {Peng}, {Bridges}, {Forbes}, \& {Harris}}]{Woodley+07_PNe_GC_mass_estimates}
{Woodley}, K.~A., {Harris}, W.~E., {Beasley}, M.~A., {et~al.} 2007, \aj, 134, 494

\bibitem[{{Woodley} {et~al.}(2010{\natexlab{b}}){Woodley}, {Harris}, {Puzia}, {G{\'o}mez}, {Harris}, \& {Geisler}}]{Woodley+10_Ages_metallicities}
{Woodley}, K.~A., {Harris}, W.~E., {Puzia}, T.~H., {et~al.} 2010{\natexlab{b}}, \apj, 708, 1335

\bibitem[{{Zemp} {et~al.}(2011){Zemp}, {Gnedin}, {Gnedin}, \& {Kravtsov}}]{Zemp+11_shapes_DM_halo_method}
{Zemp}, M., {Gnedin}, O.~Y., {Gnedin}, N.~Y., \& {Kravtsov}, A.~V. 2011, \apjs, 197, 30

\bibitem[{{Zentner} {et~al.}(2005){Zentner}, {Kravtsov}, {Gnedin}, \& {Klypin}}]{Zentner+05_Anisotropic_distribution_of_satellites}
{Zentner}, A.~R., {Kravtsov}, A.~V., {Gnedin}, O.~Y., \& {Klypin}, A.~A. 2005, \apj, 629, 219

\bibitem[{{Zhu} {et~al.}(2016){Zhu}, {Romanowsky}, {van de Ven}, {Long}, {Watkins}, {Pota}, {Napolitano}, {Forbes}, {Brodie}, \& {Foster}}]{Zhu+16}
{Zhu}, L., {Romanowsky}, A.~J., {van de Ven}, G., {et~al.} 2016, mnras, 462, 4001

\end{thebibliography}
%

\begin{appendix}

\section{GC tracer density}\label{ch:App_TD_MGEs}

For the binning, we first define bin edges based on the closest and outermost GC in the bright photometric population.
We use those as limits to generate 16 equidistant bins in log space.
We then apply these bin edges to the Subpopulation C and in each bin compute the centre as the median position of GCs in a given bin.
The uncertainties are given by the Poisson counting noise in each bin.
Bins with fewer than 3 GCs are masked out.

The priors for the fitting of the analytic profile are uniform in the following regions:
$\log R_e \in [1,3]~ {\rm arcsec}$,  $\alpha, n \in [0.2 ,6]$ and 
$\log \Sigma_S,\ \log \Sigma_P \in [-10,3]~ {\rm arcsec}^{-2}$.
These are listed in the Tables \ref{tab:app_MGE_GCs_blue} for the blue GCs and \ref{tab:app_MGE_GCs_red} for the red GCs.
The three columns show the parameters of the MGE: the central number density ($N_{\rm GC}\ {\rm pc}^{-2}$), the width along the major axis ($\sigma$) in arcsec, and the last is the flattening of the Gaussian component.

\begin{table}
\caption{MGE for the \textbf{blue} GCs of subpopulation C in the input format needed for CJAM. The first column shows the sorted number of the Gaussian components. Column~2 the central number density, Col.~3 the major axis dispersion of the Gaussian and Col.~4 the projected flattening.}\label{tab:app_MGE_GCs_blue}
\centering
\begin{tabular}{cccc}
\hline\hline
n & $N_{\rm GC}\ {\rm pc}^{-2}$ & arcsec & $q'$ \\
\hline
1 & 1.352e-07 & 70.71 & 0.82 \\
2 & 5.567e-08 & 119.24 & 0.82 \\
3 & 1.423e-09 & 192.91 & 0.82 \\
4 & 1.090e-07 & 230.29 & 0.82 \\
5 & 1.062e-08 & 436.06 & 0.82 \\
6 & 2.145e-09 & 470.45 & 0.82 \\
7 & 1.405e-09 & 779.61 & 0.82 \\
8 & 1.831e-09 & 1012.21 & 0.82 \\
9 & 1.373e-09 & 1939.49& 0.82 \\
10 & 7.152e-10 & 4612.45& 0.82 \\
11 & 3.477e-10 & 18257.41& 0.82 \\
\hline
\end{tabular}
\end{table}

\begin{table}

\caption{MGE for the \textbf{red} GCs of subpopulation C in the input format needed for CJAM. The column structure corresponds to the \autoref{tab:app_MGE_GCs_blue}.}\label{tab:app_MGE_GCs_red}
\centering
\begin{tabular}{cccc}
\hline \hline
n & $N_{\rm GC}\ {\rm pc}^{-2}$ & arcsec & $q'$ \\
\hline
1 & 2.241e-07 & 72.37 & 0.72 \\
2 & 9.751e-08 & 137.25 & 0.72 \\
3 & 3.466e-08 & 216.51 & 0.72 \\
4 & 1.886e-08 & 277.48 & 0.72 \\
5 & 9.599e-09 & 482.48 & 0.72 \\
6 & 3.424e-09 & 878.25 & 0.72 \\
7 & 7.064e-10 & 1182.39& 0.72 \\
8 & 2.624e-11 & 1635.44 & 0.72 \\
9 & 1.589e-09 & 1972.49 & 0.72 \\
10 & 8.265e-10 & 4678.81 & 0.72 \\
11 & 3.952e-10 & 18257.41& 0.72 \\
\hline
\end{tabular}
\end{table}

The median profile was computed by randomly drawing 1000 samples from the posterior, evaluating the profile given by Eq.~\ref{eq:Sersic_profile} and Eq.~\ref{eq:PowerLaw_profile} and computing the median and the percentiles.

\section{Mass - concentration relation}\label{ch:App_m200_c200}

\begin{figure}[h]
    \centering
    \includegraphics[width=\columnwidth]{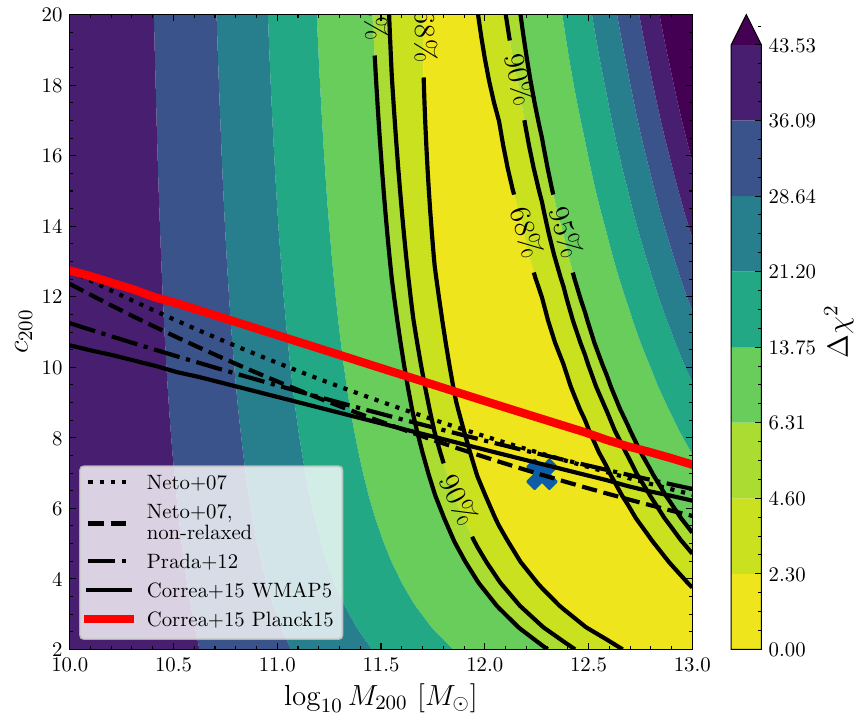}
    \caption{$\Delta \chi^2$ surface for \logM$ - c_{200}$ plane in iteration 4 (see \autoref{tab:2D_grid_parameters} for a summary of the parameters).
The black contour lines show the 68th, 90th and 95th percentile confidence levels, with minimal $\chi^2$ solution shown with the blue cross. 
Dotted, dashed, dot-dashed black lines and solid red and black lines show the different literature $M_{200}-c_{200}$.
The solid lines show the $M_{200}-c_{200}$ computed with \textsc{commah} code from \cite{Correa+15_M200-c200_relation} using the cosmological parameters from \cite{Planck+15_cosmological_parameters} in red and the cosmological parameters from \citet{WMAP+09} in black.}
    \label{fig:2D_mass_concentration_relation}
\end{figure}

In the 2D grid model set-up there is a strong degeneracy between the total DM mass and the halo concentration.
Because of it and in order to reduce the number of free parameters we introduced the literature relation of $M_{200}-c_{200}$ at iteration step 5.
The 2D $\chi^2$ surface in these two parameters is shown in Fig.~\ref{fig:2D_mass_concentration_relation} with the location of the minimum in the statistics highlighted by the blue cross.
The literature $M_{200}-c_{200}$ are shown as black dotted and dashed lines \citep[for relaxed and non-relaxed haloes respectively]{Neto+07_c200-M200_uncertainties}, dash-dotted line from \cite{Prada+12_M200-c200} and black and red solid lines from \cite{Correa+15_M200-c200_relation} using cosmological parameters from two different studies: \citet[WMAP5]{WMAP+09} and \citet[Planck15]{Planck+15_cosmological_parameters}.
We adopt the latter relation highlighted in red.

\section{Convergence of the $\chi^2$ solution}\label{ch:App_convergence_grid_set_up}

\begin{figure}[h]
    \centering
    \includegraphics[width=\columnwidth]{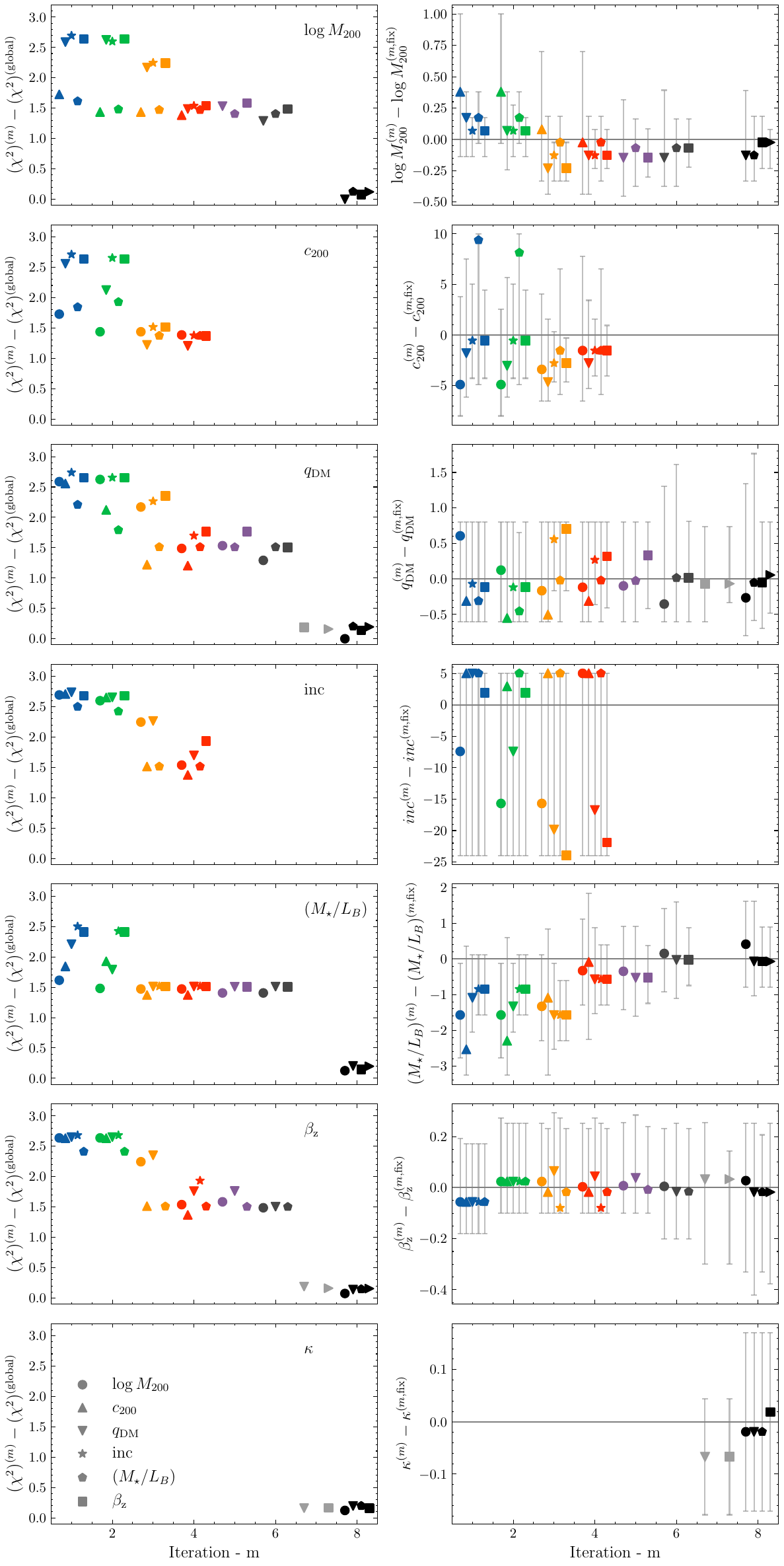}
    \caption{Convergence of the parameters and minimal $\chi^2$ for the \textbf{blue} GCs. The symbol shapes and colours correspond to those in Fig.~\ref{fig:2D_minChi2_sol_comparison} (and are also summarised in the bottom left panel). The panels in the left column show the evolution of the minimal value of the statistics and the right panel the value of the parameter at the minimum  $\chi^2$.}
    \label{fig:2D_minChi2_sol_comparison_blue}
\end{figure}

\begin{figure}[h]
    \centering
    \includegraphics[width=\columnwidth]{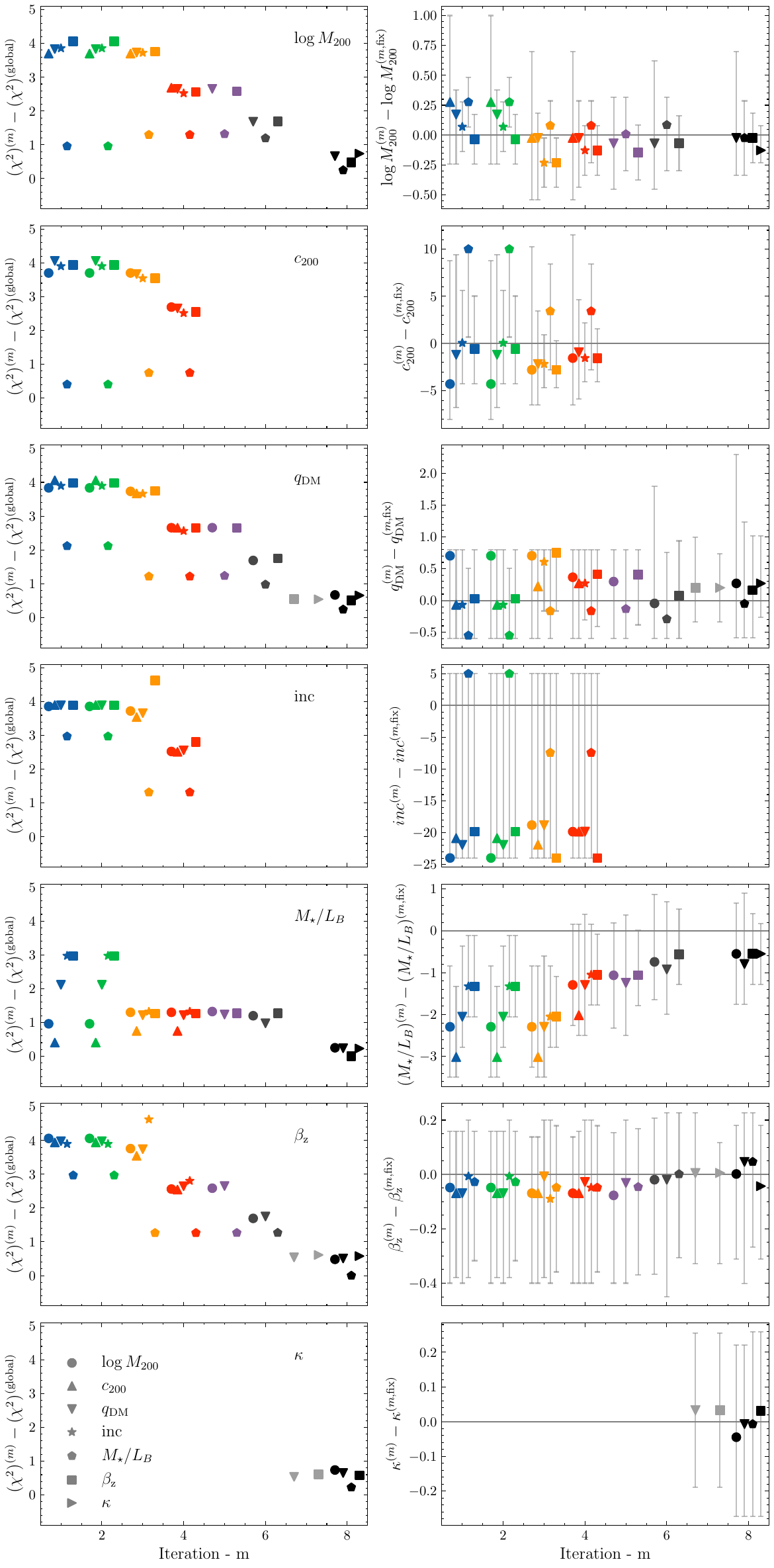}
    \caption{Convergence of the parameters and minimal $\chi^2$ for the \textbf{red} GCs. The symbols and colours correspond to those in Fig.~\ref{fig:2D_minChi2_sol_comparison} and columns are explained in Fig.~\ref{fig:2D_minChi2_sol_comparison_blue}. }
    \label{fig:2D_minChi2_sol_comparison_red}
\end{figure}

For completeness, we show the convergence of the best-fit solution for the blue GCs in Fig.~\ref{fig:2D_minChi2_sol_comparison_blue} and red GCs in Fig.~\ref{fig:2D_minChi2_sol_comparison_red}. 
\autoref{tab:2D_grid_parameters} shows how the parameters were varied between the iteration steps.


\newpage

\section{Comparing oblate and prolate models}\label{ap:2D_plots}

\begin{figure}[h]
    \centering
    \includegraphics[width=\columnwidth]{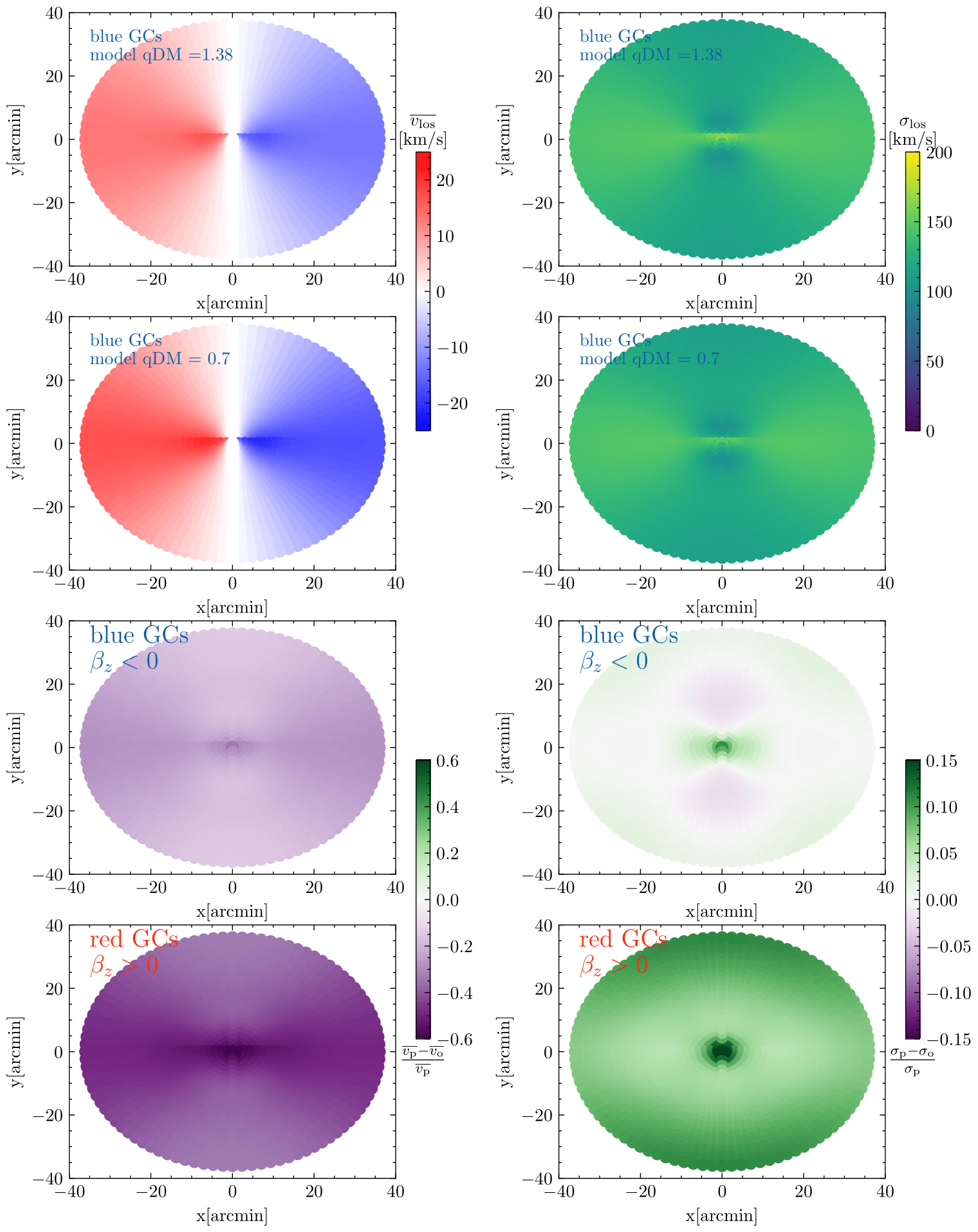}
    \caption{Comparison between the first and second moment velocity maps for axisymmetric Jeans modelling in the left and right column, respectively. 
    The first two rows show the best-fit models for blue GCs, with the best fit $q_{\rm DM} = 1.38$ in the top $q_{\rm DM} = 0.7$ below. 
    The bottom two rows show the residuals for the blue GCs on the top and red GCs on the bottom.
    The parameters used to evaluate the model were taken from \autoref{tab:final_results_marginalisation} and for $q_{\rm DM} = 0.7$ we take the other parameters at the minimal $\chi^2$ based on the 2D $\chi^2$ grids as shown in Fig.~\ref{fig:Final_corner_blue}.}
    \label{fig:Moment_maps_oblate_prolate_res}
\end{figure}

The results of the comparison between our best fit prolate model with $q_{\rm DM} = 1.38$ and an oblate model, where we enforced $q_{\rm DM} = 0.7$ are shown in Fig.~\ref{fig:Moment_maps_oblate_prolate_res} for blue and red GCs.
To make these plots we evaluated CJAM on a grid of points linearly spaced in R and $\phi$ for the best fit parameters shown in \autoref{tab:final_results_marginalisation}.
The left column shows the maps of the first moment projected along the line-of-sight and the right column shows the second moment.
The first two rows show the model results from the blue GCs best-fit solutions for the prolate (top) and the oblate (bottom) cases same as for Fig.~\ref{fig:comparing_oblate_prolate_best_fit}.
The bottom two rows show the residual of the prolate map w.r.t. the oblate case for the blue GC in the top and red GCs in the bottom.
The streaks along the major axis are an artefact of the plotting and not a real feature.

The 2D maps reveal the full complexity of the velocity moments which can be probed by discrete dynamical modelling.
The residuals both for blue (in the third row) and red GCs (in the bottom row), show that the best fit oblate model favours a stronger rotational signature. 
The residuals in the left column are higher for red than blue GCs and they consistently show higher residuals along the semimajor axis.
This means that the gradient of the mean velocity from the minor to major axis depends on the flattening of the dark matter halo.
Velocity dispersion maps in the right column show a similarly complex signature, with the opposite sign of the residuals. 
The prolate solution shows on average higher dispersion compared with the best-fit oblate solution.
These maps highlight the importance and strength of discrete dynamical modelling, where the position of each GC is used to constrain the 2D velocity distribution, which shows distinct signatures for oblate and prolate solutions.



\end{appendix}

\end{document}